\documentclass[12pt]{article}
\usepackage{amsmath}
\usepackage{amssymb,verbatim,xcolor}
\usepackage{bm,lscape}
\usepackage{graphicx}
\usepackage{wasysym}
\usepackage{theorem}
\usepackage{float}
\usepackage{multirow}
\usepackage{tabularx}
\usepackage{subcaption}
\usepackage{pdflscape}

\usepackage{natbib}

\allowdisplaybreaks

\def\bSig\mathbf{\Sigma}

\newcommand{\E}{\mathbb{E}}

\newcommand{\bw}{\mathbf{w}}
\newcommand{\bs}{\mathbf{s}}
\newcommand{\be}{\boldsymbol{\epsilon}}
\newcommand{\bo}{\boldsymbol{\omega}}

\newcommand{\ddr}{\mathrm{d}}

\newcommand{\beq}{\begin{equation}}
\newcommand{\eeq}{\end{equation}}
\newcommand{\iid}{\stackrel{\mathrm{iid}}{\sim}}
\newcommand{\ind}{\stackrel{\mathrm{ind}}{\sim}}

\newcommand{\Rea}{{\mathbb{R}}}

\DeclareMathOperator{\Var}{Var}

\theoremstyle{plain}

\allowdisplaybreaks

\begin{document}
\title{
Bayesian nonparametric temporal dynamic clustering via autoregressive Dirichlet priors}

\author{Maria De Iorio,  Stefano Favaro, Alessandra Guglielmi, and Lifeng Ye\\
Yale-NUS College \& UCL, Universit\`a di Torino,Politecnico di Milano and UCL}
\maketitle

\begin{abstract}
In this paper we consider the problem of dynamic clustering, where cluster memberships may change over time and clusters may split and merge over time, thus creating new clusters and destroying existing ones. We propose a Bayesian nonparametric approach to dynamic clustering via mixture modeling. Our approach relies on a novel time-dependent nonparametric prior defined by combining: i) a copula-based transformation of a Gaussian autoregressive process; ii) the stick-breaking construction of the Dirichlet process. Posterior inference is performed through a particle Markov chain Monte Carlo algorithm which is simple, computationally efficient and scalable to massive datasets. Advantages of the proposed approach include flexibility in applications, ease of computations and  interpretability. We present an application of our dynamic Bayesian nonparametric mixture model to the study the temporal dynamics of gender stereotypes in adjectives and occupations in the 20th and 21st centuries in the United States. Moreover, to highlight the flexibility of our model we present additional applications to time-dependent data with covariates and with spatial structure.
\end{abstract}

\medskip



\section{Introduction}
\label{sec:intro}
In this paper we consider a dynamic clustering problem where cluster memberships may change over time and clusters may split and merge over time, thus creating new clusters and destroying existing ones. We present a Bayesian nonparametric (BNP) approach to  dynamic clustering, which relies on a novel prior distribution for a sequence of discrete random probability measures (RPMs) $(G_{t})_{t\geq1}$, with $t$ being a discrete time. Our prior construction combines a copula-based transformation of a Gaussian autoregressive process of order $1$ \citep{guolo2014beta} with the stick-breaking construction of the Dirichlet process (DP) \citep{Fergus73}. The resulting $(G_{t})_{t\geq1}$ is a dependent DP \citep{maceachern2000dependent} such that: i) $(G_{t})_{t\geq1}$ has an autoregressive structure of order $1$; ii) $G_{t}$ is a DP, for any $t\geq1$. We apply the law of $(G_{t})_{t\geq1}$ as a prior for BNP mixture modeling with time-dependent data $(Y_{t1},\ldots,Y_{tn})$, $t=1,\ldots,T$, namely we assume the model
\begin{align*}
Y_{tj}   \mid  \theta_{tj} &\ind k(\cdot\,;\, \theta_{tj}) \quad j=1,\ldots , n\\
 \theta_{t1},\ldots, \theta_{tn} \mid G_t &\iid G_t \quad t=1,2,\ldots, T\notag
\end{align*}
with $k(\cdot\,;\,\theta)$ being a density function for $\theta\in\Theta\subset\Rea^p$, and $(G_{t})_{t\geq1}$ being an autoregressive DP of order $1$ (AR1-DP). This is a dynamic counterpart of the celebrated DP mixture model \citep{lo1984class}.

Our BNP dynamic mixture model clusters observations at each time point, with the clustering structure changing over time. This is an effective approach to dynamic clustering, where the clustering configuration at time $t+1$ depends on the clustering configuration at time $t$. The degree of dependence covered by the AR1-DP is flexible, ranging from independence between time periods to exact same clustering structure across time periods. The proposed BNP dynamic mixture model allows for posterior inference at a specific time point to borrow strength from the clustering distribution at different times; also, it allows to include covariate information, when available. Posterior inference can be performed through a particle Markov chain Monte Carlo (MCMC) algorithm \citep{andrieu2010particle} which is simple, computationally efficient and scalable to massive datasets. 

There exists a plethora of works on dependent RPMs in discrete time, e.g., \cite{GriSte06}, \cite{caron_etal2007}, \cite{rodriguez2008bayesian}, \cite{taddy2010autoregressive}, \cite{rodriguez2011nonparametric}, \cite{GriSte11}, \cite{nieto2012time}, \cite{DiLucca_etal13}, \cite{Bassetti_etal14}, \cite{xiao2015modeling}, \cite{gutierrez2016time} and  \cite{deyoreo2018modeling}. Among them, the RPMs proposed in \cite{taddy2010autoregressive} and  \cite{deyoreo2018modeling} are closely related to our AR1-DP. In particular, \cite{deyoreo2018modeling} developed a BNP model for ordinal regressions evolving in time. Their model relies on a dependent RPM $(G_{t})_{t\geq1}$ where, similarly to our approach, the dependence among the $G_{t}$'s is introduced by transforming Gaussian autoregressive processes of order $1$. However, in contrast with \cite{deyoreo2018modeling}, we show that our copula-based transformation allows for a more flexible degree of dependence among the $G_{t}$'s.

We present an application of our BNP dynamic mixture model to the study the temporal dynamics of gender stereotypes in adjectives and occupations. This is a important topic in many disciplines at the intersection between statistics and social sciences. Natural language processing, such as language modeling and feature learning, are standard tools to discover and understand gender stereotypes \citep{holmes2008handbook}. The recent work of \cite{Garg_etal_2018} exploits word embeddings, a common tool in natural language processing, to study the dynamics of gender and ethnic stereotypes in the 20th and 21st centuries in the United States. Word embedding map (transform) each English word into a high-dimensional real-valued vector, whose geometry captures local and global semantic relations between the words, e.g., vectors being closer together has been shown to correspond to more similar words \citep{collobert2011natural}. These models are trained on large corpora of text, such as Google News articles or Wikipedia, and are known to capture relationships not found through co-occurrence analysis. 

We follow the work of \cite{Garg_etal_2018}, and we consider word embeddings trained on Corpus of Historical American English (COHA) \citep{coha} for three decades $t=1900, 1950, 2000$. In particular, by using  these embeddings together with the list of words provided by  \cite{Garg_etal_2018}, we obtain data for (standardized) adjective's biases for women and (standardized) occupation's biases for women, for each word in the corresponding list. A negative value for the bias means that the embedding more closely associates the (occupation or adjective) word with men, because the distance between the word is closer to men than women. Hence, gender bias corresponds to either negative or positive values of the embedding bias. We apply our BNP dynamic mixture model to the time-dependent data for adjective's bias and for occupation's bias to quantify how  stereotypes towards women have evolved in the United States.


The paper is structured as follows. In Section~\ref{sec:sbar} we introduce the AR1-DP, whereas in Section~\ref{sec:mixturemodel} we apply it to BNP mixture modeling with time-dependent data, and we present the MCMC algorithm for posterior inference. Section~\ref{sec:comp} contains a discussion between the AR1-DP and the priors of \cite{taddy2010autoregressive} and  \cite{deyoreo2018modeling}. In Section~\ref{sec:genderbias} we apply our model to the study of the dynamics of gender stereotypes, in words and occupations, in the 20th and 21st centuries in the United States. Sections~\ref{sec:dose} and \ref{sec:breastcancer} contain additional applications to time-dependent data with covariates and with spatial structure. Concluding remarks and an outline of future works are in Section~\ref{sec:conclusion}. Appendix~A contains details on the particle MCMC step; Appendix~B includes the list of words for the gender bias application, and definitions of the occupational and adjective bias; Appendix~C presents an extensive simulation study;  Appendix~D includes extra plots. 


\section{Autoregressive Dirichlet processes}
\label{sec:sbar}

Let $\text{N}(\mu,\sigma^{2})$ denote a Gaussian distribution with mean $\mu$ and variance $\sigma^{2}$, let $\epsilon\sim\text{N}(0,1)$ and let $\Phi(\cdot)$ denote the cumulative distribution function (CDF) of $\epsilon$. Recall that  if $F(\cdot\,;\,a,b)$ is the CDF of a Beta random variable with parameter $(a,b)$, then $Y=F^{-1} (\Phi(\epsilon);a,b )$ is a Beta random variable with parameter $(a,b)$. Similarly to \cite{guolo2014beta}, we consider a discrete time stochastic process $\boldsymbol{\epsilon}= (\epsilon_t)_{t\geq 1}$ defined as follows
\begin{equation}\label{eq:epst}
\epsilon_1 \sim \text{N}(0,1) \quad\text{and}\quad \epsilon_t = \psi \epsilon_{t-1}+\eta_t \qquad t\geq 2 
\end{equation}
where $\psi\in(-1,1)$ and $(\eta_{t})_{t>1}$ are independent and identically distributed (i.i.d.) as $\text{N}(0,1-\psi^2)$. That is, $\boldsymbol{\epsilon}$ is an autoregressive process of order $1$ with parameter $\psi$, with $\epsilon_t\sim \text{N}(0,1)$; for brevity, $\boldsymbol{\epsilon}\sim\textrm{AR}(1;\psi)$. Let $(\boldsymbol{\epsilon}_{l})_{l\geq1}$ be i.i.d. such that $\boldsymbol{\epsilon}_{l}\sim\textrm{AR}(1;\psi)$ and let
\begin{equation}\label{eq:dyn_beta}
\xi_{tl}= F^{-1} (\Phi(\epsilon_{tl}); a,b), 
\end{equation}
for any $t\geq1$ and $l\geq1$. Because of the autoregressive structure of $\boldsymbol{\epsilon}_{l}$, $\xi_{tl}$ depends on $\xi_{(t-1) l}$, with the parameter $\psi$ controlling the dependence. In particular, for any fixed $l\geq1$, the assumption $\psi=0$ corresponds to the assumption of independence among the $\xi_{tl}$'s. Furthermore, for every $t\geq1$, $\xi_{tl}$ is independent of $\xi_{th}$ if $l\neq h$, and they have the same distribution.

Let $G$ be a DP on $\Theta \subset \Rea ^p$ with parameter $(M,G_{0})$, where $G_{0}$ is a non-atomic (base) distribution on $\Theta$ and $M>0$ is the scale parameter; for brevity $G\sim \text{DP}(M,G_0)$. It is known from \cite{Sethuraman1994} that $G=\sum_{h\geq1} w_h \delta_{\theta_h}$ where: i) $(w_{h})_{h\geq1}$ are such that $w_1= \xi_1$ and $w_h = \xi_h \prod_{1\leq l\leq h-1}(1-\xi_l)$ for $h>1$, with $(\xi_l)_{l\geq 1}$ being i.i.d. as a Beta distribution with parameter $(1,M)$; ii) $(\theta_{h})_{h\geq1}$ are i.i.d. as $G_{0}$, and independent of $(\xi_l)_{l\geq 1}$. This is the stick-breaking construction of $G\sim\text{DP}(M,G_0)$. We make use of \eqref{eq:dyn_beta} to extend the stick-breaking construction of the DP and defining a sequence of dependent RPMs $(G_t)_{t\geq1}$, with $t$ being a discrete time, such that $G_{t}\sim\text{DP}(M,G_0)$. As such, let
\begin{equation}\label{eq:GT}
G_t =\sum_{h\geq1} w_{th} \delta_{\theta_h},
\end{equation}
where $w_{t1}= \xi_{t1}$ and $w_{th} = \xi_{th} \prod_{l=1}^{h-1} (1-\xi_{tl} )$, for  $h>2$, with $(\xi_{tl})_{t\geq1,l\geq1}$ and $(\theta_{h})_{h\geq1}$ such that: i) $(\xi_{tl})_{t\geq1,l\geq1}$ are distributed as in   \eqref{eq:dyn_beta}, with parameters $a=1$ and $b=M$; ii) $(\theta_{h})_{h\geq1}$ are i.i.d. with common distribution $G_{0}$, and independent of  $(\xi_{tl})_{t\geq1,l\geq1}$. The time-dependent discrete RPM $(G_t)_{t\geq1}$ is referred to as the autoregressive DP of order $1$; for brevity, $(G_t)_{t\geq1}\sim\text{AR1-DP}(\psi,M,G_{0})$. $(G_t)_{t\geq1}$ is a dependent DP in the sense of \cite{maceachern2000dependent}. It is also an example of the  single atom dependent DPs of \cite{barrientos_etal2012}. 

Since $\epsilon_{tl}\sim\text{N}(0,1)$ and $F$ is the CDF of a Beta random variable with parameter $(1,M)$, then $\xi_{tl}$ is a Beta random variable with parameter $(1,M)$. This implies that if $(G_t)_{t\geq1}\sim\text{AR1-DP}(\psi,M,G_{0})$ then $G_{t}\sim\text{DP}(M,G_0)$ for any $t\geq1$. Observe that   \eqref{eq:dyn_beta}, whose dynamics in $t\geq1$ is driven by \eqref{eq:epst}, induces a dynamics in the sequence of RPMs $(G_t)_{t\geq1}$.  Most importantly, for every $l\geq1$ the stochastic process $(\xi_{tl})_{t\geq1}$ inherits the same autoregressive (order $1$) Markov structure of each stochastic process $\boldsymbol{\epsilon}_{l}$. Therefore $(G_t)_{t\geq1}\sim\text{AR1-DP}(\psi,M,G_{0})$ has an autoregressive (order $1$) Markov structure, with $\psi$ controlling the dependence among the $G_{t}$'s. The assumption $\psi=0$ corresponds to independence among the $G_{t}$'s. A natural extension of $(G_t)_{t\geq1}\sim\text{AR1-DP}(\psi,M,G_{0})$ arises by setting $F$ to be the CDF of a Beta random variable with parameter $(a_{t},b_{t})$, for any $t\geq1$. 

\begin{figure}[ht!]
\begin{center}
\includegraphics[width=0.4\textwidth, height=0.4\textwidth]{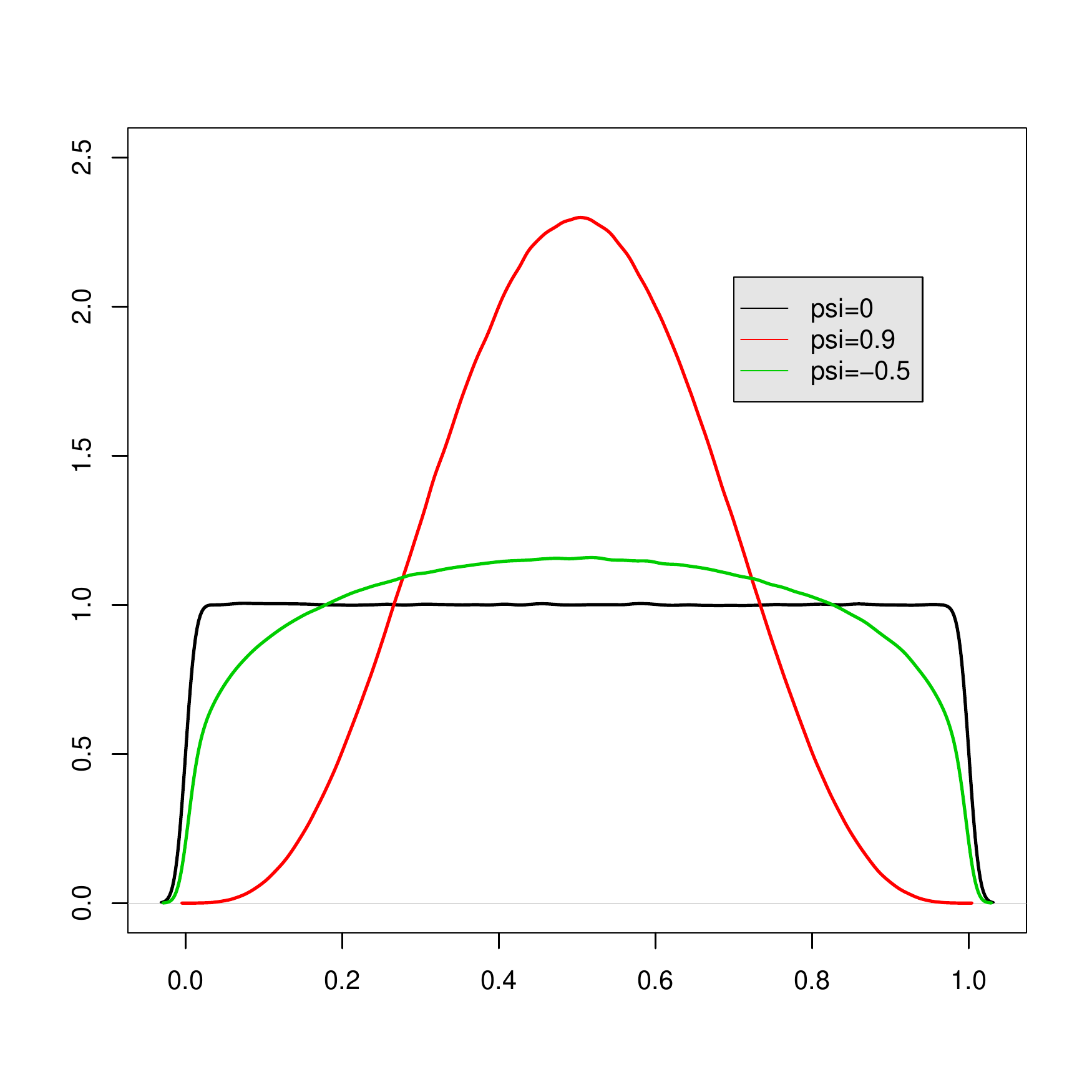}
\includegraphics[width=0.4\textwidth, height=0.4\textwidth]{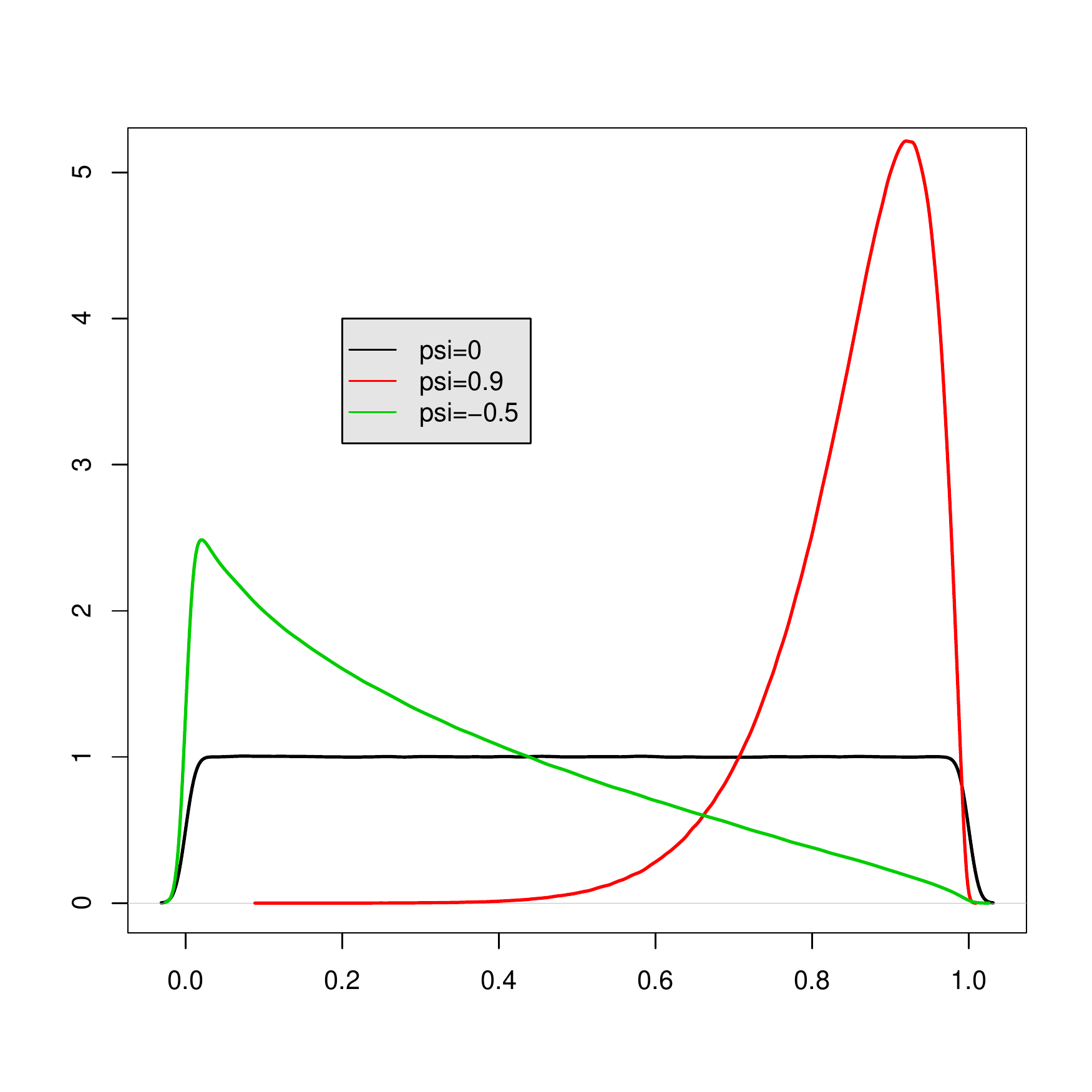}
\end{center}
\vspace{-0.5cm}
 \caption{\scriptsize{Conditional density of $\xi_2$ given $\xi_1=0.5$ (left) and given $\xi_1=0.9$ (right) for some values of $\psi$ and $M=1$. }}
 \label{fig:pippo}
\end{figure}

For ease of explanation we drop the sub-index $l$. From $\xi_1=1-\left( 1- \Phi(\epsilon_1)\right) ^{1/M}$ (see Equation \eqref{eq:dyn_beta}), we write $\epsilon_1 = \Phi^{-1}\left(1-(1-\xi_1)^M \right)$. From $ \xi_2=1-\left( 1- \Phi(\epsilon_2)\right) ^{1/M}$, using \eqref{eq:epst} and the expression of $\epsilon_1$, we have $\xi_2=  1-\left[ 1- \Phi(\psi\Phi^{-1}\left( 1-(1-\xi_1)^M\right)+\eta_2)\right] ^{1/M}$, where $\eta_2\sim  \text{N}(0,1-\psi^2)$.
Hence, the conditional distribution of $\xi_{2}$ given $\xi_{1}$ coincides with the distribution of $1 - \left(1-\Phi(Z)\right)^{1/M}$, where $Z\sim \text{N}\left( \psi\Phi^{-1}\left( 1-(1-\xi_1)^M\right),1-\psi^2\right)$. Along similar lines, the conditional distribution of $\xi_{t}$ given $\xi_{t-1}$ coincides with the distribution of
\begin{equation}\label{eq:YtgivenY{t-1}} 
1 - \left(1-\Phi(Z)\right)^{1/M},
\end{equation}
where $Z\sim \text{N}\left( \psi\Phi^{-1}\left( 1-(1-\xi_{t-1})^M\right),1-\psi^2\right)$. Equation \eqref{eq:YtgivenY{t-1}} is crucial for sampling from $(G_{t})_{t\geq1}\sim\text{AR1-DP}(\psi,M,G_{0})$. Figure \ref{fig:pippo} displays the conditional density function of $\xi_2$ given $\xi_1=0.5$ (left) and given $\xi_1=0.9$ (right) for different values of $\psi$ and $M=1$. Our construction is flexible, allowing for different shapes of the distribution. For $\psi=0$ the conditional distribution coincides with the marginal distribution and it is a Uniform distribution. 


\section{Autoregressive Dirichlet process mixture models}
\label{sec:mixturemodel}


Let $k(\cdot\,;\,\theta)$ denote a density function for $\theta\in\Theta\subset\Rea^p$, and $(G_{t})_{t\geq1}\sim\text{AR1-DP}(\psi,M,G_{0})$. We define a temporal dynamic version of the DP mixture model of \cite{lo1984class}. Precisely, we assume that data at time $t=1,2,\ldots,T$ are (conditionally) i.i.d. as the random density function
\begin{equation}
\label{eq:rand_dens}
f_t(y)=\int k(y\,;\, \theta)G_t(d\theta).
\end{equation}
Equivalently, we assume
\begin{align}
\label{eq:m_data}
Y_{tj}   \mid  \theta_{tj} &\ind k(\cdot\,;\, \theta_{tj}) \quad j=1,\ldots , n\\
\theta_{t1},\ldots, \theta_{tn} \mid G_t &\iid G_t \quad t=1,2,\ldots, T \label{eq:m_latent}\\
(G_t)_{t\geq1} &\sim \text{AR1-DP} (\psi,M,G_0).\label{eq:m_sbar}
 \end{align}
The model is completed with the specification of a prior on $\psi$ and $M$, whereas $G_{0}$ is typically fixed a priori. The model \eqref{eq:m_data} - \eqref{eq:m_sbar} is referred to as the AR1-DP mixture model.

Recall that a random sample $(\theta_{1},\ldots,\theta_{n})$ from $G\sim\text{DP}(M,G_{0})$ features $1\leq K_n\leq n$ distinct types, labelled by $(\theta_{1}^{\ast},\ldots,\theta_{K_n}^{\ast})$, with frequencies $(N_{1,n},\ldots,N_{K_n,n})$ such that $1\leq N_{i,n}\leq n$ and $\sum_{1\leq i\leq K_n}N_{i,n}=n$. In other terms, $(\theta_{1},\ldots,\theta_{n})$ induces a random partition $\Pi_{n}$ of $\{1,\ldots,n\}$ into $K_n$ blocks with sizes $(N_{1,n},\ldots,N_{K_n,n})$. The probability of any partition of $\{1,\ldots,n\}$ having $l$ blocks with corresponding frequency counts $(n_{1},\ldots,n_{l})$ is 
\begin{equation}\label{ewe}
p_{l,n}(n_{1},\ldots,n_{l})=\frac{M^{l}}{\prod_{i=0}^{n-1}(M+i)}\prod_{i=1}^{l}(n_{i}-1)!.
\end{equation}
See \cite{Antoniak74}. In our context, a random sample $(\theta_{t1},\ldots, \theta_{tn})$ from $G_{t}$, for $t=1,\ldots,T$ and with $(G_t)_{t\geq1}\sim \text{AR1-DP} (\psi,M,G_0)$, induces a collection $\{\Pi_{1,n},\ldots,\Pi_{T,n}\}$ of dependent random partitions of $\{1,\ldots,n\}$. Due to the definition  of $\text{AR1-DP} (\psi,M,G_0)$, $\Pi_{t,n}$ is distributed as \eqref{ewe}, for any $t=1,\ldots,T$. The number of blocks/clusters of $\Pi_{t,n}$, denoted by $K_{t,n}$, at each time $t$ is a random variable that is stationary, so that its prior marginal distribution will not change with $t$. We skip the subscript $t$, but also the subscript $n$ when there is no chance of misunderstanding, from this notation. Recall that the prior mean of the number $K$ of clusters for any $t$, given the total mass $M$, is $\sum_{1\leq i\leq n} M/(M+i-1)$.
 
We consider the problem of sampling from the posterior distribution of the AR1-DP mixture model. The design of a Gibbs sampler for such a problem is straightforward, once we truncate the infinite series \eqref{eq:GT} to $J$ terms and introduce allocation variables $s_{tj}$ for each of the latent variable $\theta_{tj}$. By using a latent variable representation, we can write
\begin{align}
\label{eq:trunc_model}
Y_{tj}   \mid  s_{tj}, \theta_{tj} &\ind k(\cdot\,;\,\theta_{s_{tj}}) \quad j=1,\ldots , n \nonumber\\
s_{tj}\mid  \bw_t  &\iid \sum_{h=1}^J w_{th}\delta_{h} \quad  j=1,\ldots,n\\
\theta_{tj} &\iid G_0,\nonumber 
\end{align}
where, for any $t=1,2,\ldots,T$, 
\begin{equation}
\label{eq:pesetti}
w_{t1}= \xi_{t1}\quad\text{and}\quad w_{tj} = \xi_{tj} \prod_{h=1}^{j-1} (1-\xi_{tl}) \quad j=2,\ldots,J-1,
\end{equation}
with $w_{tJ} = 1- \sum_{h=1}^{J-1} w_{th}$ and 
\begin{align}
&\xi_{tj} = 1-\left( 1- \Phi(\epsilon_{tj})\right)^{1/M} \quad   j=1,\ldots,J-1 \label{eq:nonlin}	\\
&\epsilon_{1j} \iid\text{N} (0,1) \quad   j=1,\ldots,J-1\\
\label{eq:trans}
\epsilon_{tj}\mid \epsilon_{t-1,j} , \psi  &\ind \text{N}(\psi \epsilon_{t-1 ,j},1 - \psi^2) \quad j=1, \ldots, J-1.
 \end{align}
Here $\bs_t=(s_{t1},\ldots,s_{tn})$ is the component of the mixture to which $(Y_{t1},\ldots,Y_{tn})$ are allocated at time $t$, $\bm\epsilon_t=(\epsilon_{1t},\ldots,\epsilon_{t J-1})$ is the latent autoregressive process, ${\bm\theta}_t=(\theta_{t1},\ldots,\theta_{tJ})$ are the component-specific parameters and $\bm w_t=(w_{1t},\ldots,w_{t J-1})$ are the weights of the components in $G_t$ (see Section~\ref{sec:sbar}). Thus the unknown parameters are $\{\bm\theta_t, \bs_t, \bw_t, \bm\epsilon_t, \psi, M\}$. 

We outline the MCMC scheme for sampling from the posterior distributions of $\{\bm\theta_t, \bs_t, \bw_t, \bm\epsilon_t, \psi, M\}$. Further details of the algorithm can be found in Appendix~A.
Note that while $\bm\theta$ is the same for each time period, the vector $\bs_t$ changes over time so that individuals can change clusters. The main steps of the MCMC algorithm are the following: 
\begin{enumerate}
\item sampling $\bm\theta$ given the rest: sample the non-allocated values of $\bm\theta$ from the base distribution $G_{0}$, i.e. $\theta_l \iid G_0$, and then update the allocated $\theta_l$ from the conditional distribution
$$p(\theta_{l}\,|\,\text{rest})\propto G_0(\ddr \theta)  \prod_{(t,j):s_{tj}=l} f(y_{tj} \,;\, \theta_{tj})  $$
 
\item sampling $(\bs_1,\ldots,\bs_T)$ given the rest: this distribution   can be factorized into the product of the distributions of each $\boldsymbol{s}_{t}$ given the rest; this is because, given $(\bw_1,\ldots,\bw_T)$, the allocations at different times are conditionally independent; for each $t, j$, 
$$p(s_{tj}\,|\,\text{rest}) \propto w_{tl} f(y_{tj}\,;\, \theta_l) $$
 
\item sampling $\{\psi, \bw_1,\ldots, \bw_T\}$ given the rest: this requires to sample from the conditional distribution of $\{\psi, \bw_1,\ldots, \bw_T\}$ given $(\bs_1 \ldots, \bs_T)$; this step requires to make use of a particle MCMC update, and it will be discussed more in details in Appendix~A. 
 
\end{enumerate}


\begin{figure}[ht]
\begin{center}
\includegraphics[width=0.24\textwidth]{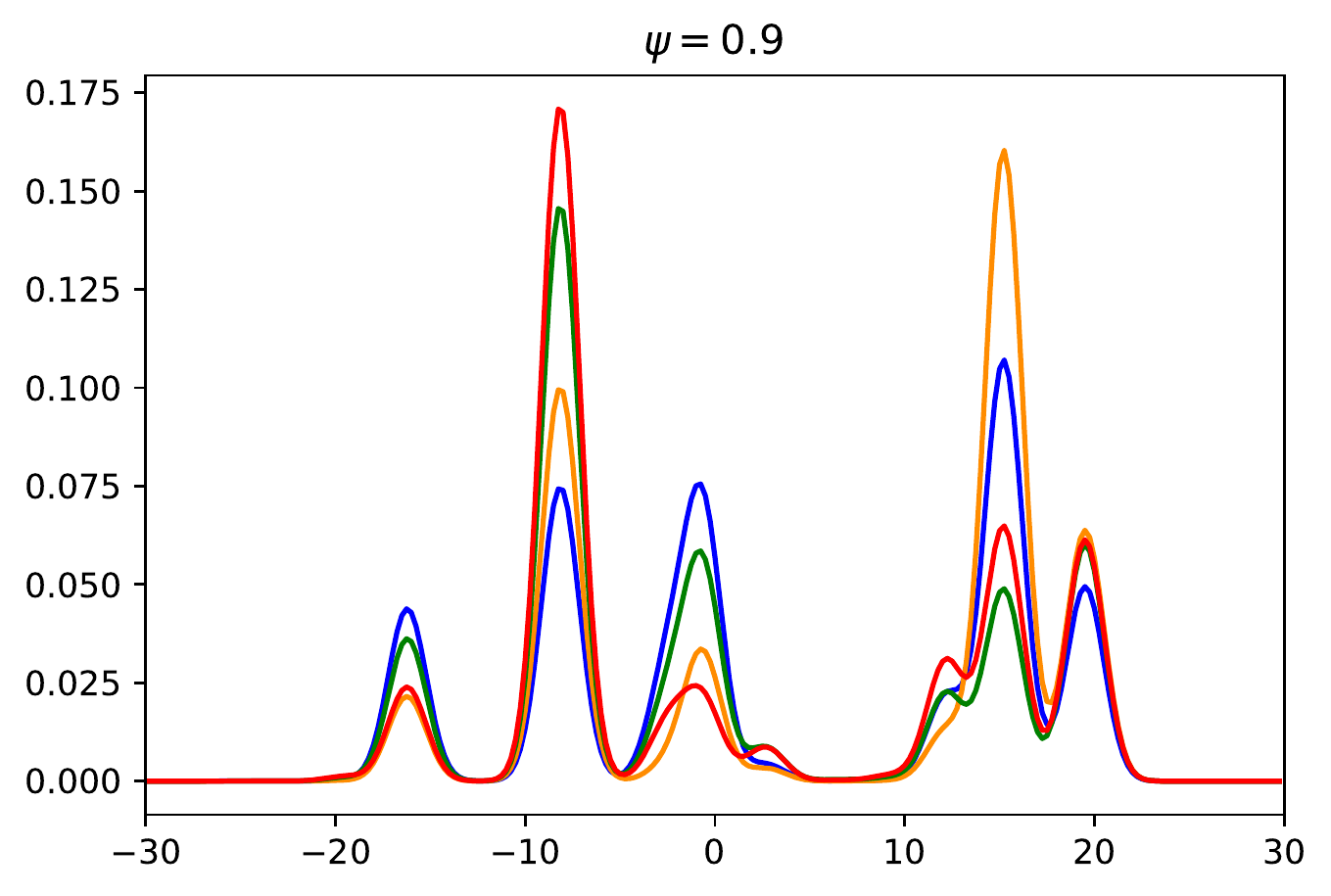}
\includegraphics[width=0.24\textwidth]{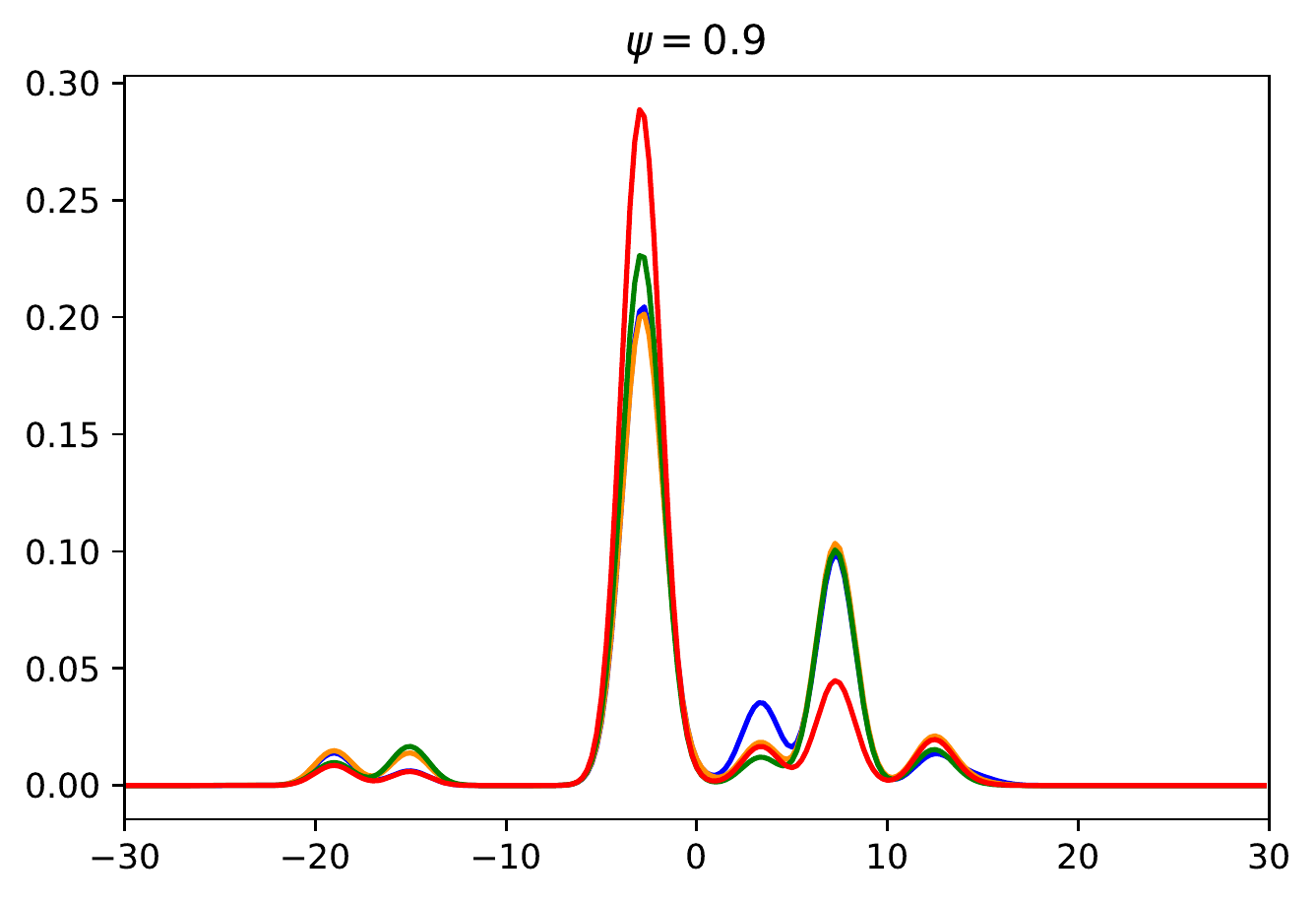}
\includegraphics[width=0.24\textwidth]{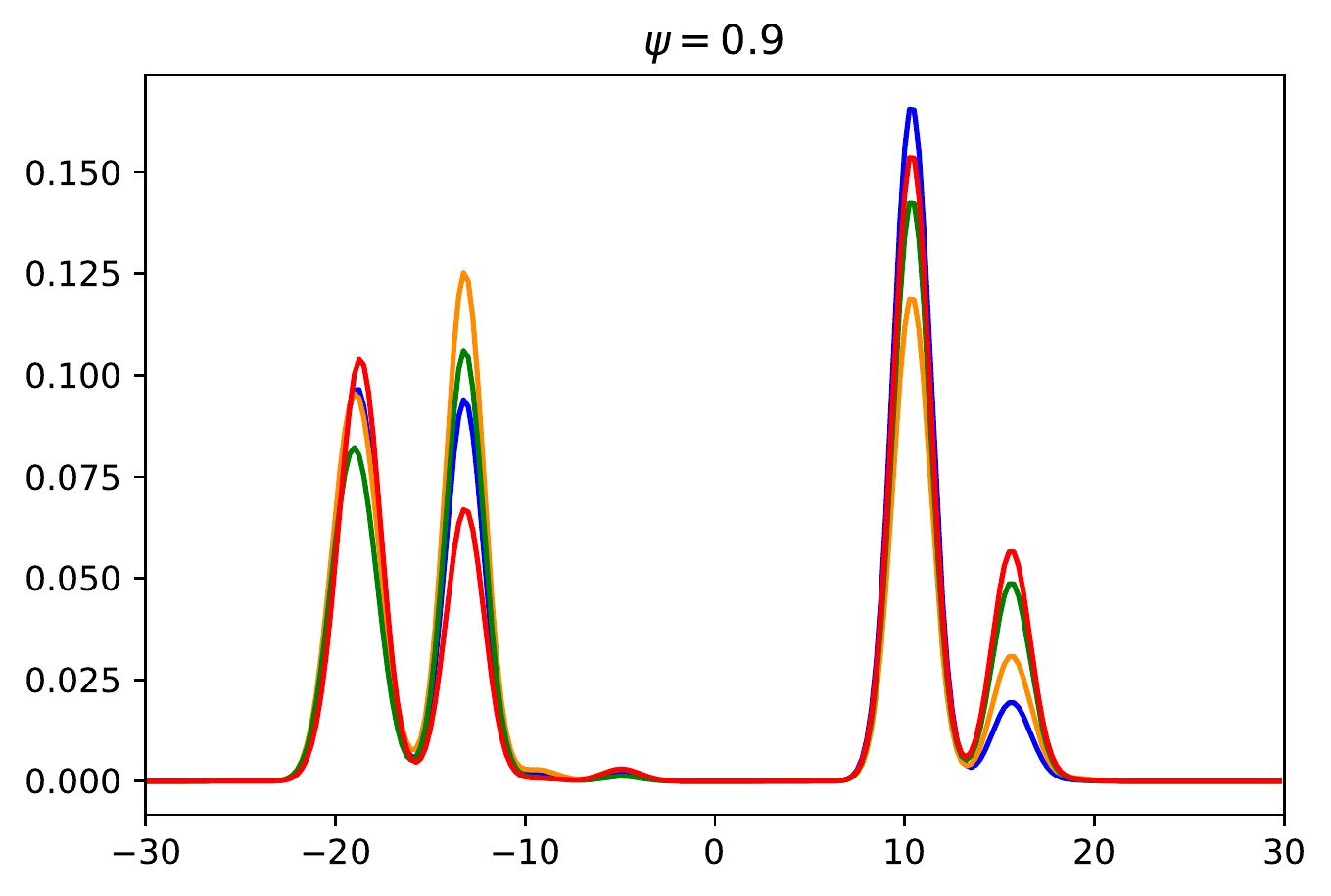}\\
\includegraphics[width=0.24\textwidth]{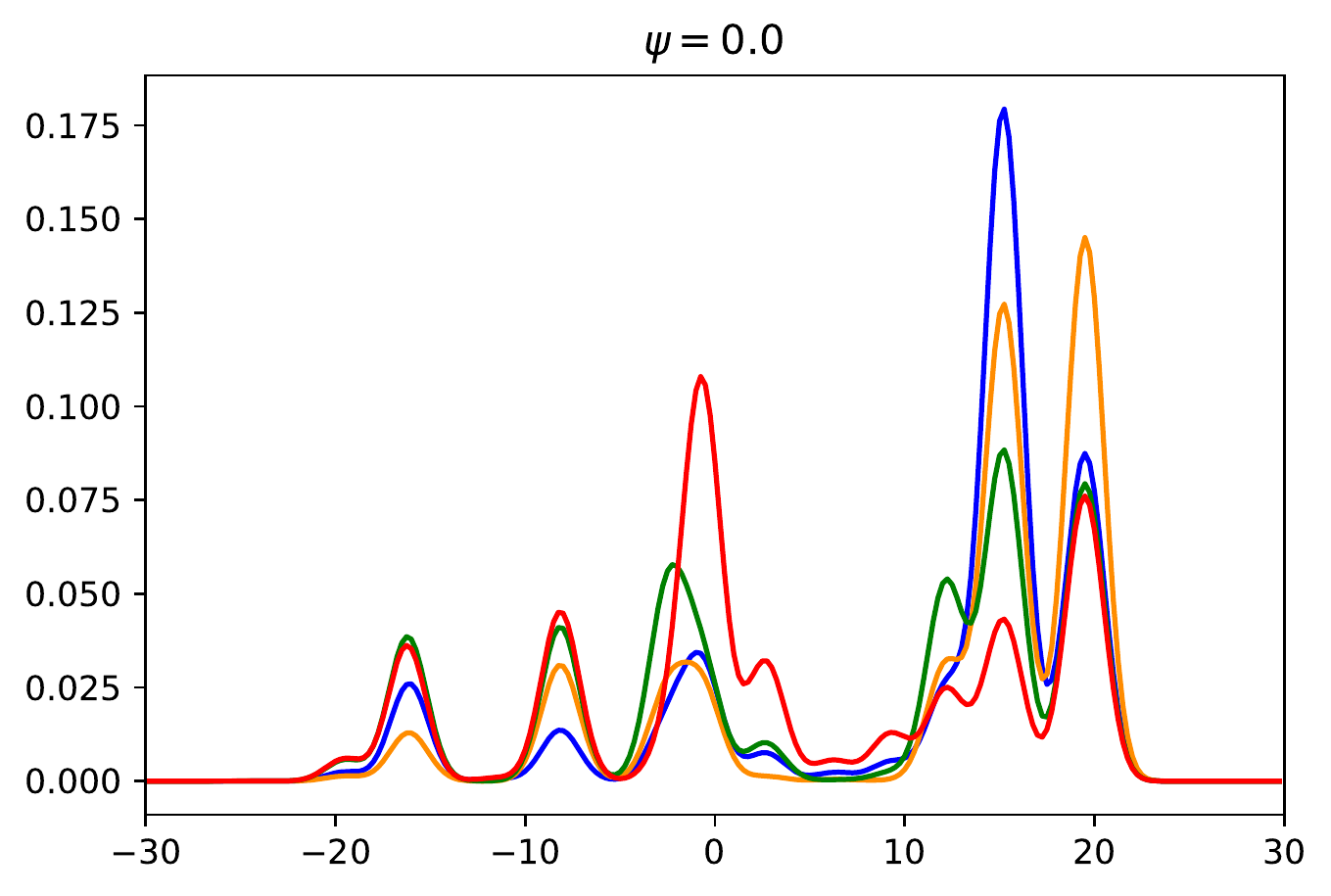}
\includegraphics[width=0.24\textwidth]{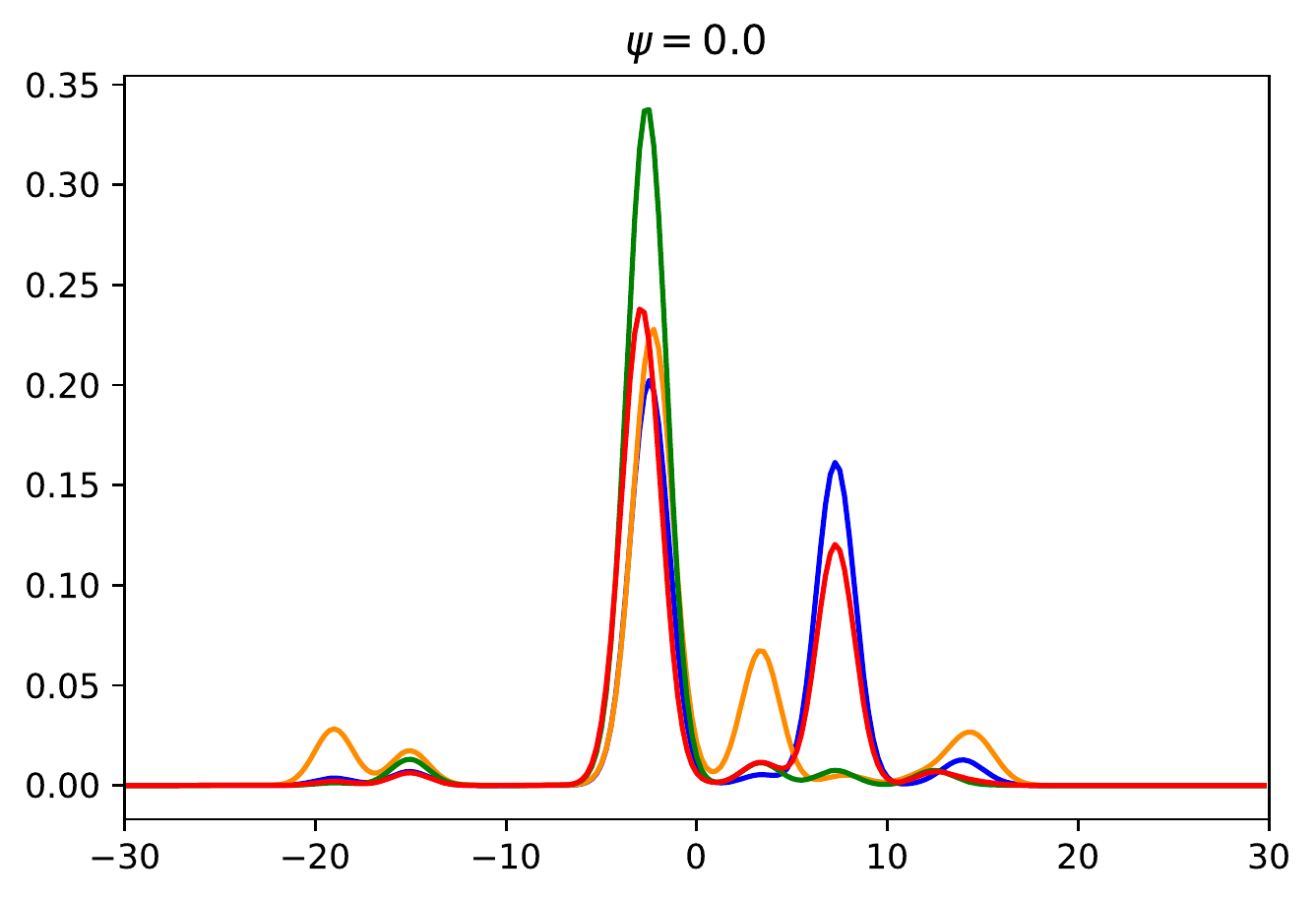}
\includegraphics[width=0.24\textwidth]{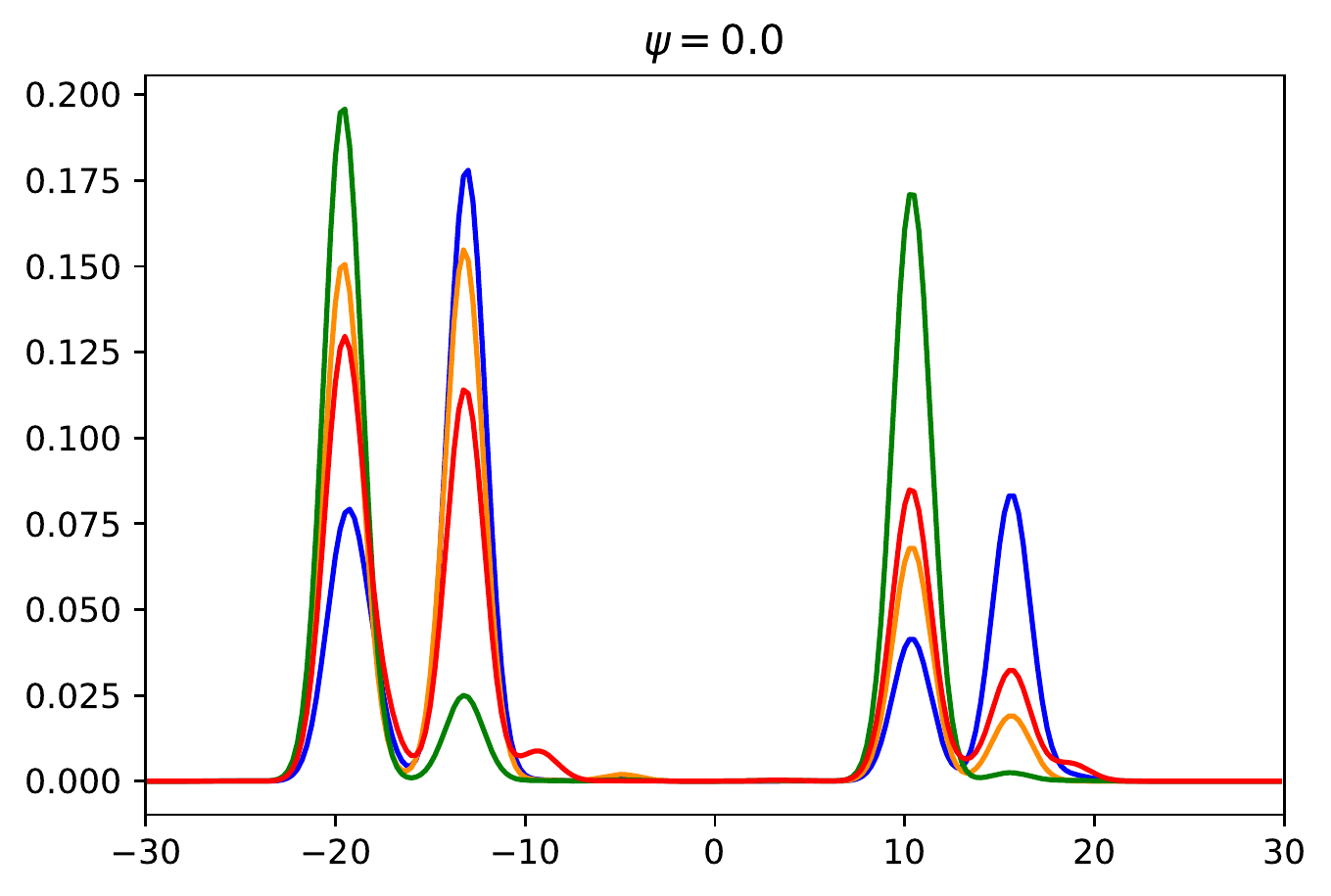}\\
\includegraphics[width=0.24\textwidth]{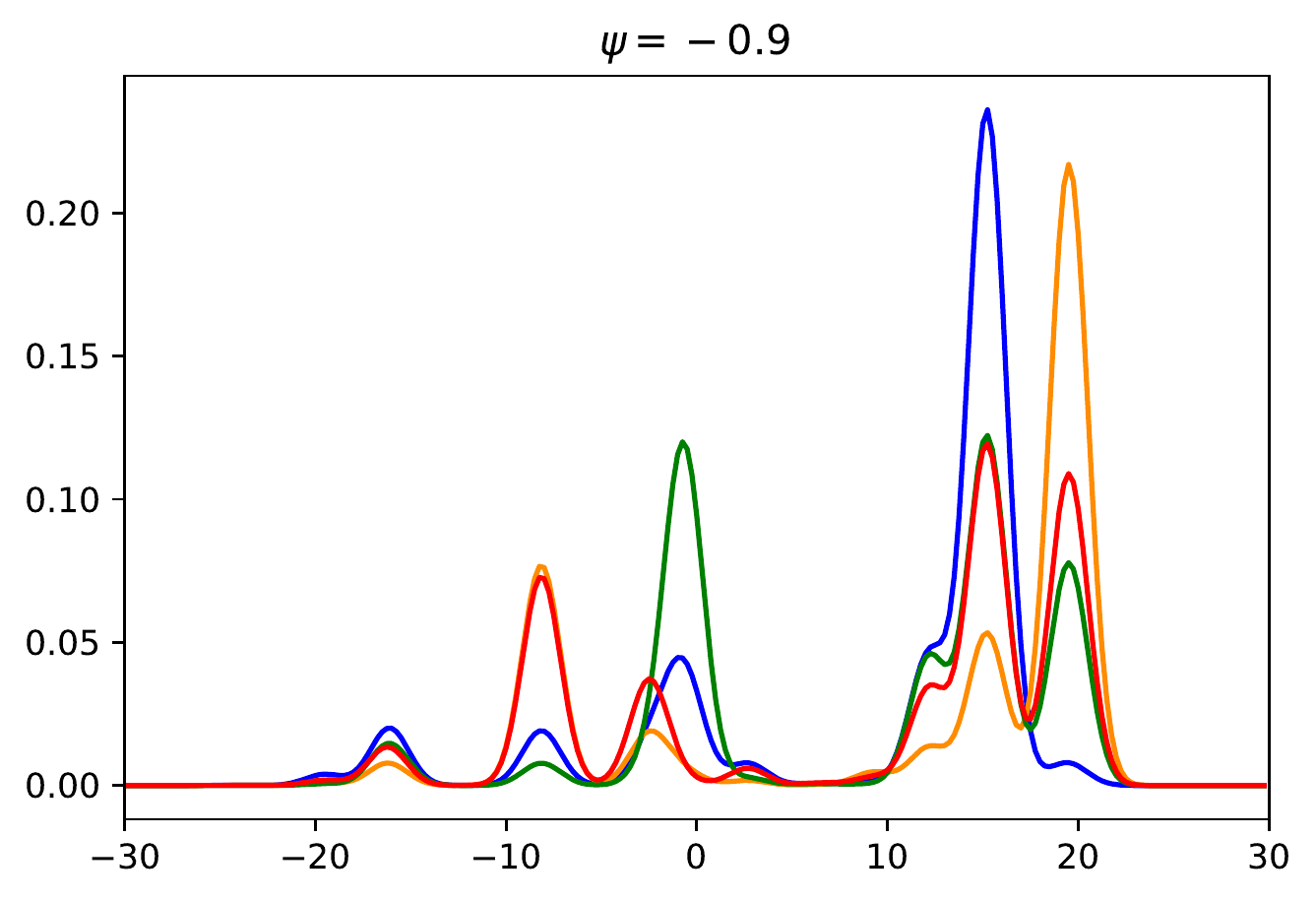}
\includegraphics[width=0.24\textwidth]{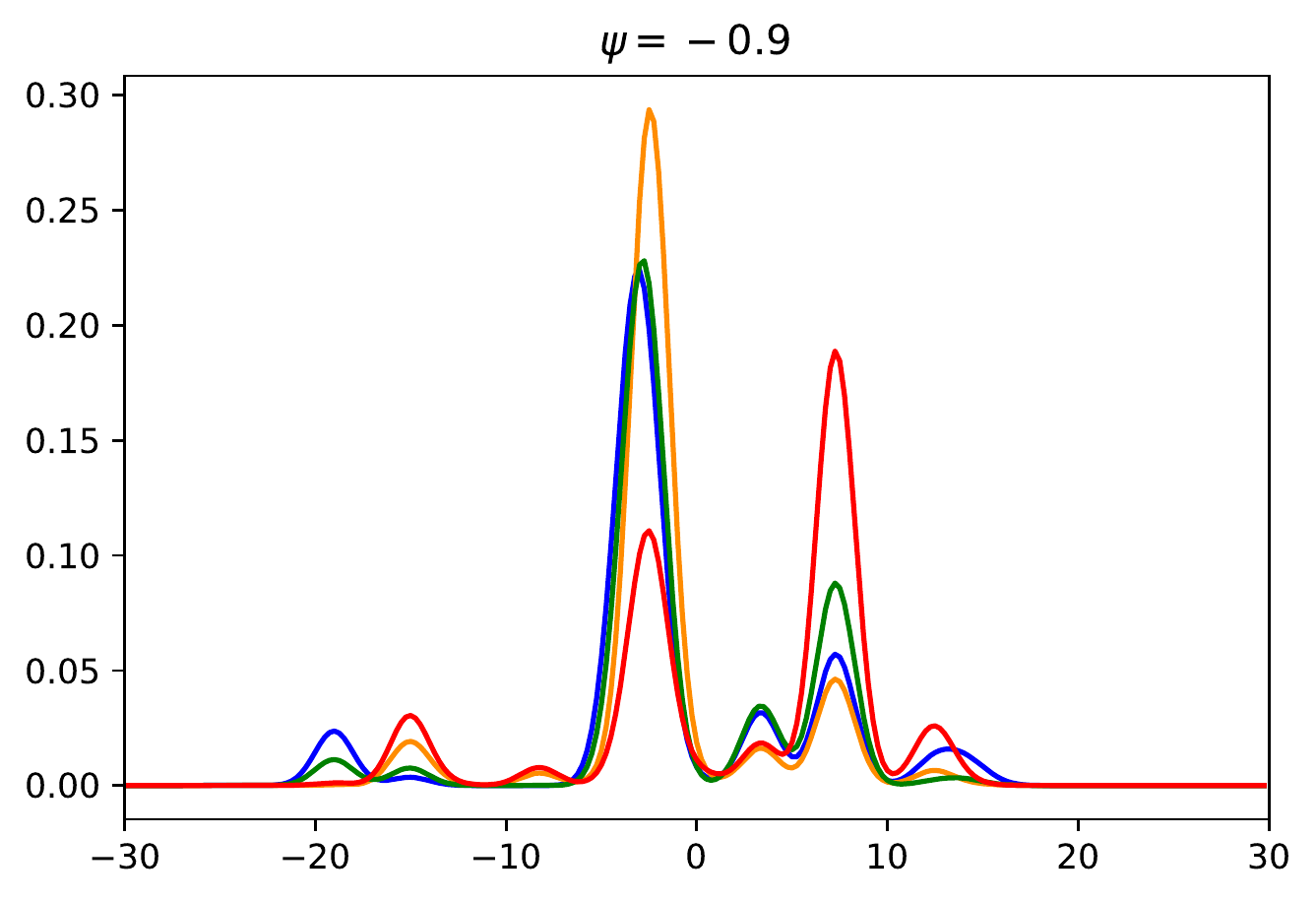}
\includegraphics[width=0.24\textwidth]{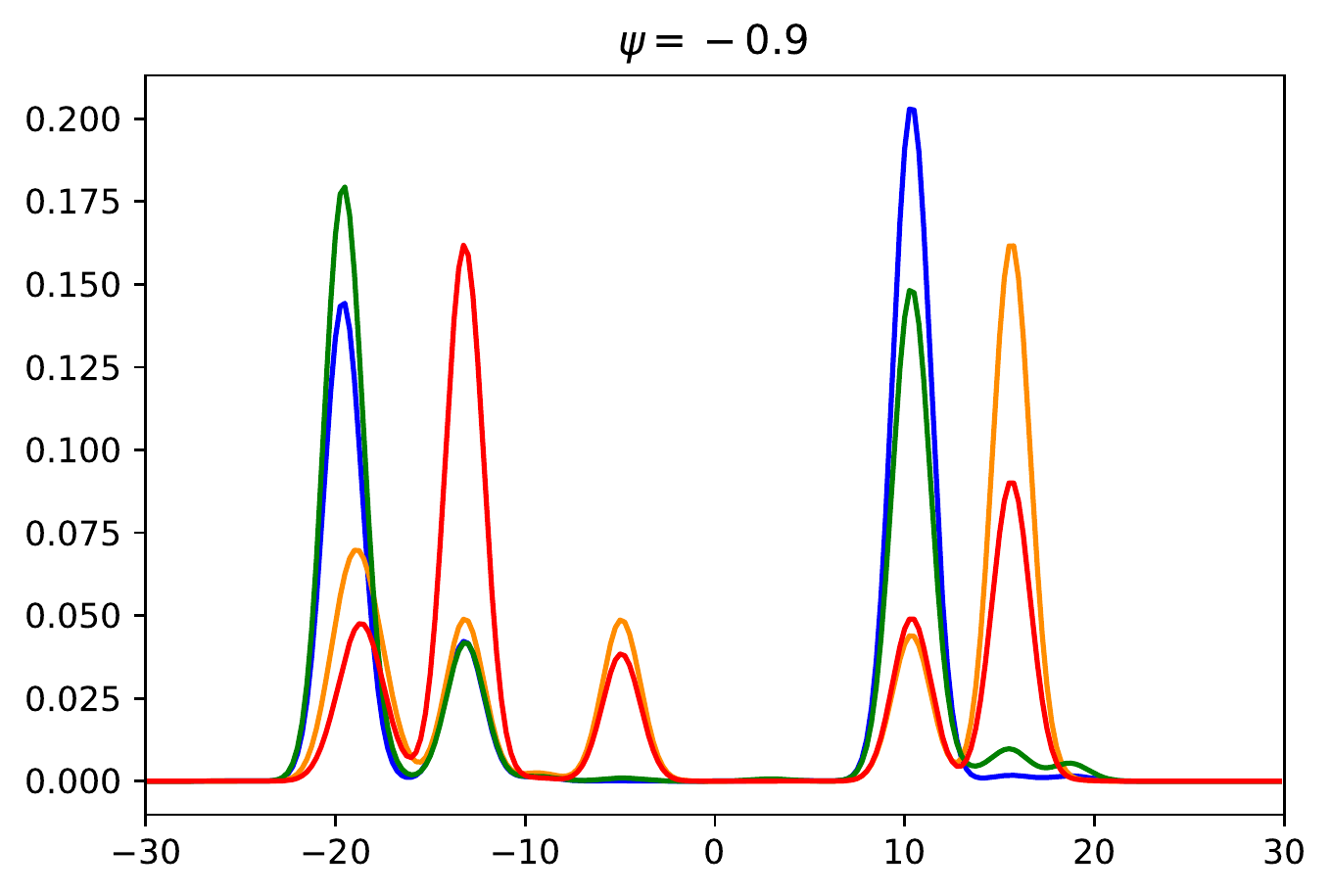}\\
\end{center}
\vspace{-0.5cm}
 \caption{\scriptsize{Realizations of $f_t$, for $t=1,2,3,4$ (blue, orange, green, red lines, respectively) and for some values of $\psi$.}}
 \label{fig:marginal_dens}
\end{figure}

Model \eqref{eq:m_data}-\eqref{eq:m_sbar} accommodates a variety of temporal behaviour, defining a large class of time-evolving random density functions. Figure \ref{fig:marginal_dens} displays realizations of the  $\text{AR1-DP}(\psi,M,G_{0})$ mixture model with a Gaussian kernel, for  different values of $\psi$, when $T=4$, $M= 2$, $G_0$ is a Uniform distribution with parameter $(-20,20)$, and fixing the  truncation  \eqref{eq:trunc_model} with $J = 100$. See also Figure \ref{fig:hellinger} in Appendix~D for a plot of the distribution of the Hellinger distances between  $f_t, t=2,3,4$ and $f_1$, defined as in \eqref{eq:rand_dens}. 
The Hellinger distance shows a time-dependent behaviour which is what we expect from an AR1 process. 

%
%


\section{Competitor models}
\label{sec:comp}

\cite{taddy2010autoregressive} introduces a dependent DP prior for modeling (discrete) time series of marked spatial point patterns. This dependent DP prior is the distribution of a sequence of discrete RPMs 
$(G_{t})_{t\geq1}$ with $G_{t}=\sum_{h\geq1} \xi_{th} \prod_{1\leq l\leq h-1}(1-\xi_{tl}) \delta_{\theta_h}$ is such that
\begin{equation}\label{eq:taddy}
\xi_{tl} = 1 - u_{tl}(1-w_{tl} \xi_{(t-1)l})\quad t\geq1,
\end{equation}
where: i) $u_{tl}$ and $w_{t}$ are Beta random variables with parameter $(M,1-\psi)$ and $(\psi, 1-\psi)$, respectively, for any $t\geq1$, with $M>0$ and $0<\psi<1$; ii) $(\theta_{h})_{h\geq1}$ are i.i.d. with common distribution $G_{0}$, and independent of $(\xi_{tl})_{t\geq1,l\geq1}$. Then $\xi_{tl}$ results distributed as a Beta distribution with parameter $(1,M)$ and hence $G_{t}\sim\text{DP}(M,G_{0})$. The prior of  \cite{taddy2010autoregressive} introduces only the additional prior parameter $0<\psi <1$, which accounts for modeling dependence over time.  The correlation between $\xi_{tl}$ and $\xi_{(t-k)l}$  is $(\psi M/(1+M-\psi))^k>0$, which rules out negative correlation among the RPMs $G_t$'s. 


\cite{deyoreo2018modeling} develop a dependent DP prior for temporal dynamic ordinal regressions. 
Differently from our AR1-DP prior, the prior of \cite{deyoreo2018modeling} has both random atoms and random weights depending on $t\geq1$. Specifically, the prior is assumed to be the distribution of $(G_{t})_{t\geq1}$ with $G_{t}=\sum_{h\geq1} \xi_{th} \prod_{1\leq l\leq h-1}(1-\xi_{tl}) \delta_{\theta_{th}}$ such that
\begin{equation}
\label{eq:Kottas_bit}
\xi_{tl}=1-\exp \left\{-\frac{\zeta^2_l+\eta^2_{tl}}{2M}\right\},
\end{equation}
where $(\eta^2_{tl})_{t\geq1}\iid\text{AR}(1,\psi)$, with $\psi\in(-1,1)$, $M>0$ and $\zeta_l\iid N(0,1)$. This implies that $1-\exp \{-\frac{\zeta^2_l+\eta^2_{tl}}{2M}\}$ is distributed as a Beta random variable with parameter $(1,M)$, and hence $G_{t}\sim\text{DP}(M,G_{0})$. \cite{deyoreo2018modeling} show that, since $\eta_{t l}$ enters squared in the stick-breaking weights $\xi_{tl}$'s, the correlation between $\xi_{t-k l}$ and $\xi_{tl}$ depends on the factor $(\psi^2)^k$ and the correlation between $G_t$ and $G_{t+1}$ is always positive. Furthermore, $M\geq 1$ implies that $0.5$ is a lower bound on the correlation between $\xi_{tl}$ and $\xi_{ t-k l}$, for any $\psi \in(-1,1)$, and that such a peculiar issue may be overcome with time-varying locations.

\cite{taddy2010autoregressive} and  \cite{deyoreo2018modeling} focus on dynamic density estimation, and they do not consider the problem of dynamic clustering. In particular, \cite{taddy2010autoregressive} states that the dynamic clustering produced by his model is not robust, and very different clustering may corresponds to relatively similar predictive distributions.  However, priors developed in \cite{taddy2010autoregressive} and  \cite{deyoreo2018modeling}, as well as our prior, are stationary, so that, once sample from $G_t$, features like the number of unique values in the sample, will have the same marginal distribution under the different priors. 

\section{Application to gender stereotypes}\label{sec:genderbias}

We study how gender stereotypes, with respect to adjectives and occupations, change over time in the 20th and 21th centuries in the United States. We use word embeddings  \citep{Garg_etal_2018} to measure the gender bias. We consider embeddings trained on Corpus of Historical American English (COHA) \citep{coha} for three decades $t=1900, 1950, 2000$. These embeddings are applied to lists of words from \cite{Garg_etal_2018}, representing each gender (men and women) and neutral words (occupations and adjectives). This leads to data for (standardized) adjective's biases and occupation's biases for women, for each word in the corresponding list. For each time $t$, we obtain two (unidimensional) datasets: i) the occupation's biases $\{y_{tj}, j=1,\ldots, n_O\}$, with $n_O=76$; ii) the adjective's biases $\{z_{tl}, l=1,\ldots, n_A\}$, with $n_A=230$). See  Appendix~B for details.  A negative value of the bias means that the embedding more closely associates the word with men, because of the distance closer to men than women. Hence, gender bias corresponds to either negative or positive values of the embedding bias. We  write that there is a ``bias against women'' when the value of the embedding bias is negative. 

Hereafter we model occupation's biases $\{y_{tj}, j=1,\ldots, n_O\}$ and adjective's biases $\{z_{tl}, l=1,\ldots, n_A\}$ with the AR1-DP mixture model with a Gaussian kernel. Specifically, we assume
\begin{equation}
\begin{split}
\label{eq:occupation_model}
Y_{tj}   \mid (\mu_{tj}, \tau_{tj}) &\ind \textrm{N}(\mu_{tj}, \left(\lambda \tau_{tj})^{-1}\right)\quad j=1,\ldots , n_{O}\\
(\mu_{tj}, \tau_{tj}) \mid G_t &\iid G_t \quad t=1900, 1950, 2000\\
(G_t)_{t\geq1} &\sim \text{AR1-DP} (\psi,M,G_0).
\end{split}
\end{equation}
where $G_0$ is a Gaussian-Gamma distribution with parameter $(\mu_0,\lambda,\alpha,\beta)$, i.e. $\mu|\tau \sim \textrm{N}(\mu_0,\frac{1}{\lambda\tau})$ and  $\tau \sim \textrm{Gamma}(\alpha,\beta)$ with $\E(\tau)=\alpha/\beta$. We assume the same model as in \eqref{eq:occupation_model} for the $Z_{tl}$'s. 
We set  $\mu_0 = 0$, $\lambda = 0.01$,  $\alpha =2$, $\beta = 1$ and $M\sim\textrm{Gamma}(4,4)$. The prior on $M$ implies that the (prior) expected number $K$ of clusters at time $t$ is $4$ for occupations and $5$ for adjectives, with support between $2$ clusters and $10$ clusters. Posterior inference is obtained via the MCMC algorithm described in Section 3. In particular, we set the number of iterations equal to 20,000, with a burn-in period of 10,000 iterations and thinning equal to 10. Finally, the truncation level $J$ is fixed to $50$.

\subsection{Posterior inference: occupational embedding bias}
\label{sec:occupational}

\begin{figure}[h!]
\begin{center}
\includegraphics[width=.4\textwidth]{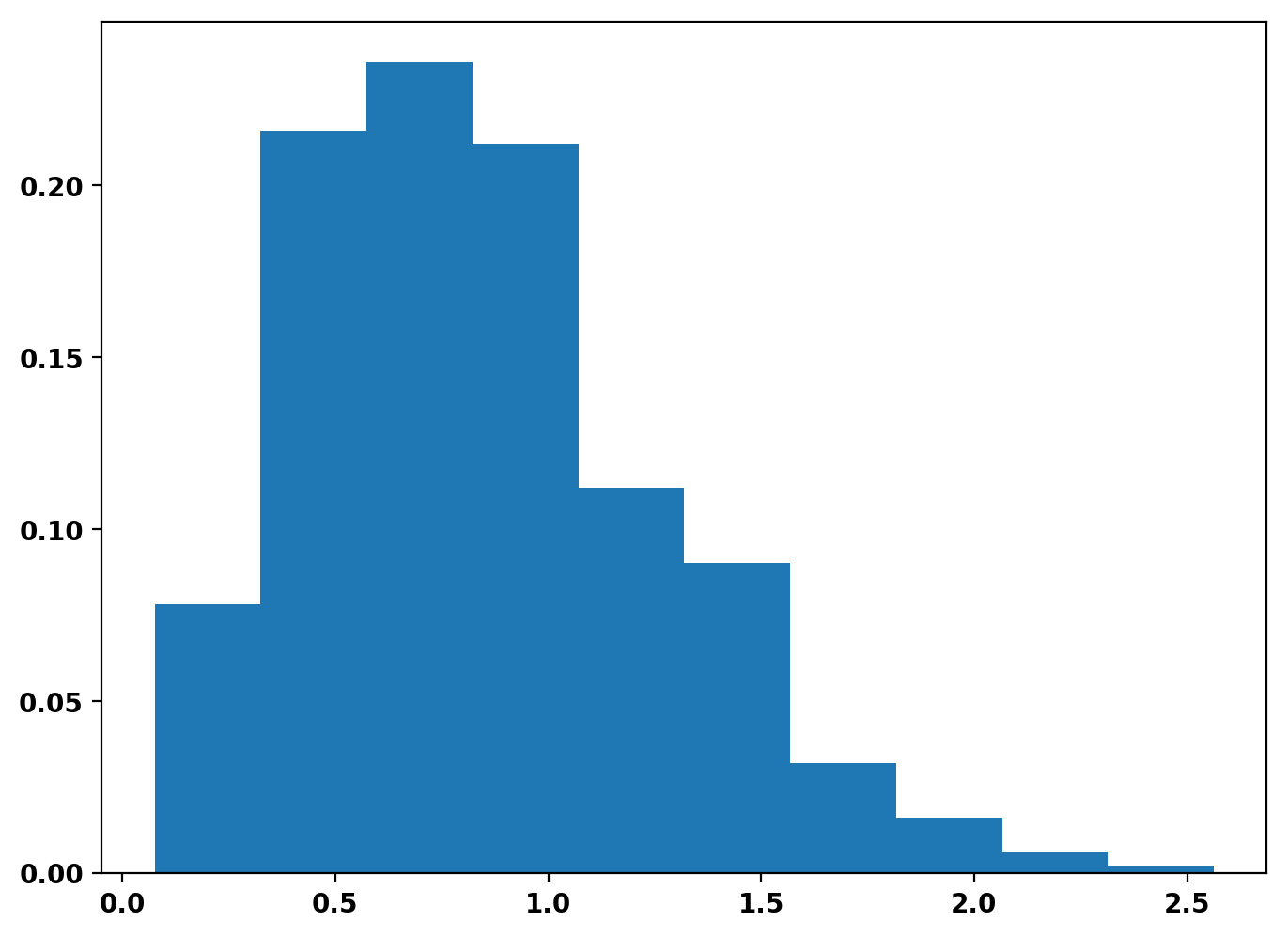}
\includegraphics[width=.4\textwidth]{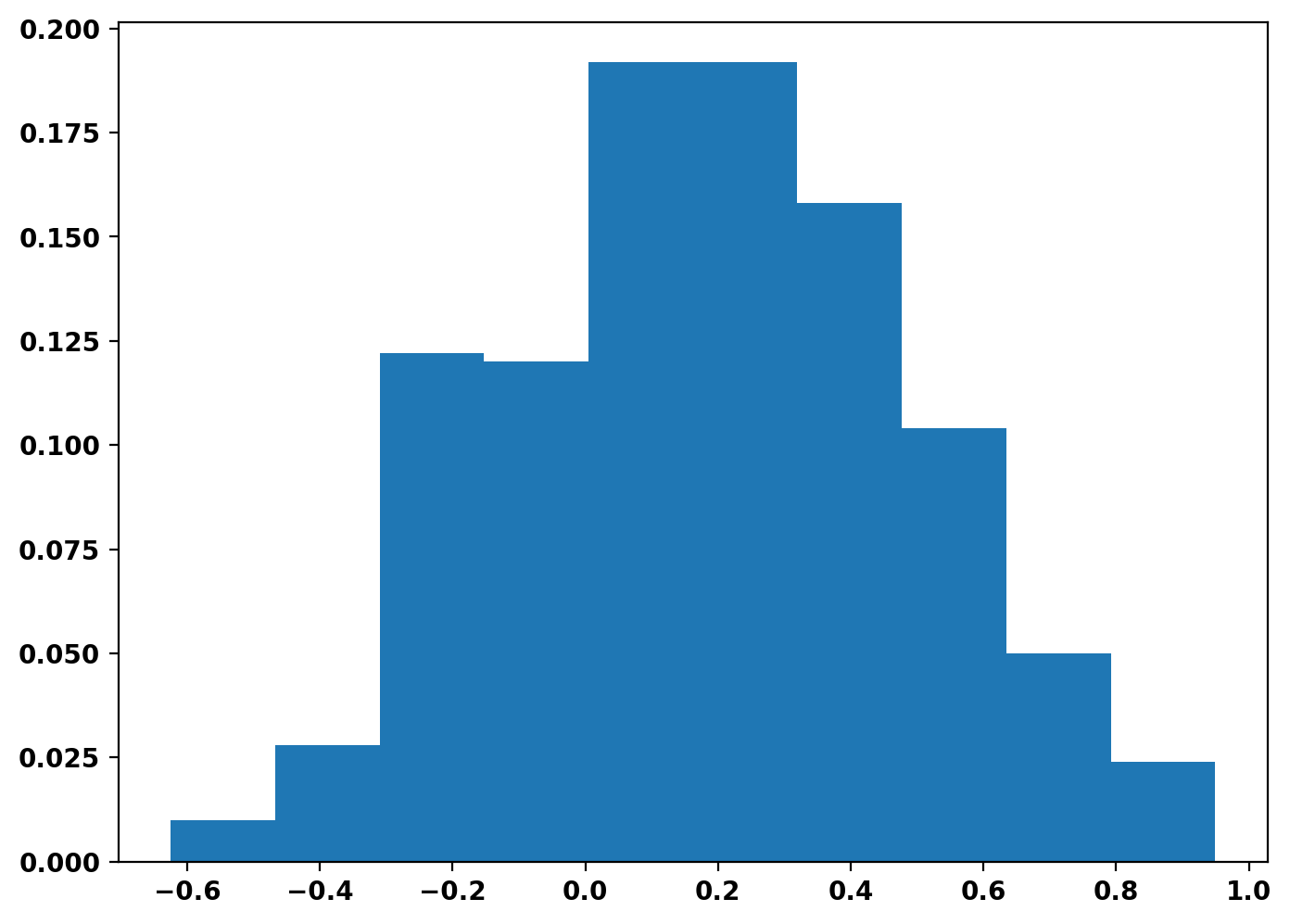}
\end{center}
\vspace{-0.5cm}
\caption{\scriptsize{Marginal posterior distributions of $M$ (left panel) and $\psi$ (right panel) for occupational bias data.} }
\label{fig:occupatbias_Mpsi}
\end{figure}

\begin{figure}[h]
\begin{center}
\includegraphics[width=.32\textwidth]{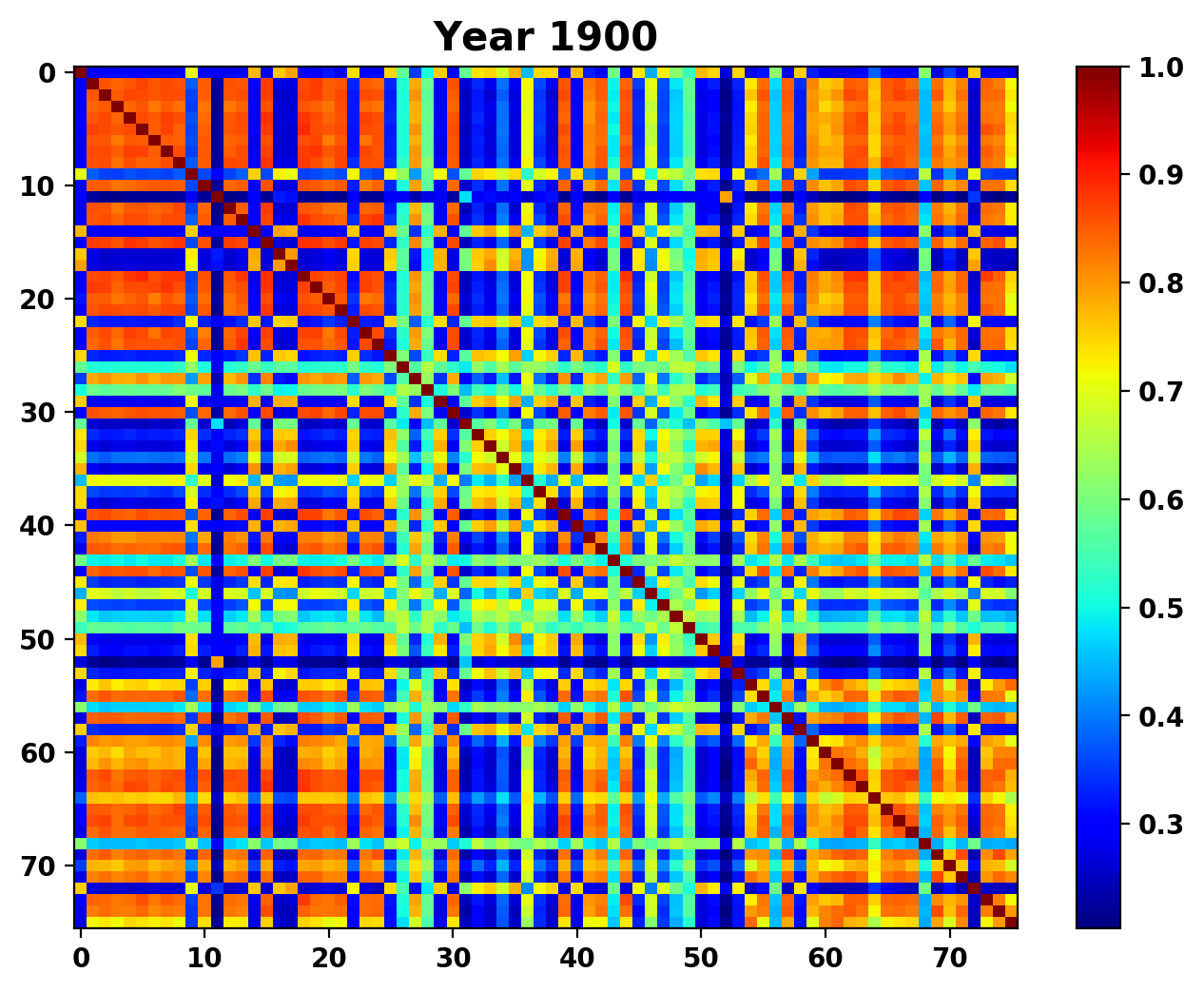}
\includegraphics[width=.32\textwidth]{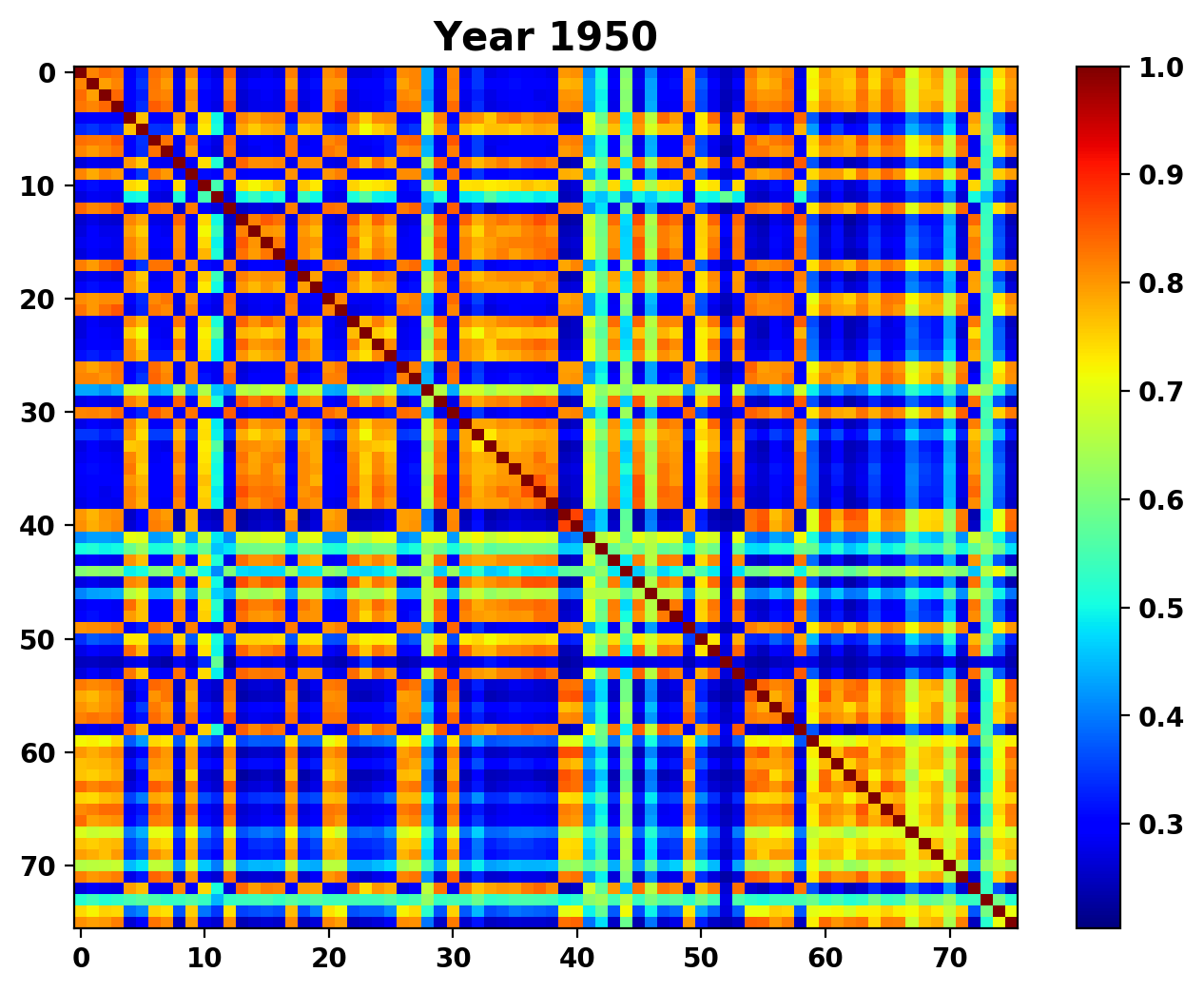}
\includegraphics[width=.32\textwidth]{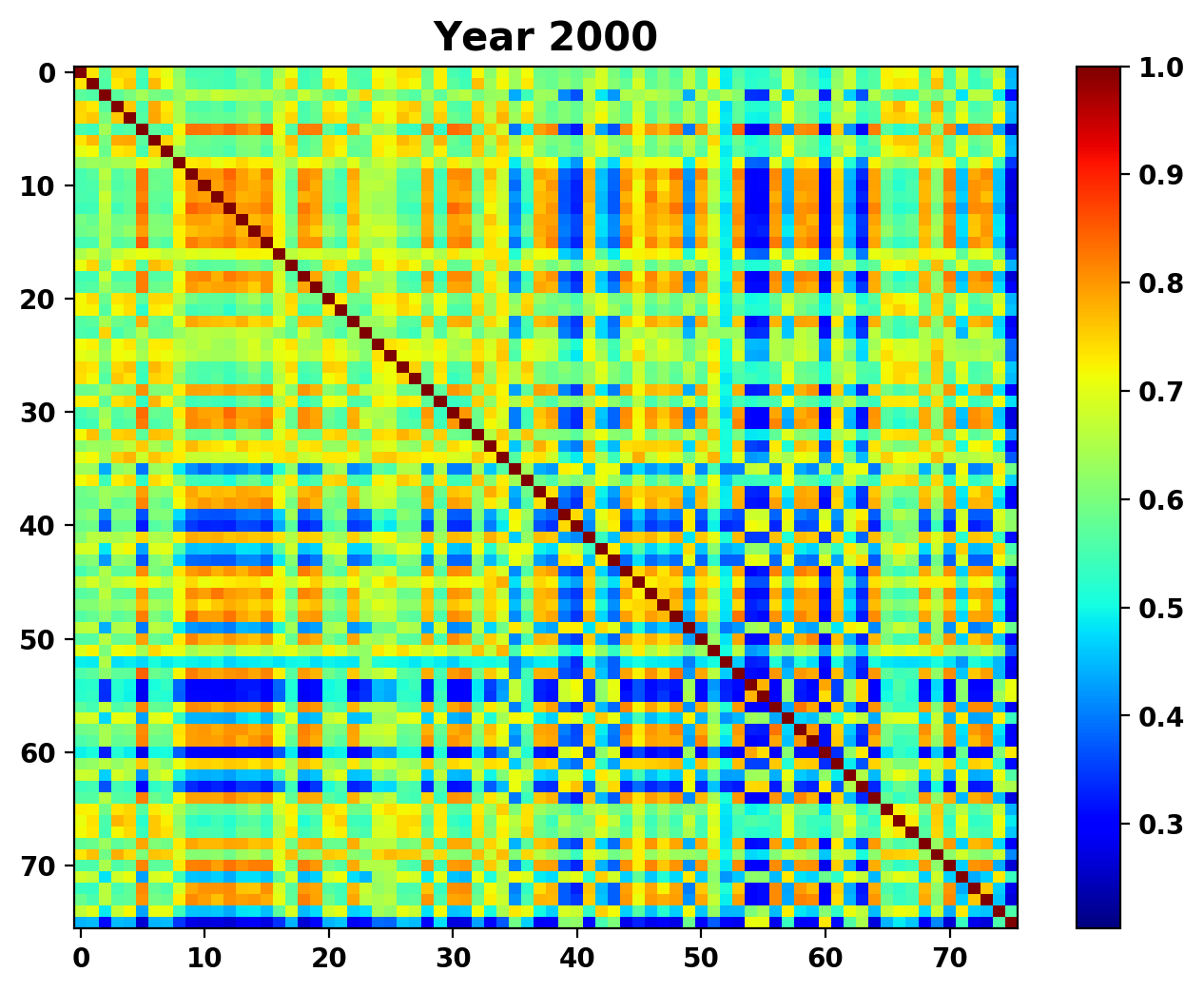}
\end{center}
\vspace{-0.5cm}
\caption{\scriptsize{Co-clustering for occupational bias: $t=1990$ (left panel), $t=1950$ (center panel), $t=2000$ (right panel).} }
\label{fig:occupatbias_coclust}
\end{figure}

Figure~\ref{fig:occupatbias_Mpsi} (right-panel) shows that the posterior distribution of $\psi$ puts most of its mass on the positive real line; the posterior probability that $\psi>0$ is $0.729$. From Figure~\ref{fig:occupatbias_coclust}, the predominant colors in 1900 and 1950 are blue (low posterior co-clustering probability) and orange (high posterior co-clustering probability) but in different proportions, whereas the predominant colors in 2000 are yellow and green, indicating a posterior co-clustering probability around $0.5$. This decrease over time in the bias against women can be seen in Figure~\ref{fig:pred_genderbias2} (left), where we plot posterior predictive densities of the occupational embedding bias.

From Figure~\ref{fig:pred_genderbias2} (left), the estimated density function at time $t=1900$ has two peaks: a higher peak centered at a negative (gender) bias location, and a lower peak centered at a positive (gender) bias location. This means that in 1900, the fraction of man-biased jobs was larger than the fraction of woman-biased jobs. In 1950, the estimated density function still has two peaks. While the locations of these two peaks  remain the same, the fractions of man-biased jobs and woman-biased jobs change. In particular, the fraction of man-biased jobs decreases with respect to the fraction of woman-biased jobs. That is, more occupations become gender neutral or woman-biased. This behaviour is also reflected in the predominance of two colors in the left and central panels of Figure~\ref{fig:occupatbias_coclust}. However, for 2000, the estimated density function in Figure~\ref{fig:pred_genderbias2} (left) has a single peak, on the positive real line, showing that  most of jobs are neutral. Posterior predictive means decrease over time, i.e.  -0.196, -0.009, 0.180 in 1900, 1950 and 2000, respectively.


Table~\ref{tab:occutable} show estimated cluster configurations at each time $t$. It reports the clustering that minimizes the posterior expectation of Binder's loss  \citep{Binder78} under equal misclassification costs (R package \texttt{mcclust}). For all pairs of subjects, Binder's loss measures the distance between the true co-clustering probability and the estimated cluster allocation. We obtain three clusters in 1900 and 1950, and one cluster in 2000. Clusters are interpreted by assigning them a ``label'' as follows: i) ``man-cluster" (i.e. occupations in the cluster are biased against women) if the empirical mean of all the data points in the cluster is negative and zero is not within one (empirical) standard deviation from the mean; ii) ``woman-cluster" (i.e. the occupations are biased in favor of women, or against men) if the empirical mean of all the data points in the cluster is positive and zero is not within one standard deviation from the mean itself.  A cluster is ``neutral" if zero is within one standard deviation from the empirical mean of all the data in the cluster.  


\begin{center}
Table \ref{tab:occutable} about here
\end{center}

From Table~\ref{tab:occutable}, the number of occupations in the ``man-cluster'' decreases from 44 to 36 from 1900 to 1950, while the frequency counts in the ``neutral-cluster'' increases from 30 to 38 and the number of occupations in the ``woman-cluster'' is approximately constant. From 1900 to 1950, the empirical mean of the non-standardized embeddings in the ``man-cluster'' increases from -0.107 to -0.085, the mean of the ``neutral-cluster" increases from -0.015 to -0.007, and the mean of ``woman-cluster" decreases from 0.082 to 0.072.  Note that, in 2000, there is only one cluster with empirical mean -0.035 and standard deviation 0.040. This shows that the occupational bias against women decreases from 1900 to 2000, although the overall empirical mean of the cluster in 2000 is still negative.

With regards to single occupation words, inclusion of some of them in the ``man-cluster''  makes sense as well. For example, we see that  in 1900, the occupation ``athlete" is associated with men, while in 1950, it belongs to the ``neutral-cluster''. Of course, women athletes were very few in 1900, but their number started to increase during the 20th century. For instance, the number of Olympic women athletes increased from 65 at the 1920 Summer Olympics to 331 at the 1936 Summer Olympics. Data have been retrieved from the official website of the Olympic Games (\texttt{https://www.olympic.org/}). On the other hand, occupations requiring a great amount of physical strength, such as for example ``farmer", ``soldier" and ``guard", remain associated with men from 1900 to 1950. Note also that, both in decades 1900 and year 1950, ``housekeeper" and ``nurse" are associated with women.
However, sometimes the individual occupation word in  the cluster is counter-intuitive. For instance, the occupation word ``midwife" is strongly associated with men in both years, which is the  opposite of what we should have expected. 


\subsection{Posterior inference: adjective embedding bias}
\label{sec:adjective}
%
%
%

\begin{figure}[H]
\begin{center}
\includegraphics[width=.4\textwidth]{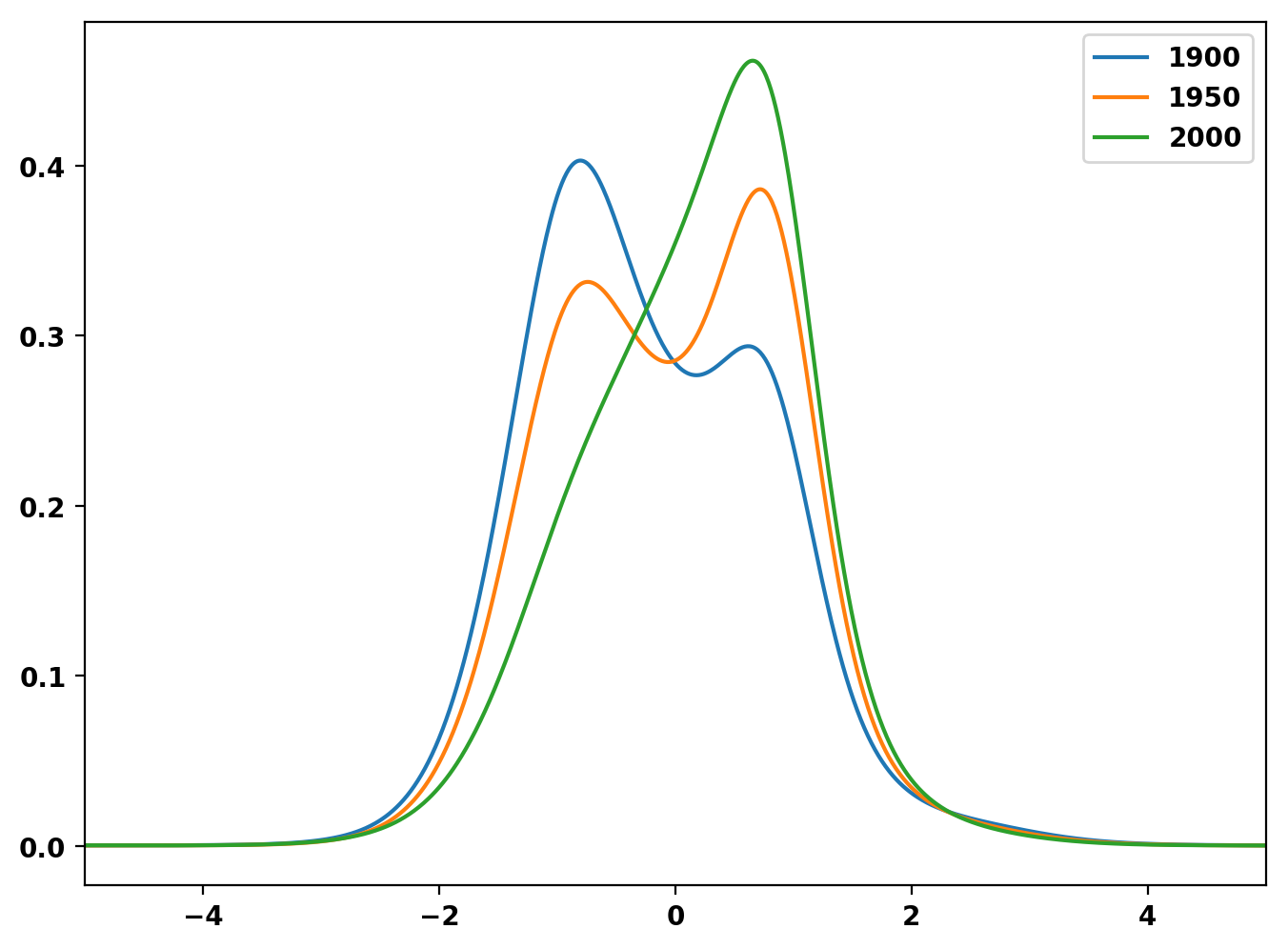}
\includegraphics[width=.4\textwidth]{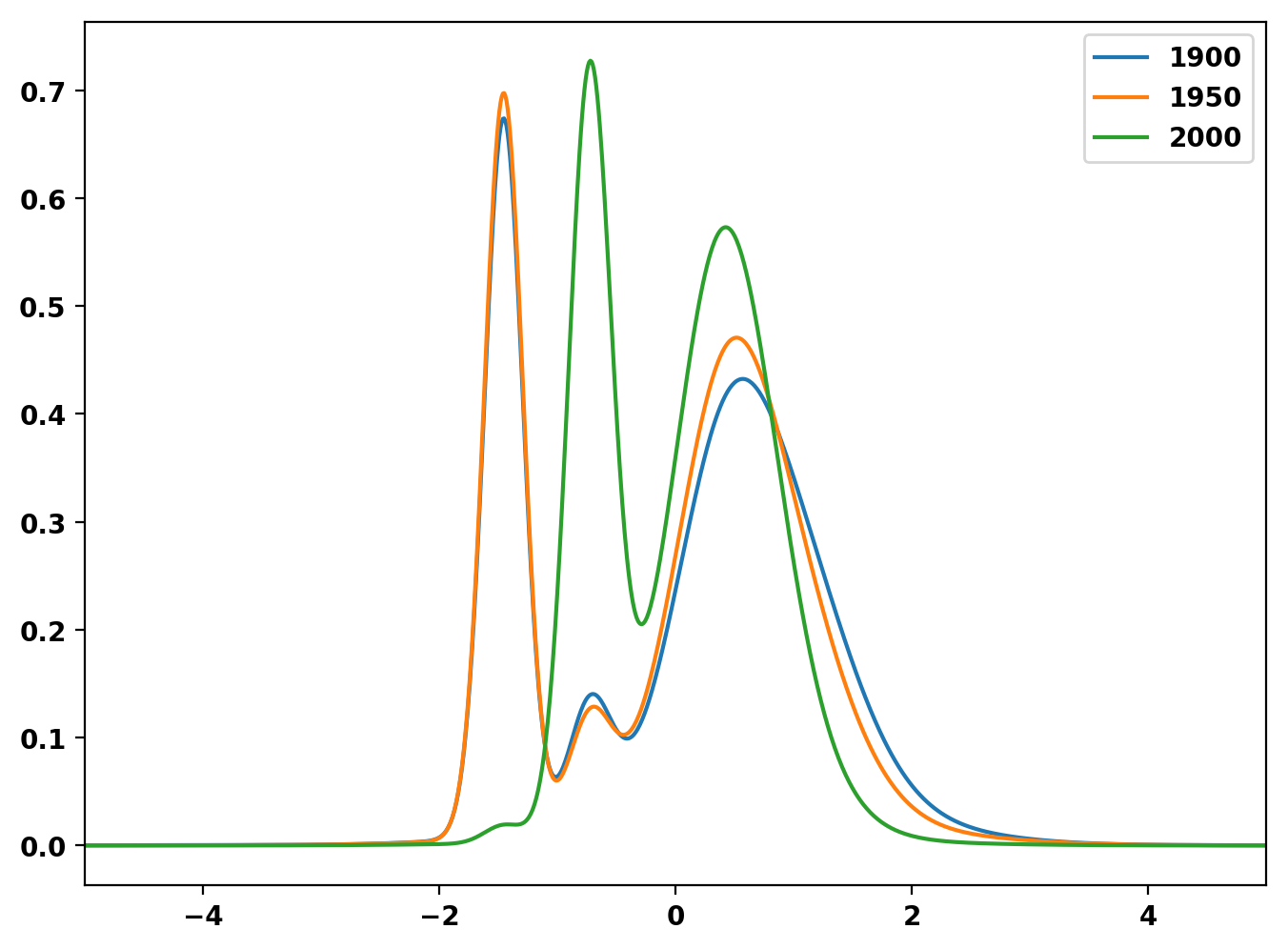}
\end{center}
\vspace{-0.5cm}
\caption{\scriptsize{Posterior predictive densities of the occupational bias $Y_t^{new}$ (left) the adjective bias $Z^{new}_t$ (right).}}
\label{fig:pred_genderbias2}
\end{figure}

From Figure \ref{fig:pred_genderbias2} (right), the estimated density function for the adjective embedding bias at time $t=1900$ has three peaks: the highest peak is centered near to the value -2, a small peak centered around the value -0.5 and a third peak centered at a positive value for the gender bias. This means that in 1900, the fraction of man-biased adjectives is larger than the fraction of neutral or woman-biased adjectives. In 1950 and 2000, the estimated density functions still have three peaks. While the location of these three peaks remain the same, the fractions of man-biased jobs and woman-biased jobs change. In particular, while the estimated density function in 1950 is very similar to that of $t=1900$, the estimated density function in 2000 is different. Indeed in 2000, the peak centered near the value -2 becomes very small, while the peak centered around the value -0.5 increases notably and the peak on the positive location grows moderately. Posterior predictive means are 0.027, -0.039, 0.058 for $t=1900, 1950, 2000$, respectively.

Table~\ref{fig:adjtable} shows nine clusters in 1900, seven clusters in 1950 and five clusters in 2000. In 1900, the cluster that is most biased against women (i.e., the cluster with the smallest mean) is Cluster 2, which contains 70 adjective words and has an average non-standardized gender bias of -0.088, while the most biased cluster associating adjective to women (the cluster with the largest mean) is Cluster 7, which contains e.g. ``attractive", ``charming" and ``feminine" and has an average gender bias of 0.118. In 1950, the cluster most biased against women is Cluster 2, with 74 adjective words and has an average gender bias equal to -0.081, while the cluster with the largest positive average (0.112) is Cluster 5, which contains ``attractive", ``charming" and ``feminine" for decade 1900. We conclude that the adjective bias does not change significantly from 1900 to 1950. However, in 2000, the cluster most biased against women is Cluster 5, which contains one adjective word and has an average gender bias equal to -0.076. In 2000, the second most man-biased cluster is Cluster 2, which contains 77 adjective words and has an average gender bias -0.051. In 2000, the most woman-biased cluster is Cluster 3, which contains ``attractive" and ``feminine" and has an average gender bias 0.078. This implies that adjective words become less gender-biased from 1950 to 2000 although bias still exists.

\begin{center}
Table \ref{fig:adjtable} about here
\end{center}

To simplify the  interpretation of our results, for each year we have merged clusters into three groups.  We first label all the estimated clusters as in the occupational embedding example, e.g. a cluster is defined as 
``man-cluster'' (meaning the adjective in this cluster is biased against women) if the empirical mean of all the data points in the cluster is negative and zero is not within one (empirical) standard deviation from the mean.  
Then we  merge all clusters with the same label, obtaining three groups, to which we refer as ``Man'', 
``Woman'' and  ``Neutral'' groups, which are shown in Table~\ref{fig:adjtable2}. On the whole, the size of the ``neutral'' group increases over time, while that of the ``woman'' group decreases. This result seems to support the fact that adjective bias is decreasing over time. For some of the adjective words, the inclusion in one group is very meaningful. For example, the words ``feminine" and ``attractive" are always closely associated with women, which is intuitive, while words ``effeminate", ``enterprising", ``autocratic'' are always strongly associated with men. Moreover, words such as ``affected", ``civilized", ``complicated", ``formal", ``healthy", ``informal", ``natural", ``serious" always belong to the neutral group. It is also clear that some words are clearly gender-stereotyped: ``adaptable", ``dependable", ``inventive", ``methodical", ``resourceful" and ``wise" are always associated with men. To conclude our analysis of the adjective embedding bias, and similarly to what we did for occupational embedding bias, we compute the marginal posterior of $\psi$ and $M$, as well as the posterior co-clustering probabilities, but we move them in Appendix~D (Figures \ref{fig:adjectivebias_Mpsi} and Figure \ref{fig:adjectivebias_coclust}, respectively). 

\begin{center}
Table~\ref{fig:adjtable2} about here
\end{center}


\section{A dose-escalation study}
\label{sec:dose} 
\begin{figure}[h]
\begin{subfigure}[t]{0.5\textwidth}
\includegraphics[width=0.9\textwidth, height=.2\textheight]{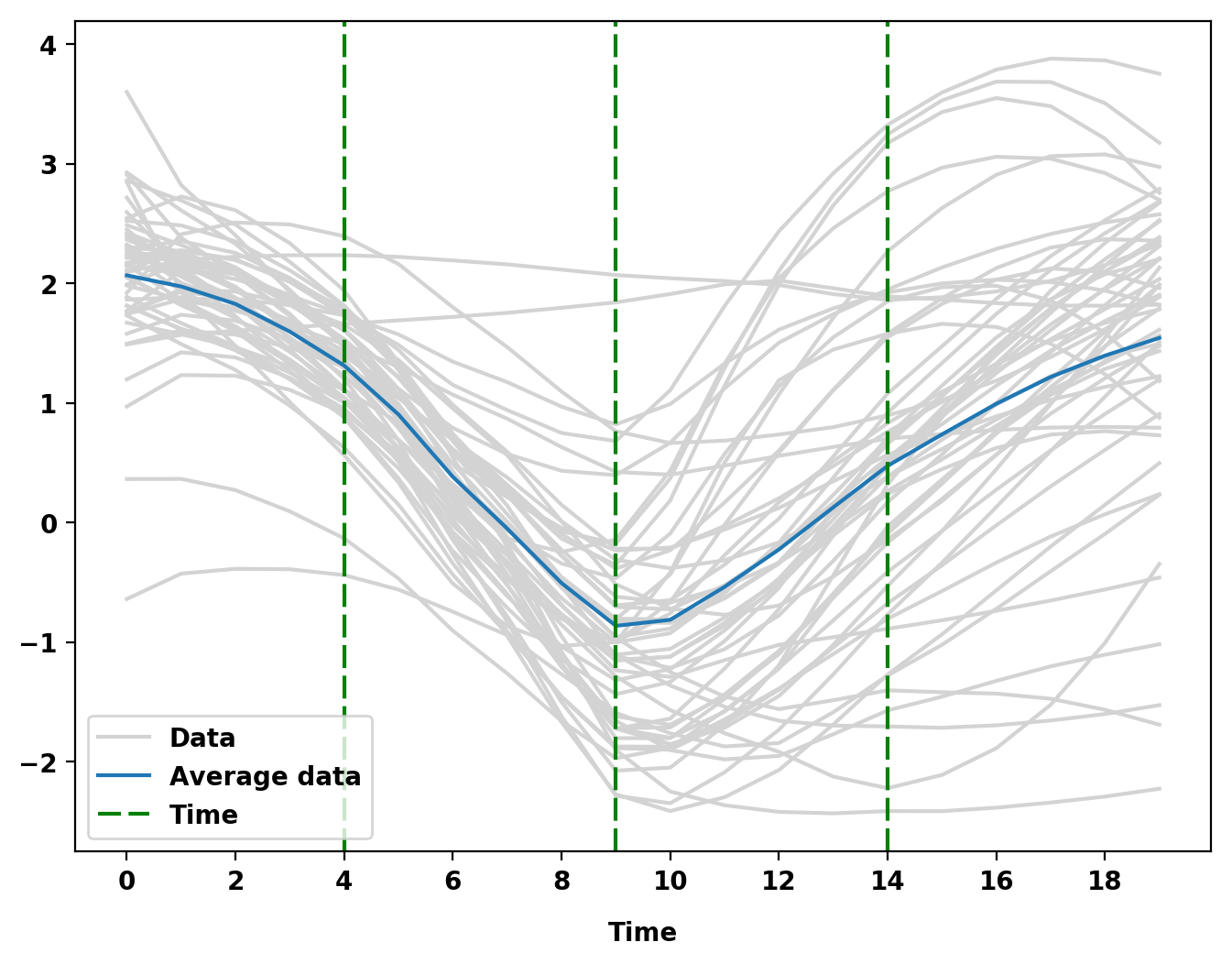}
\subcaption{}
\label{fig:wbc_data}
\end{subfigure}
\begin{subfigure}[t]{0.5\textwidth}
\includegraphics[width=0.9\textwidth, height=.2\textheight]{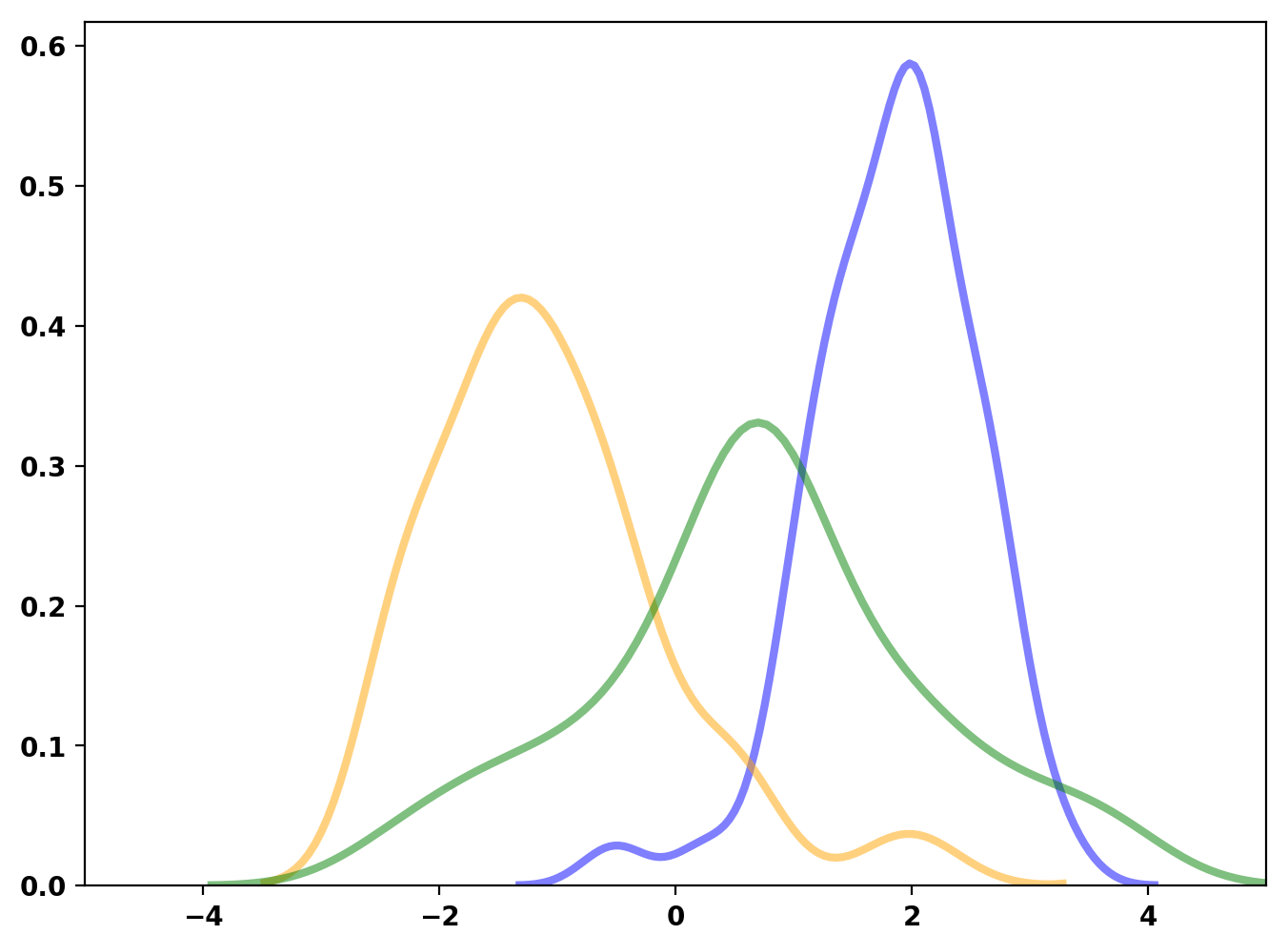}
\subcaption{}
\label{fig:datadensity}
\end{subfigure}
\caption{\scriptsize{Time profiles of the WBC (a) where the vertical green dashed lines denote the time points considered in the analysis; Kernel density plot of the WBC data (b) for day 4 (blue), day 9 (orange) and day 14 (green).}}
\end{figure}

We consider haematologic  data from a dose-escalation study  \citep{lichtman1993phase,muller1997bayesian}. Data are daily white blood cell counts (WBC) for $n=52$ patients over time receiving high doses of cancer chemotherapy. The anticancer agent CTX (cyclophosphamide) is known to lower a person's WBC as the dose increases, and a combination of drugs (GM-CSF) is given to mitigate some of the side-effects of the chemotherapy.  Figure \ref{fig:wbc_data} displays the daily WBC profiles over time.  After an initial baseline, the WBC in each patient has a sudden decline at the beginning of chemotherapy;  then a slow S-shaped recovery takes place and the WBC counts reaches approximately the baseline value at the end of treatment cycle.  Although a longitudinal analysis would be more appropriate for this application, to illustrate important features of our model, we consider three equally spaced time points: day 4, at the beginning of the therapy; day 8, close to the nadir; day 14, during recovery. Figure \ref{fig:datadensity} shows the kernel density plot of the daily WBC profiles for days 4, 9 and 14, which will be assumed as response. 

Let $N_{r}(\boldsymbol{\mu},\boldsymbol{\Sigma})$ denote the $r$-dimensional Gaussian distribution with mean vecotor $\boldsymbol{\mu}$ and variane-covariance matrix $\boldsymbol{\Sigma}$. Moreover, let $Y_{tj}$ denote the WBC count of patient $j$ at time $t$, for $j=1,\ldots,n$ and for $t=4,9, 14$-th day, and let assume the following model
\begin{eqnarray}
 Y_{tj}\mid (\mu_{tj}, \tau_{tj}) & \ind &\text{N}(m_{tj} + \mathbf{x}_{tj}\boldsymbol{\beta}_t, (\lambda  \tau_{tj})^{-1}), \ j=1,\ldots,n   \label{eq:lik_wbc} \\
 (m_{tj}, \tau_{tj}) \mid G_t &\iid&  G_t  \ \textrm{ for all } t\nonumber\\
 (G_t)_{ t\geq 1} &\sim& \text{AR1-DP} (\psi,M,G_0), \nonumber
\end{eqnarray}
where $G_0$ is a Gaussian-Gamma distribution with parameter $(\mu_0,\lambda,\alpha,\beta)$, $\lambda$ is fixed to $0.1$, ${\bm \beta}_t \iid \text{N}_2(\bm 0, 10 \ \mathbb{I}_2)$, for $t=4,9, 14$. The $\lambda$ parameter in \eqref{eq:lik_wbc} is fixed to $0.1$. The covariate vector ${\bm x}_{tj}$ is two-dimensional, with components CTX and GM-CSF, respectively, over time for each patient $j$. Moreover, we assume $\psi\sim\textrm{Uniform}(-1,1)$ and $M\sim\textrm{Gamma}(4,4)$ so that the (prior) expected number $K$ of clusters is equal to 4, according to our prior belief. Finally, we  set $\mu_0=0$, $1/\kappa_0=0.1$, $\alpha=\gamma=1$ as a trade-off between non-informativeness and robustness  ($\E(m)=0$, $\Var(m)=+\infty$, $\E(\tau)=\Var(\tau)=1$). 

\begin{figure}[h]
\begin{center}
\includegraphics[width=.32\textwidth]{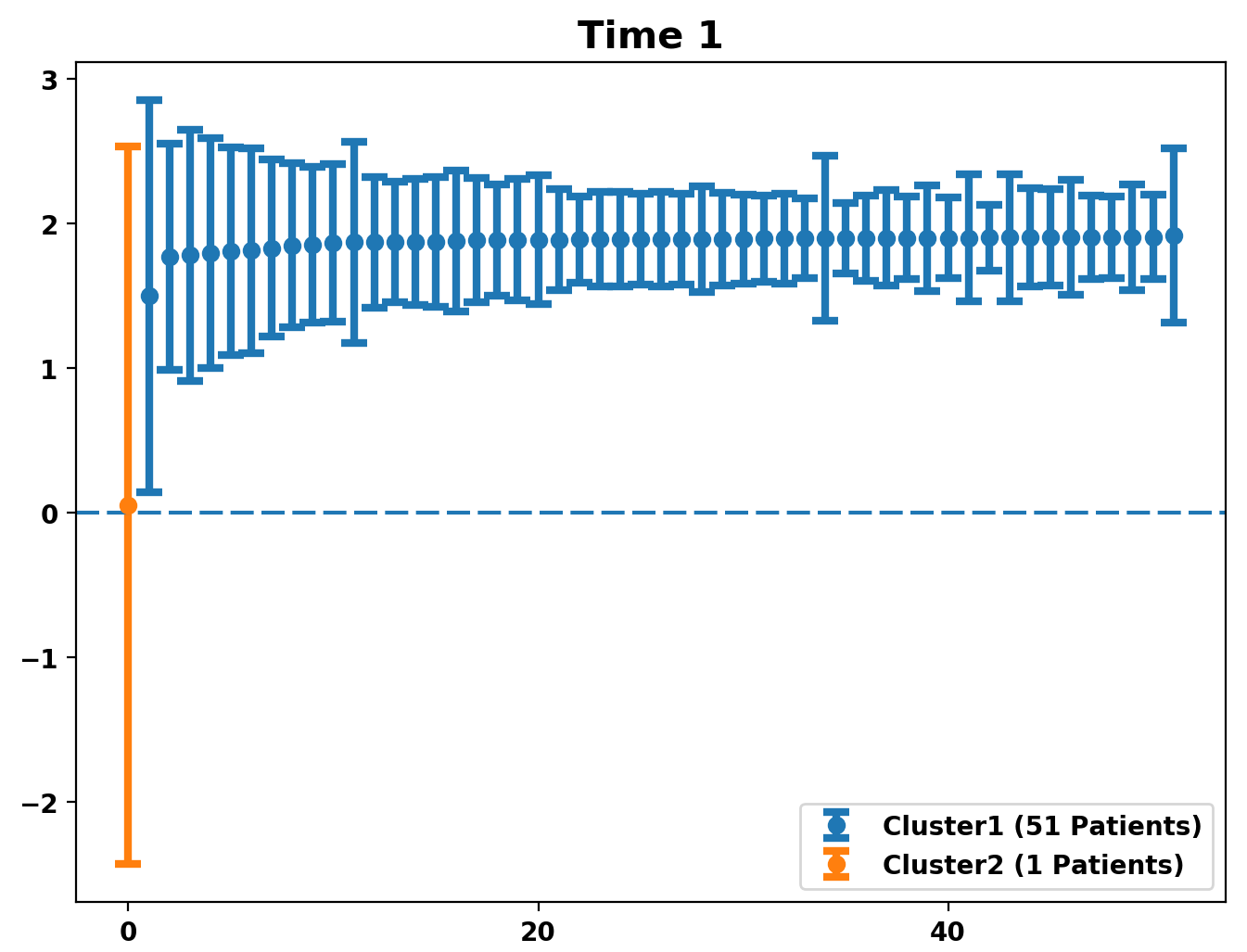}
\includegraphics[width=.32\textwidth]{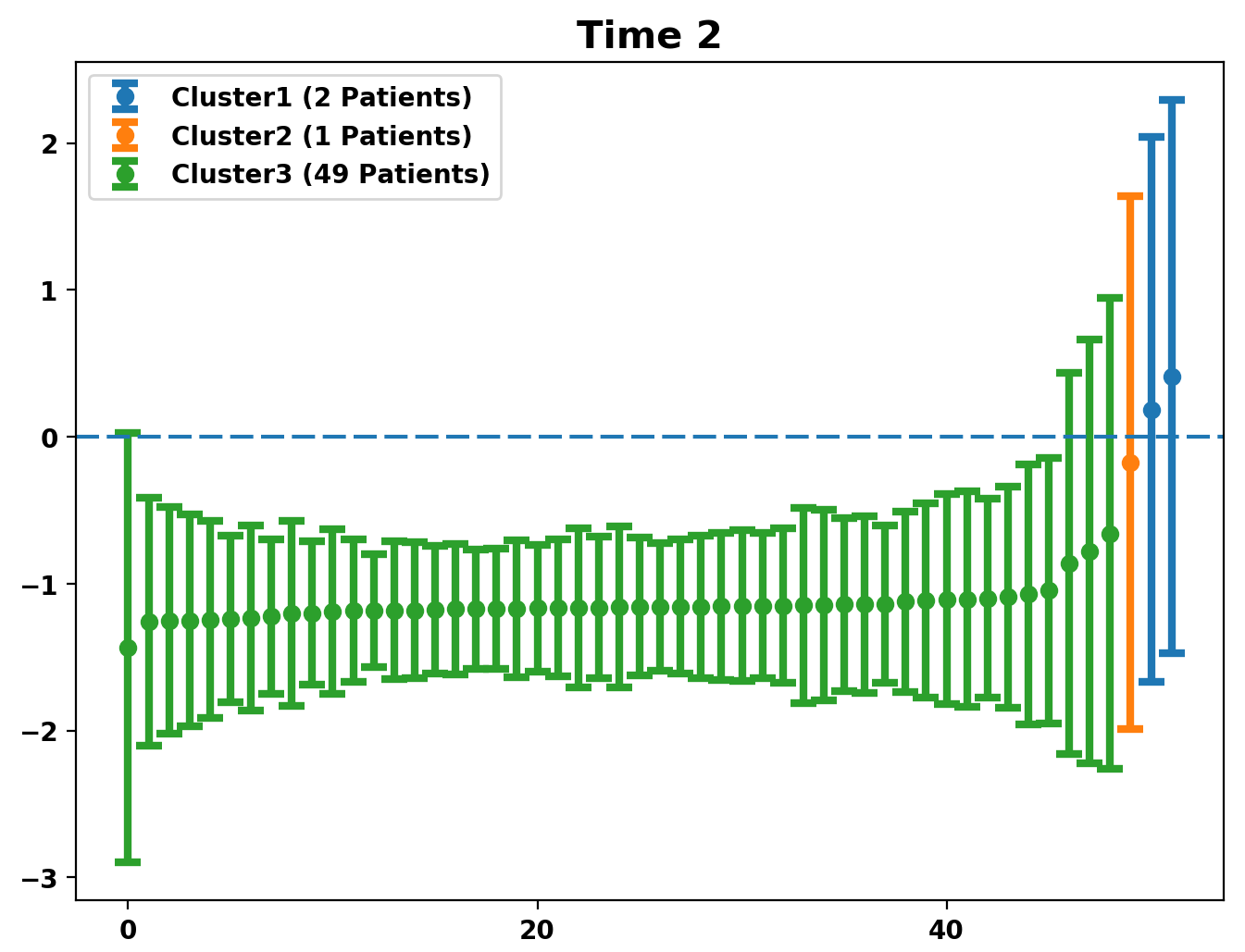}
\includegraphics[width=.32\textwidth]{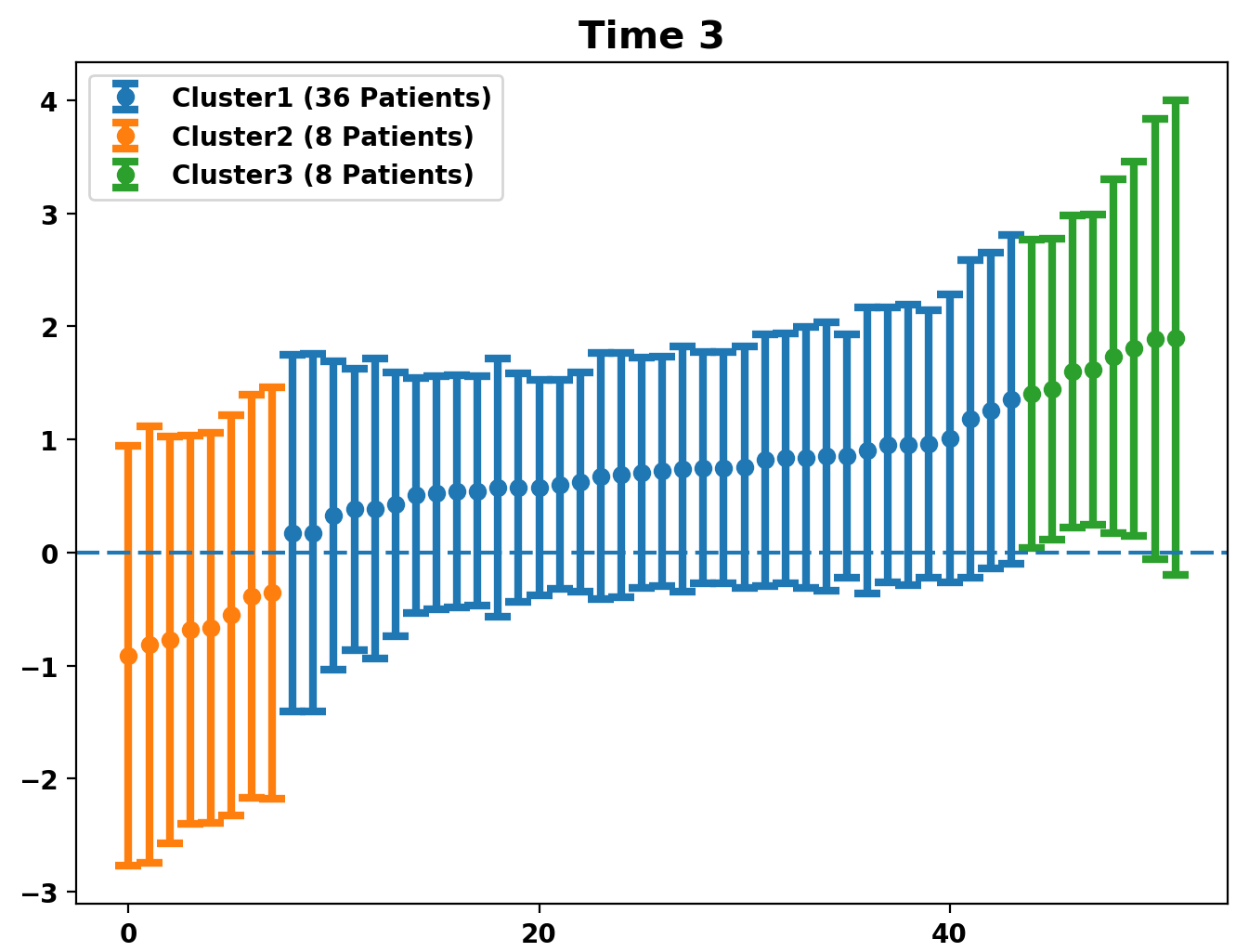}
\end{center}
\vspace{-0.5cm}
\caption{\scriptsize{Marginal posterior distributions of the  $m_{tj}$s for $t=4$-th (left panel), 9-th (center), 14-th day (right).}}
\label{fig:WBCclust}
\end{figure}

We fix $J= 50$ in \eqref{eq:trunc_model}. We run the MCMC algorithm described in Section \ref{sec:mixturemodel}  for 20,000 iterations, discarding the first 10,000 iterations as burn-in, and thinning every 10 so that the final sample size is 1,000. We check through standard diagnostic criteria  available in the R package CODA \citep{CODA}, that convergence of the chain is satisfactory. The posterior distribution of $\psi$ places more than half probability mass on the interval $(-1,0)$  with $P(\psi<0|data)=0.591$ and its 95\% posterior credible interval is (-0.395,0.340), giving support to negative correlation among clustering at different times. This result is in line with Figure~\ref{fig:WBCclust}, which shows the posterior distribution of $m_{tj}$ in \eqref{eq:lik_wbc} (posterior mean and 95\% credible intervals) for each time $t$. These parameters are the average level of WBC for each patient, accounting for the effect of CTX and GM-CSF, over time. The estimates are colored according to the cluster each patient belongs to. Posterior clustering allocation is estimated by minimizing the  Binder's loss function. From Figure \ref{fig:WBCclust} it is evident that clustering is driven by the value of the posterior mean of $m_{tj}$  for each time points.  At time 1 (day 4), most patients (except 1) have a baseline WBC counts approximately equal to 2 and as such they are clustered together. After the start of the chemotherapy ($t=2$), they  experience a drop in WBC counts, except few patients who seem to be less responsive.   As expected, more variation in patients's profiles is evident in the recovery period, as some patient recover much faster.

\begin{figure}[h]
\begin{subfigure}[t]{0.50\textwidth}
\centering
\includegraphics[width=0.9\textwidth]{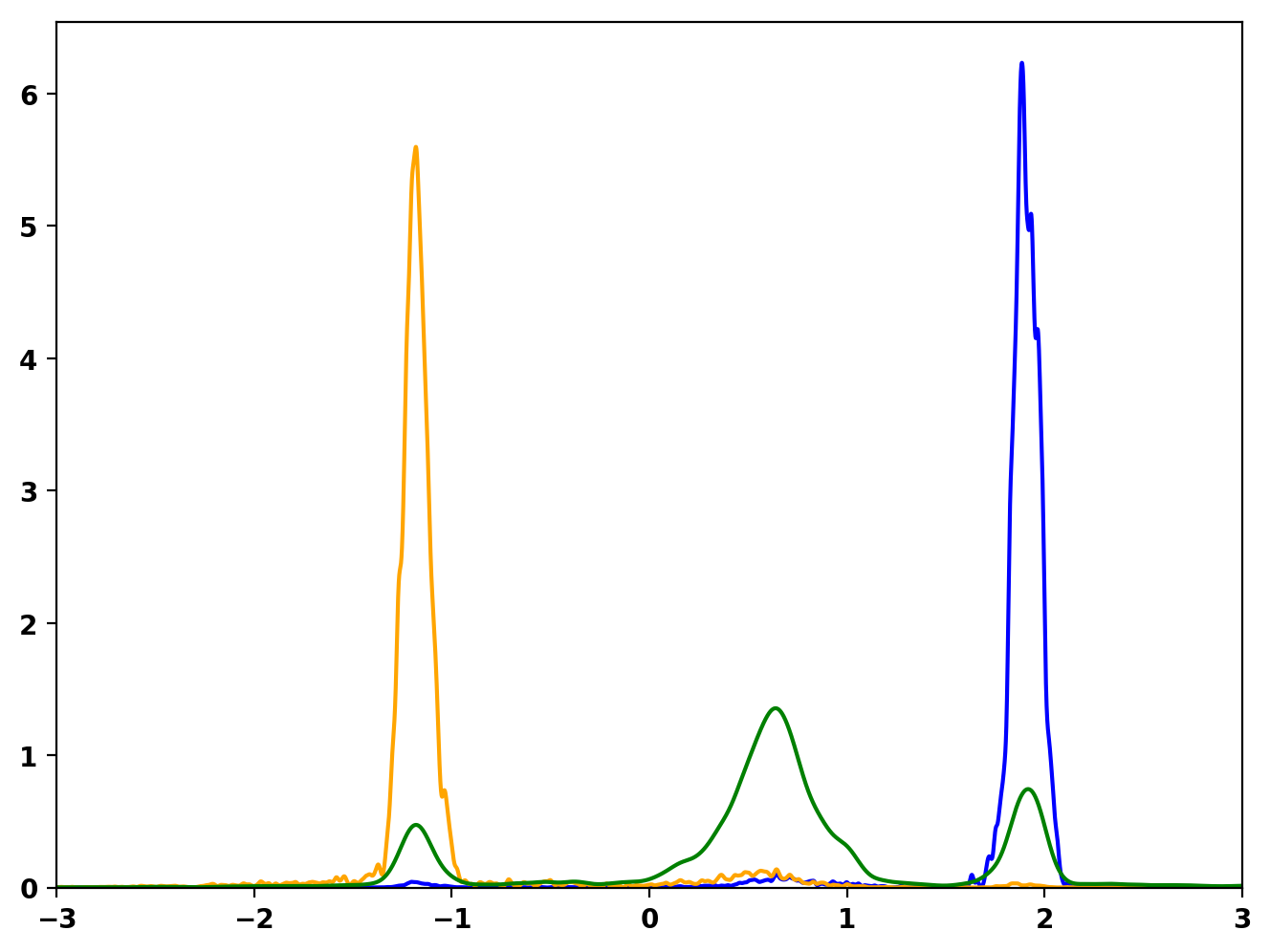}
\subcaption{AR1-DP}
\label{fig:postdens}
\end{subfigure}
\begin{subfigure}[t]{0.50\textwidth}
\centering
\includegraphics[width=0.9\textwidth]{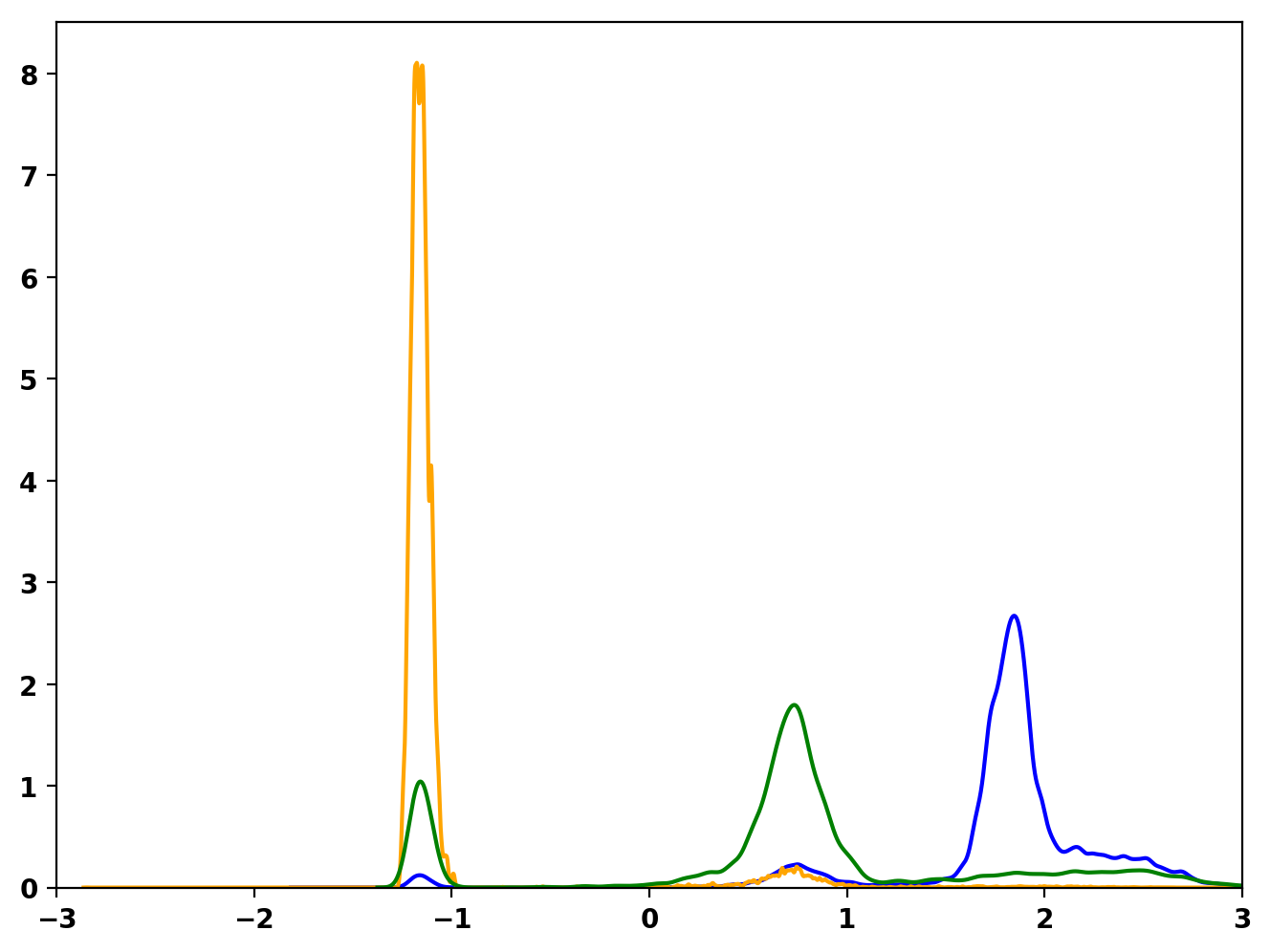}
\subcaption{Taddy}
\label{fig:taddydens}
\end{subfigure}
\begin{subfigure}[t]{0.50\textwidth}
\includegraphics[width=0.9\textwidth]{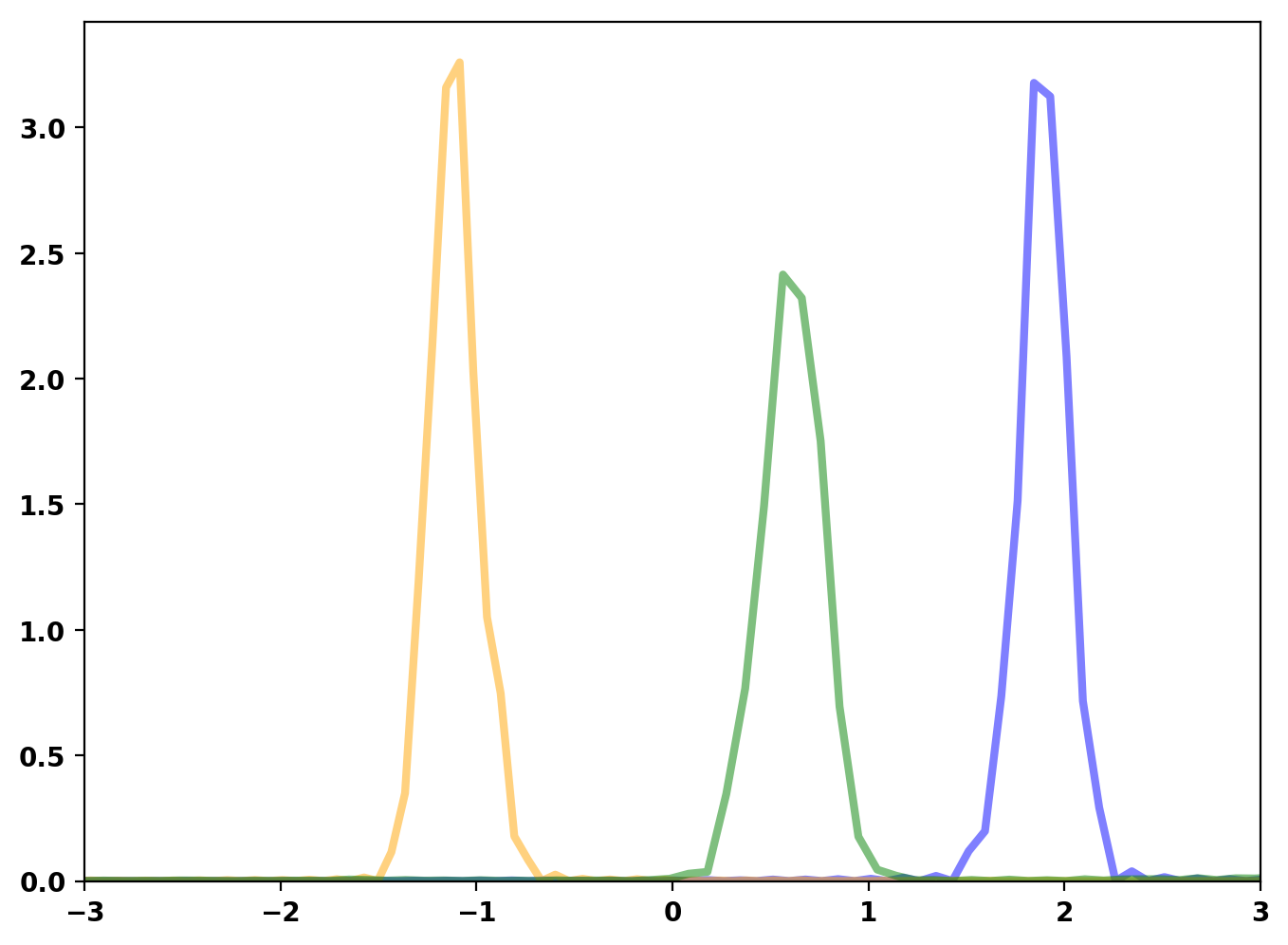}
\subcaption{DeY-K, $\theta \sim \text{U}(0,1)$}
\label{fig:kottas_posi}
\end{subfigure}
\begin{subfigure}[t]{0.50\textwidth}
\includegraphics[width=0.9\textwidth]{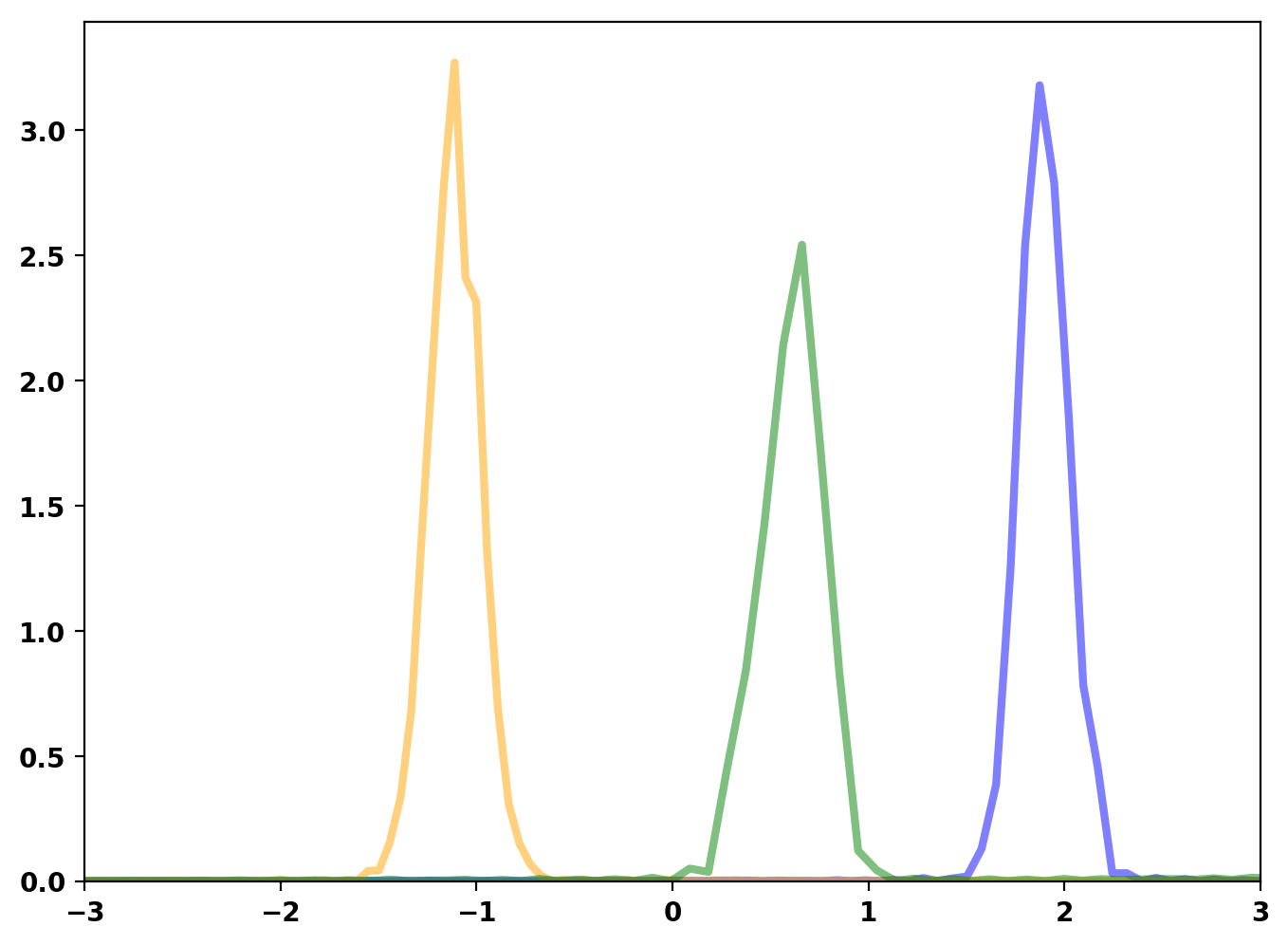}
\subcaption{DeY-K, $\theta \sim \text{U}(-1,1)$}
\label{fig:kottas_nega}
\end{subfigure}
\caption{\scriptsize{Posterior predictive density of  $m_t^{new}$  for $t=4$ (blue lines), 9 (orange), 14 (green) under our model (a), 
Taddy (b), and the model of \cite{deyoreo2018modeling} with $\theta\sim\textrm{Uniform}(0,1)$ (c) and $\theta \sim \textrm{Uniform}(-1,1)$ (d).}}
\label{•}
\end{figure}

Figure~\ref{fig:postdens} reports the posterior predictive  densities of  $m_t^{new}$ for the three time points. The posterior distribution at day 14, in green, shows clearly the presence of three clusters, and is centred between the other two time periods as expected. At day 14, eight patients retain the same WBC level as day 9, 36 subjects show positive fit the mixture model in \eqref{eq:lik_wbc} replacing  our AR1-DP with: i) the prior proposed by \cite{taddy2010autoregressive} by replacing \eqref{eq:nonlin} and \eqref{eq:trans} with \eqref{eq:taddy}; ii) the prior proposed by \cite{deyoreo2018modeling} by replacing \eqref{eq:nonlin} and \eqref{eq:trans} with\eqref{eq:Kottas_bit}. In the former case, we have specified a $\text{U}(0,1)$ as  prior distribution for $\psi$ since $\psi$ is a parameter in $(0,1)$. For the  latter case, we assume that, for  $t=4,9, 14$-th day, observations $Y_{tj}$, for $j=1,\ldots,n$, are i.i.d. $f(y_{tj}\mid G_t)$, where we set
\begin{eqnarray*}
f(y_{tj}|G_t) &=& \sum^{\infty}_{m=1} p_{m,t} \text{N}(y_{tj}|\mu_{m,t}+ \mathbf{x}_{tj}\boldsymbol{\gamma}_t ,\tau^{-1}_{m})\\
\mu_{m,0}\sim \text{N}(0,10), &\;\;\;\;&  \mu_{m,t}|\mu_{m,t-1} \sim \text{N} (\theta \mu_{m,t-1},10)\\
\tau_{m} &\iid& \text{Gamma}(1,1)\\
{\bm \gamma}_t &\iid& \text{N}_2(\bm 0, 10 \ \mathbb{I}_2), \qquad t=4,9, 14.
\end{eqnarray*}
The prior specification is completed by assuming that $M\sim\textrm{Gamma}(4,4)$ and $\psi\sim\textrm{Uniform}(0,1)$. The parameter $\theta$ is also assumed to be random. We  assume a priori independence, when not differently specified, among blocks of parameters. Computations are performed by truncating the infinite stick-breaking representation, as  for our model, at a truncation level $J=50$.  Observe that \cite{deyoreo2018modeling} assume a positive random $\psi$ and that $\theta$ represents the autocorrelation parameter between the cluster locations; see the discussion in Section \ref{sec:comp}. For a fair comparison, we  match the prior specification for the two models  in such a way that a priori the expected number of clusters is $4$ for each $t$. 

\begin{figure}[t]
\begin{subfigure}[t]{1\textwidth}
\centering
\captionsetup{width=.8\linewidth}
\includegraphics[width=.39\textwidth]{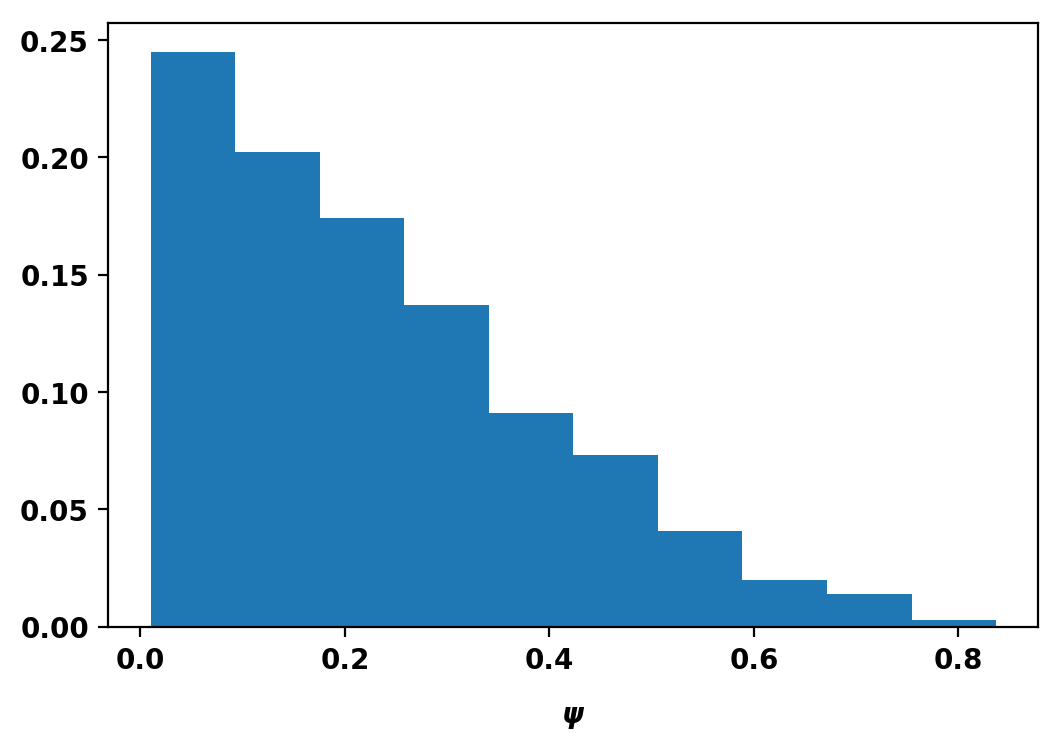}
\includegraphics[width=.39\textwidth]{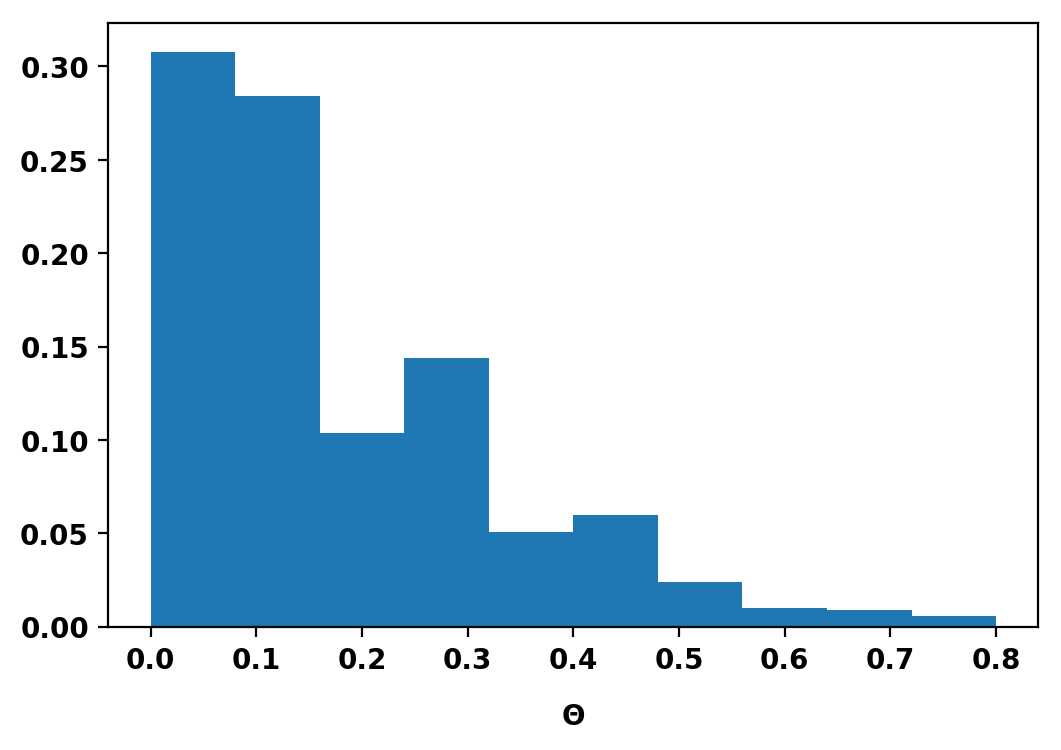}
\subcaption{DeY-K, $\theta \sim \text{U}(0,1)$}
\label{fig:post_kottas_posi}
\end{subfigure}
\begin{subfigure}[t]{1\textwidth}
\centering
\captionsetup{width=.8\linewidth}
\includegraphics[width=.39\textwidth]{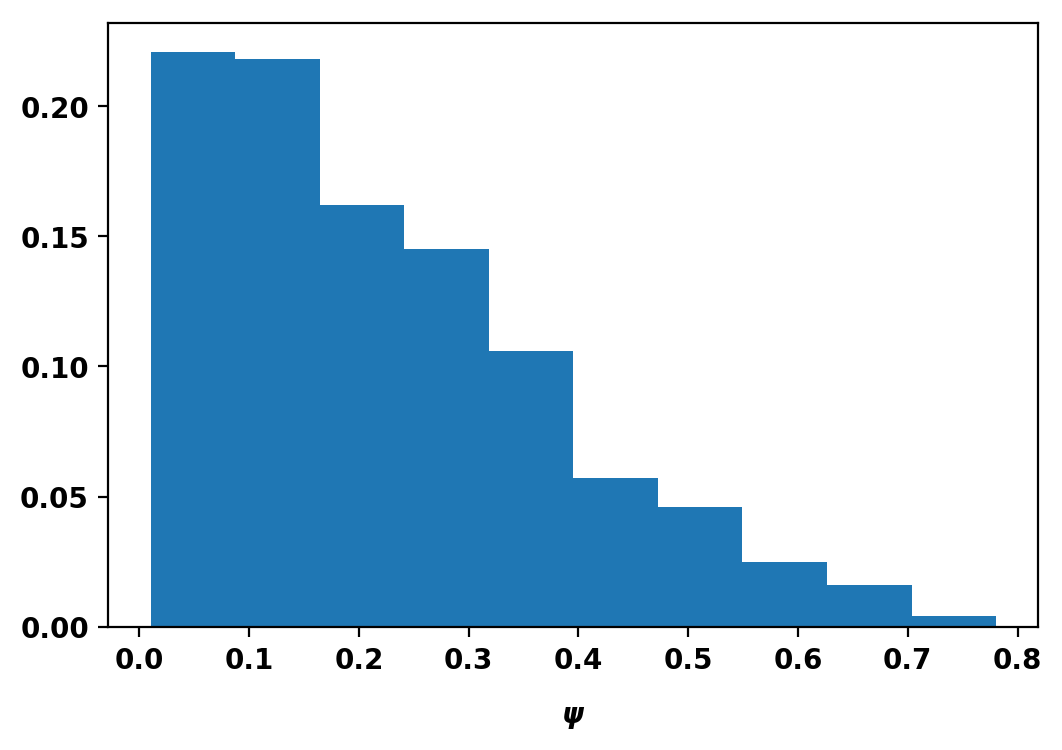}
\includegraphics[width=.39\textwidth]{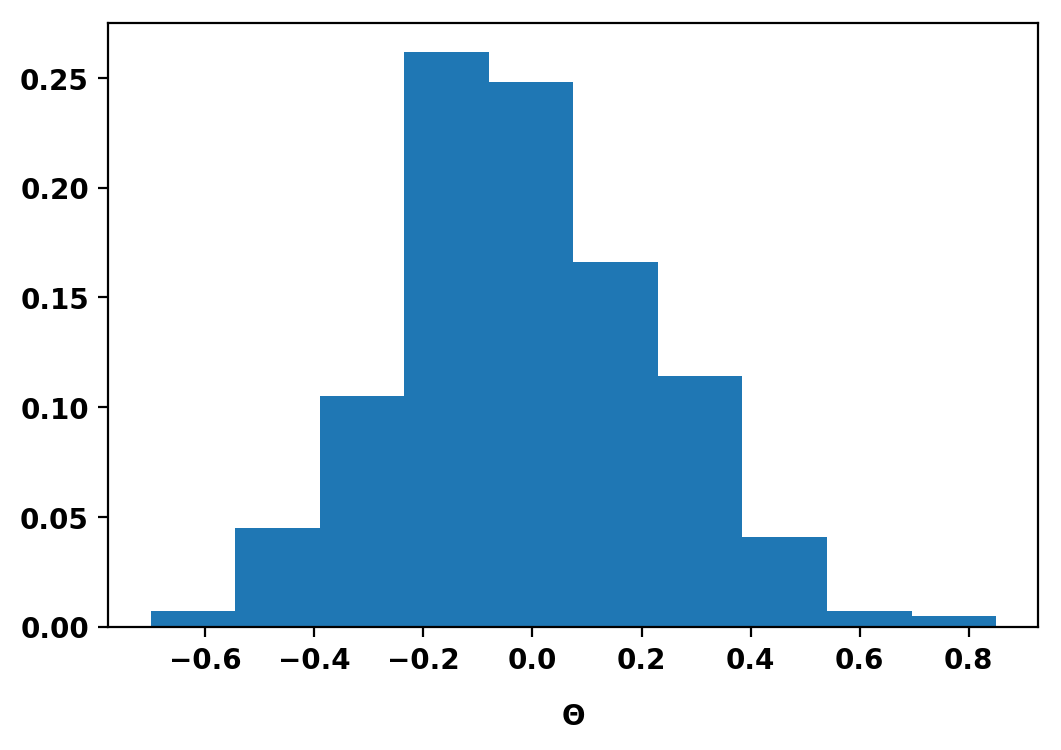}
\subcaption{DeY-K, $\theta \sim \text{U}(-1,1)$}
\label{fig:post_kottas_nega}
\end{subfigure}
\caption{\scriptsize{Marginal posterior distributions of $\psi$ and $\theta$ obtained under the model by \cite{deyoreo2018modeling}.}}
\label{fig:post_kottas}
\end{figure}

Figure \ref{fig:taddydens}, Figure \ref{fig:kottas_posi} and Figure \ref{fig:kottas_nega} display the posterior predictive distributions under the stick-breaking prior specified in the model of \cite{taddy2010autoregressive} and the model of  \cite{deyoreo2018modeling} for $\theta\sim \textrm{Uniform}(0,1)$ and  $\theta \sim \textrm{Uniform}(-1,1)$, respectively. By comparing results in \cite{taddy2010autoregressive} (Figure~\ref{fig:taddydens}) to the posterior predictive density of  $m_t^{new}$ under our model (Figure~\ref{fig:postdens}), we see that the two  distributions are similar, both recovering three clusters for $t=14$, though 
the convergence is worse for the model of \cite{taddy2010autoregressive}, as it is evident from the presence of more ``bumps'' in Figure \ref{fig:taddydens} than Figure \ref{fig:postdens}. The model of \cite{taddy2010autoregressive} has more parameters. Figure~\ref{fig:kottas_posi} and Figure~\ref{fig:kottas_nega} show that  the dynamic autoregressive DP by  \cite{deyoreo2018modeling} under both priors leads to posterior predictive densities similar to ours (but scales are different), but no clustering structure is evident (the posterior distribution of the number of clusters has mode at 1). However, since the cluster variance does not have autocorrelation parameters, the predictive densities of Day 14 of the model by \cite{deyoreo2018modeling}  does not capture the increased variance of the cluster. In particular, the model in \cite{deyoreo2018modeling} with $\theta\sim \textrm{Uniform}(0,1)$ cannot describe negative correlation of the $(\mu_{mt})$. Accordingly, the marginal posterior distributions of both $\psi$ and $\theta$ have a large density mass near $0$ (Figure \ref{fig:post_kottas_posi}), the boundary of the prior distribution, which means that the model tries to explain the negative correlation as ``independence". 

The same comment applies to the prior model introduced in \cite{taddy2010autoregressive}. In particular, the marginal posterior distribution of $\psi$, not shown here, has a large density mass near $0$, the boundary of the prior distribution. If, with the dynamic DP mixture model of \cite{deyoreo2018modeling} we assume $\theta\sim\textrm{Uniform}(-1,1)$, it is possible to describe negative correlation through $\theta$.  As a result, the marginal posterior distribution of $\psi$ has still a large density mass near $0$ (Figure \ref{fig:post_kottas_nega}), the boundary of the prior distribution, which means that the model 
fails to capture temporal dependence between the time points, while negative correlation is detected by the posterior distribution of  $\theta$.  However, the dynamic DP mixture model by \cite{deyoreo2018modeling} is more computationally efficient, because  they employ a simpler transformation  (see \eqref{eq:Kottas_bit}) which does not require the use of a particle MCMC algorithm, unlike our and the model of \cite{taddy2010autoregressive}.

\section{Breast cancer incidence data}
\label{sec:breastcancer}
We consider data on breast cancer annual incidence of 100 metropolitan statistical areas (MSA) in the United States from 1999 to 2014 (see CDC WONDER Database at \texttt{https://wonder.cdc.gov/wonder/help/cancermort-v2005.html}). The  outcome of interest is the incidence (total number of deaths per population area) in each area and  time period with  the aim of  identifying spatial and spatio-temporal patterns of disease. We model incidence data through a Poisson distribution and include spatial and temporal random effects. In particular, let $Y_{tj}$,  $N_{tj}$ denote  the number of breast cancer deaths and the population in area $j$ at time $t$ respectively. We assume the following model
\begin{align*}
 Y_{tj}|N_{tj}, \lambda_{tj} &\ind \text{Poisson}(\lambda_{tj} N_{tj})  \quad j = 1,\ldots,n, \\
 \log(\lambda_{tj}) &= \mu_{tj} + \phi_j,\\
 \mu_{tj} |G_t &\iid G_t \ \textrm{ for each } t = 1,\ldots,T \\
 ( G_t)_{ t\geq 1} &\sim\text{AR1-DP} (\psi,M,G_0).
\end{align*}
Here  $n = 100$, $T = 11$, and $\lambda_{tj}$ is the expected rate (risk) in area $j$ at time $t$, which is modelled by $\mu_{tj}$ and a random effect $\phi_{j}$. We assume that $G_{0}$ is a Gaussian distribution with parameter $(0,10)$, and we assume a conditional autoregressive (CAR)  prior distribution for the vector of the random effects $\bm\phi=(\phi_1,\ldots,\phi_n)'$. Precisely, we assume that
\begin{equation}
\label{eq:CARprior_breastcancer}
{\bm\phi}|\mathbf{W},\tau^2,\rho \sim \text{N}\left(\mathbf{0}, \tau^2\left[\rho(\text{diag}(\mathbf{W}\bm 1)-\bm W)+(1-\rho)\mathbf{I}_n \right]^{-1} \right), 
\end{equation}
where $\bm W$ is the $n\times n$ matrix which defines the neighbourhood  structure of the $n$ area units. Specifically, the $(j,k)$-th element of $\bm W$ is equal to 1 if area $j$ and $k$ share a common border, and it is equal to 0 otherwise. Finally, $\bm 1$ is a vector with $n$ elements equal to 1 \citep{Leroux_etal_2000}. In particular, the variance parameter $\tau^2$ controls the amount of variation between the random effects, while $\rho$ controls the strength of the spatial association between the area-specific random effects with $\rho=0$ corresponding to independence and $\rho$ close to one corresponding to increasingly strong spatial association. Typically, $\rho$ is assumed to be positive \citep{Wall2004}, and we assume $ \rho \sim \textrm{DiscreteUniform}(0,0.05,0.10,0.15,\ldots,0.95)$ for computational reasons, so that the number of distinct matrices in \eqref{eq:CARprior_breastcancer} to be inverted in the MCMC algorithm is limited. Finally we assume that $\tau^2$, $\rho$ and $M$ are a priori independent, with $\tau^2 \sim\textrm{Uniform}(0,10)$,  and $M \sim \textrm{Gamma}(4,4)$. This implies that the prior mean of the number $K$ of clusters is equal to $5$.  Also, an inverse-gamma prior distribution for $\tau^2$ can also be considered. Still  the posterior of $\tau^2$ is gamma-shaped, giving mass to values between 0.015 and 0.045. 

\begin{figure}[h!]
\centering
\includegraphics[width=.39\textwidth]{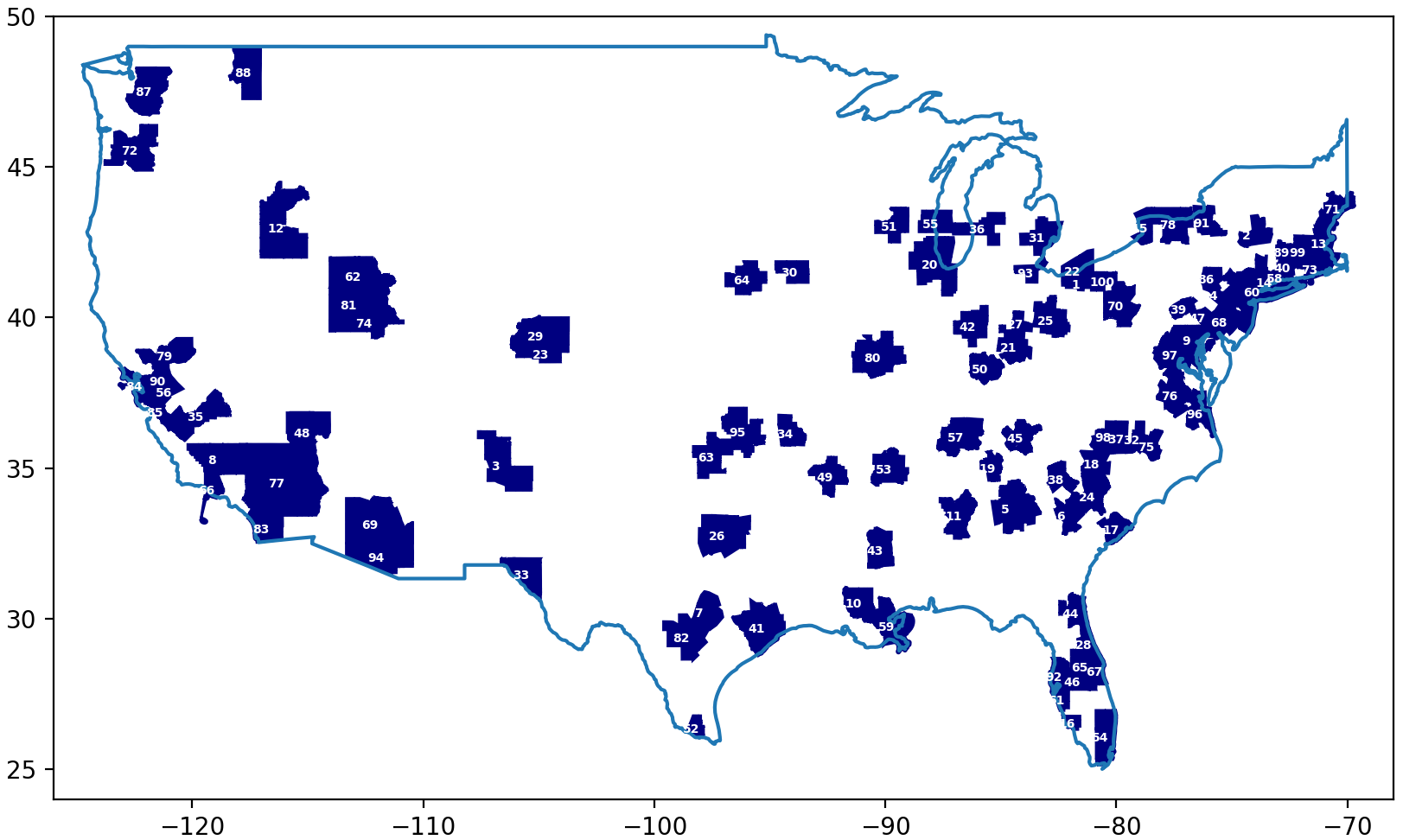}
\includegraphics[width=.39\textwidth]{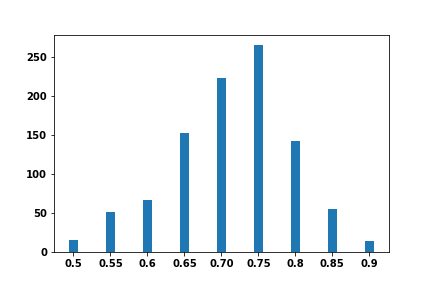}
\vspace{-5mm}
\caption{\scriptsize{Location across USA of the MSAs with labels (left) and posterior distribution of the spatial correlation parameter $\rho$ (right).}}
\label{fig:rho_breastcancer}
\end{figure}

\begin{figure}[H]
\centering
\includegraphics[width=.38\textwidth]{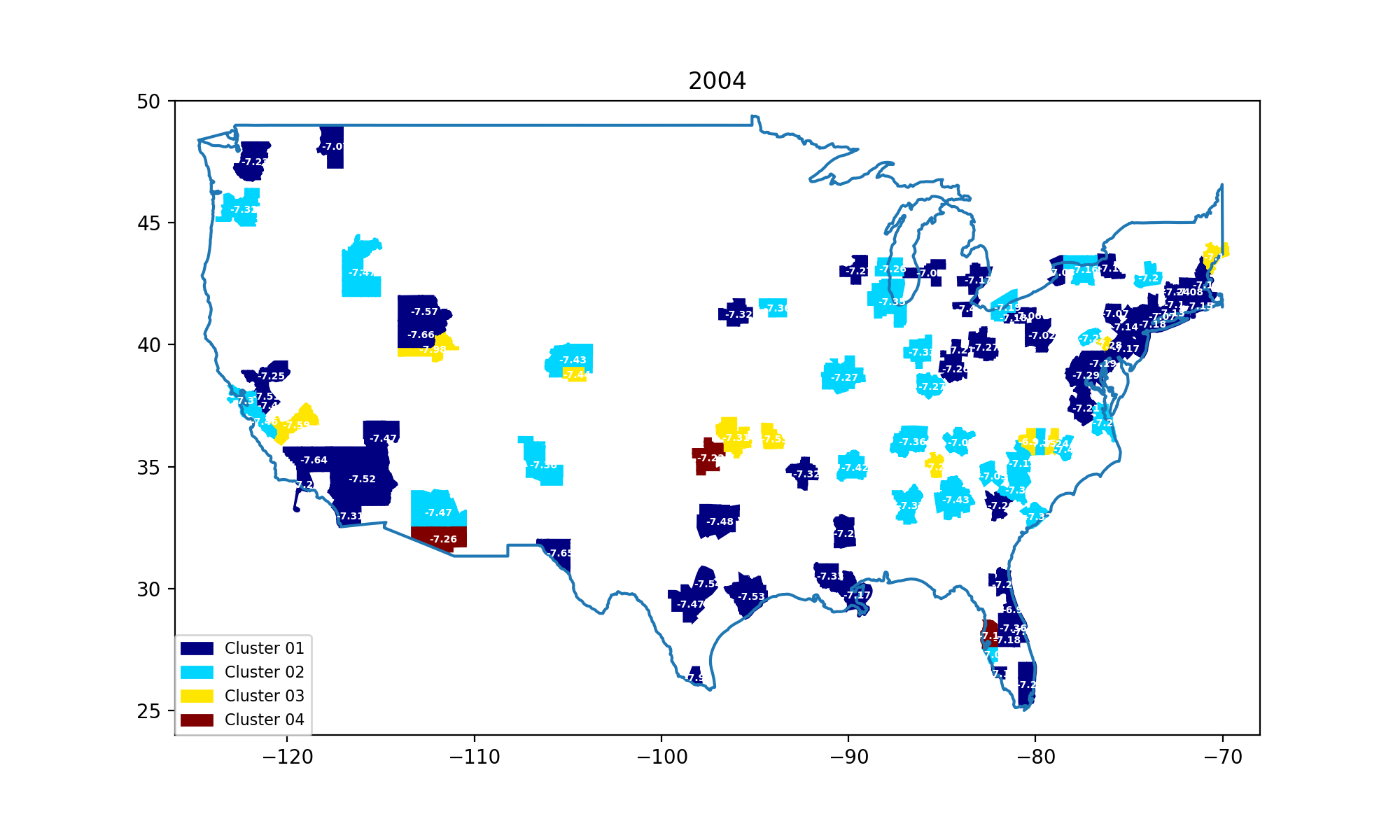}
\includegraphics[width=.38\textwidth]{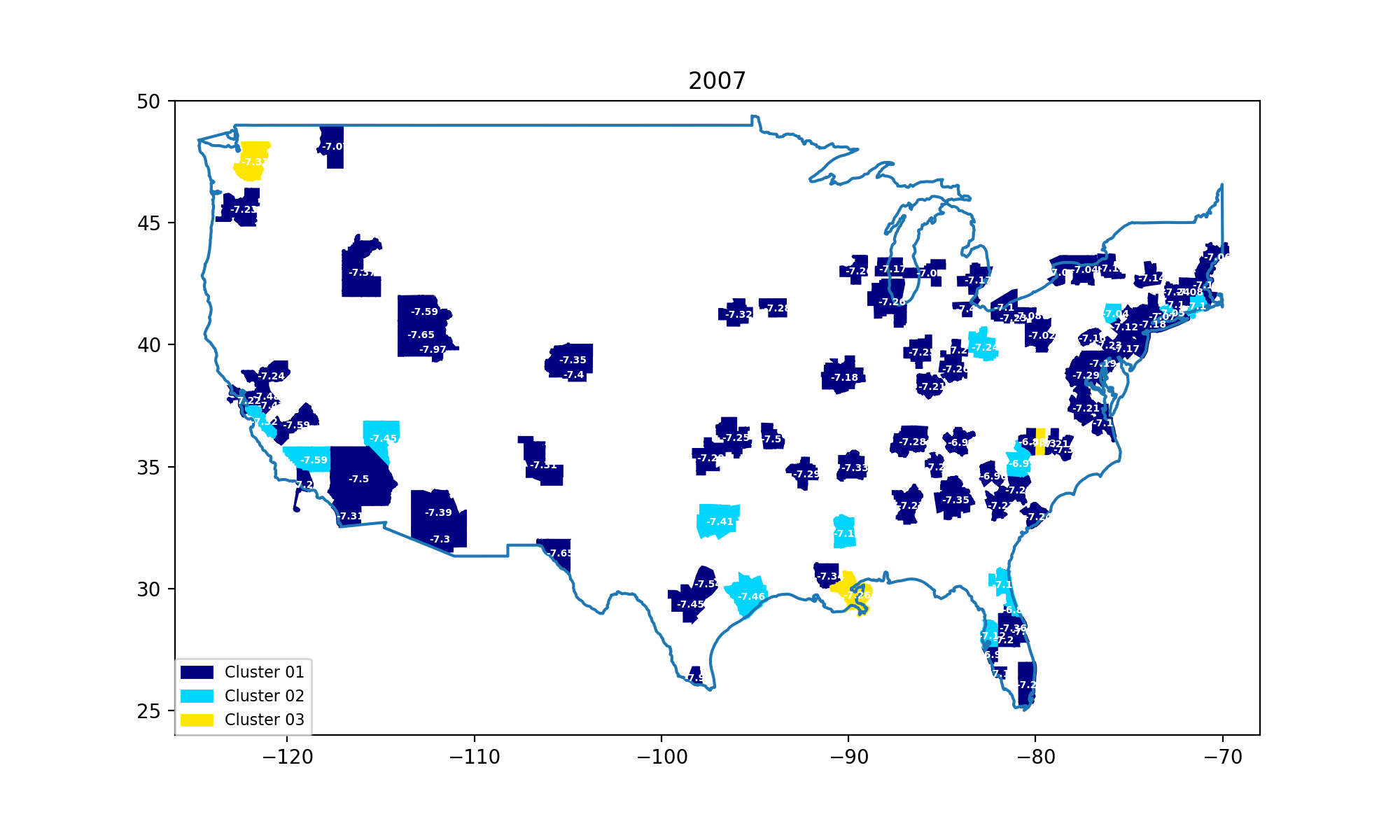}
\includegraphics[width=.38\textwidth]{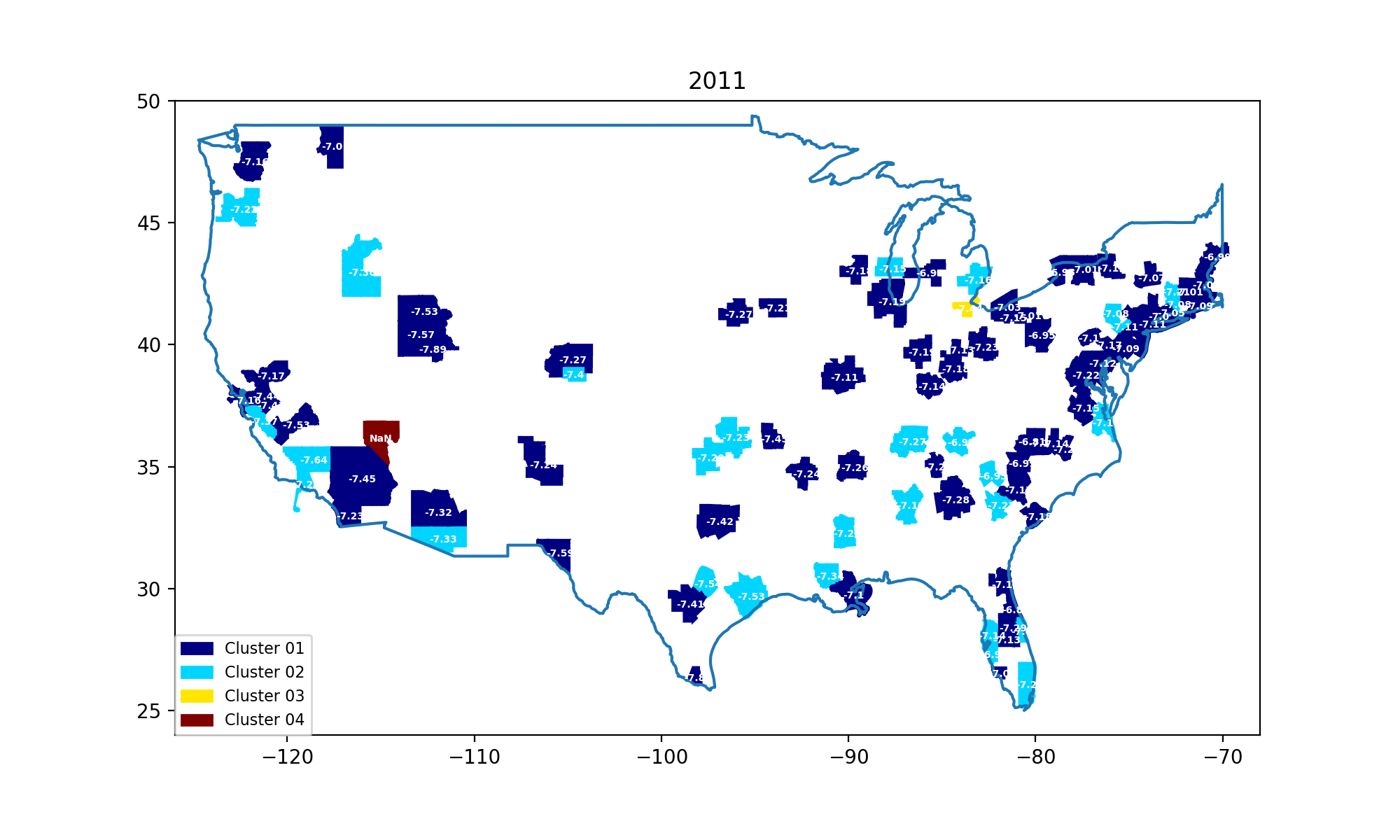}
\includegraphics[width=.38\textwidth]{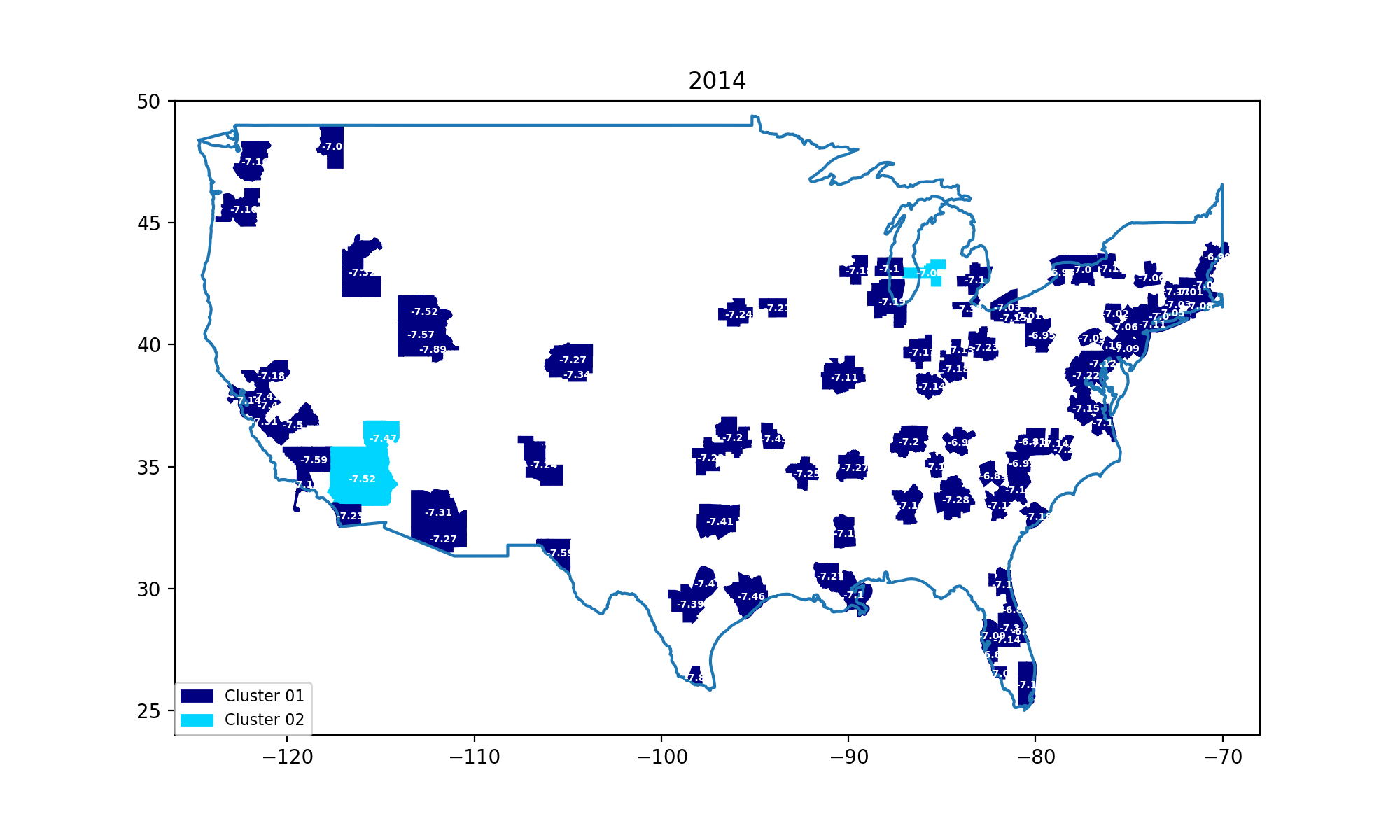}
\vspace{-5mm}
\caption{\scriptsize{Posterior clustering estimates  of the MSAs in 2004 (top left), 2007 (top right), 2011 (bottom left) and 2014 (bottpm right). Each colour denotes a  cluster.}}
\label{fig:cluster_breastcancer}
\end{figure}

Population data $\{N_{tj} \}$ of each MSA from 2000 to 2010 are available from the U.S. Census Bureau at \texttt{https://www.census.gov/topics/population.html}, while the population data for 2011-2014 have been estimated by means of a linear  interpolation. We modify the MCMC algorithm described in Section \ref{sec:mixturemodel} in order to take into account the spatial correlation.  We run the algorithm fixing $J=50$  for 20,000 iterations, discarding the first 10,000 iterations as burn-in, and thinning every 10. The final sample size is 1,000. Figure~\ref{fig:rho_breastcancer} (left) displays the MSAs over the US map. The marginal posterior distribution of  $\rho$ is shown in Figure~\ref{fig:rho_breastcancer} (right). The posterior median is 0.75 and there is evidence of spatial association among the MSAs, although not very strong  \citep[see][]{Wall2004}. Figure~\ref{fig:cluster_breastcancer} reports the estimated clustering structure of the MSAs  for $t=2004$, 2007, 2011 and 2014, with 4, 3, 3 and 2 estimated clusters, respectively.

\begin{figure}[ht!]
\centering
\includegraphics[width=0.7\textwidth]{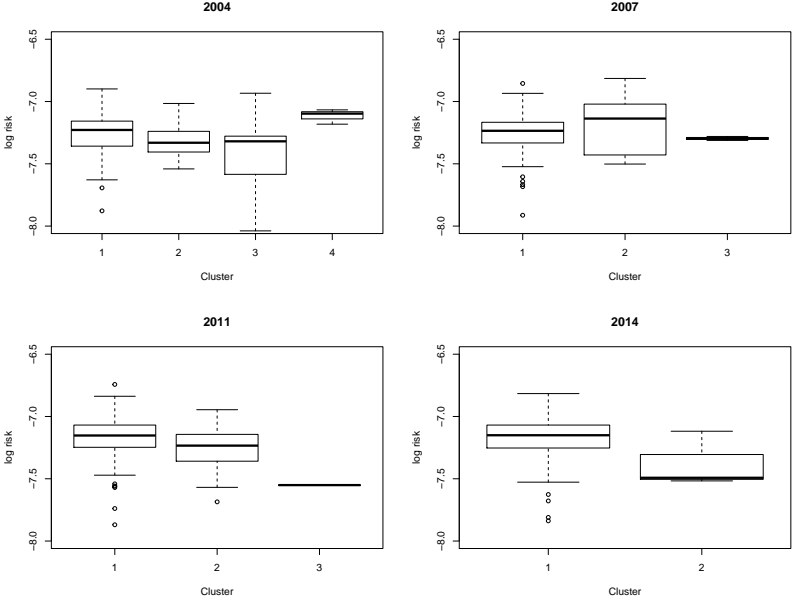}
\caption{\scriptsize{Empirical distribution of the log of the risk of breast cancer per cluster for $t= 2004, 2007,2011,2014$.}} 
\label{fig:postinc}
\end{figure}

The marginal posterior distribution of the autocorrelation parameter $\psi$, not shown here, is concentrated on values close to 1, i.e. the posterior median is 0.93 and the 95\% posterior credible interval is (0.88, 0.96), indicating a strong dependence from one year to the next one. As such  the cluster estimates do not change substantially from one year to the next, as it is evident from Figure~\ref{fig:cluster_breastcancer}. In the year 2004 there are four clusters, while toward the end of the period (2011 and 2014) the number of clusters (not singletons) decreases to two, showing more homogeneity among the MSA. In Figure \ref{fig:postinc} we report the boxplot of the empirical distribution of the  log-risk per cluster for $t = 2004,2007,2011,2014$, which shows that  the clustering structure captures differences in the empirical log/risk, with the risk slightly increasing over time.   

\section{Conclusions}
\label{sec:conclusion}

In this work we have introduced a prior distribution for a collection of discrete RPMs which allows for autoregressive (of order $1$) temporal dependence across distributions. The strength of the dependence is controlled by an autocorrelation parameter, which can capture a variety of dependence structure: from complete independence to identical distribution over time. We have shown that our construction allows for more flexible modeling than competitors models proposed in \cite{taddy2010autoregressive} and \cite{deyoreo2018modeling}. Since our model is based on the DP, we can perform time-dependent clustering, where both the number of clusters and cluster membership can change over time and the cluster configuration at time $t$ may depend on the one at time $t-1$. Once again at time $t$ we could have the same clustering structure as at time $t-1$ or we could have a completely different allocation. We have shown the wide-applicability of our model though widely differing applications and an extensive simulations study. Posterior inference is performed through a particle MCMC algorithm.  Future work will involve the investigation of higher order of dependence and more complex dependence structure such as spatial.    

 \section*{Acknowledgements}

We thank the CALGB for permission to use the data in Section \ref{sec:dose}. 
\bibliography{references_SBAR1}
\bibliographystyle{plainnat}

\begin{table}[ht!]
\begin{center}
\includegraphics[width=0.8\textwidth]{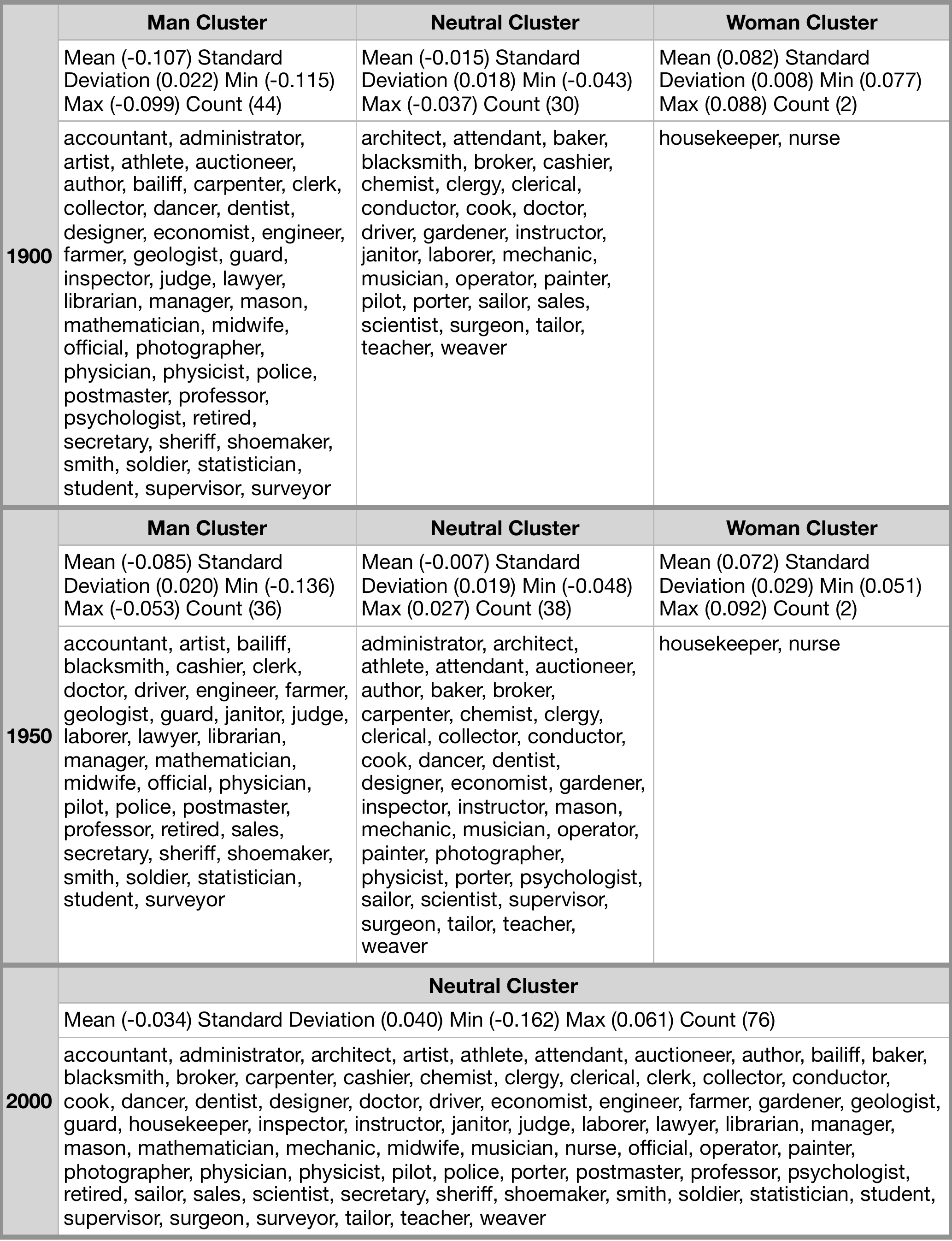}
\end{center}
\caption{\scriptsize{Cluster estimates of the occupational bias data for $t=1900, 1950, 2000$.}}
\label{tab:occutable}
\end{table}

\begin{landscape}
\begin{table}[ht!]
\begin{center}
\includegraphics[width=1.5\textwidth]{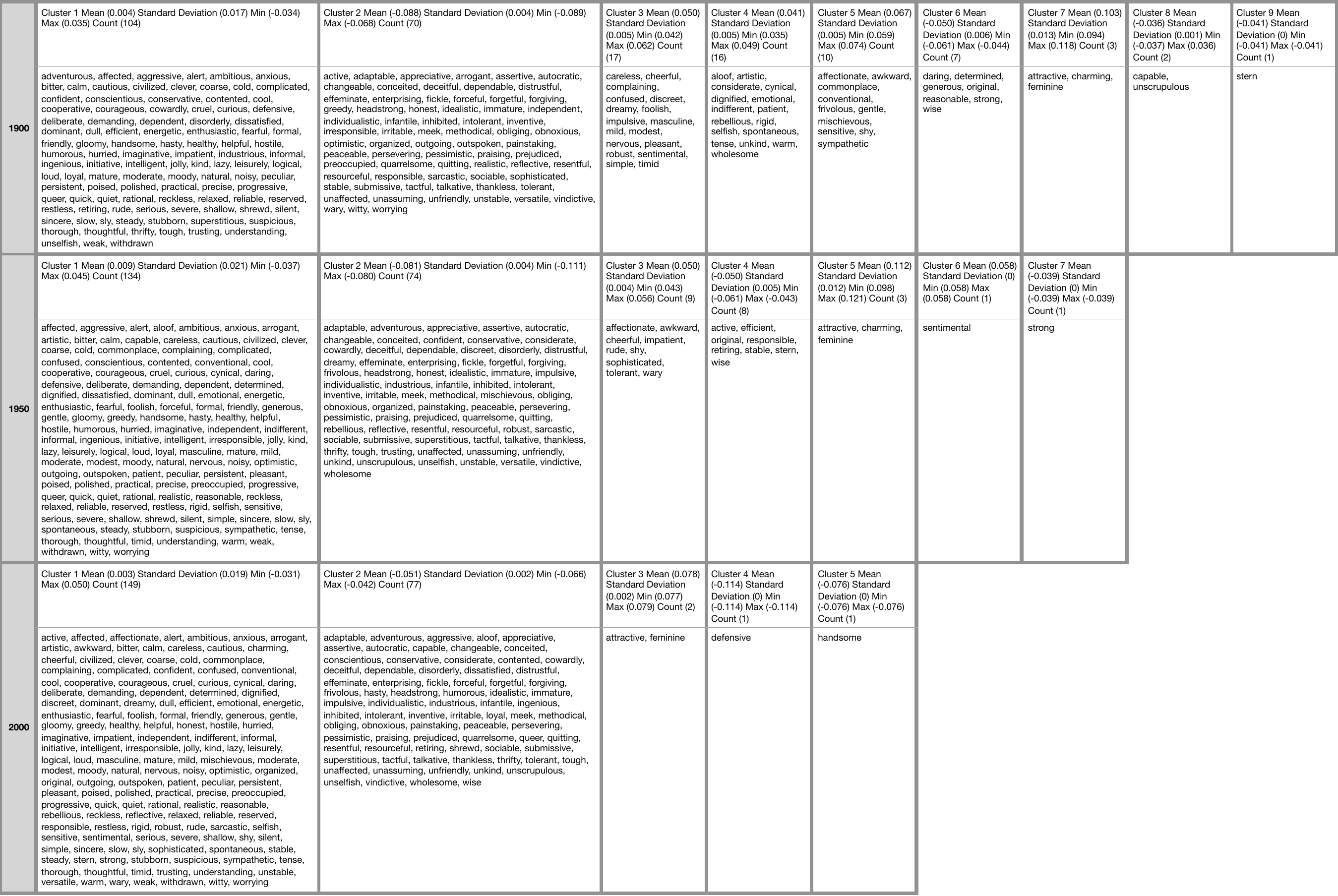}
\end{center}
\caption{\scriptsize{Cluster estimates of the adjective bias data for $t=1900, 1950, 2000$.}}
\label{fig:adjtable}
\end{table}
\end{landscape}

\begin{table}[H]
\begin{center}
\includegraphics[width=\textwidth]{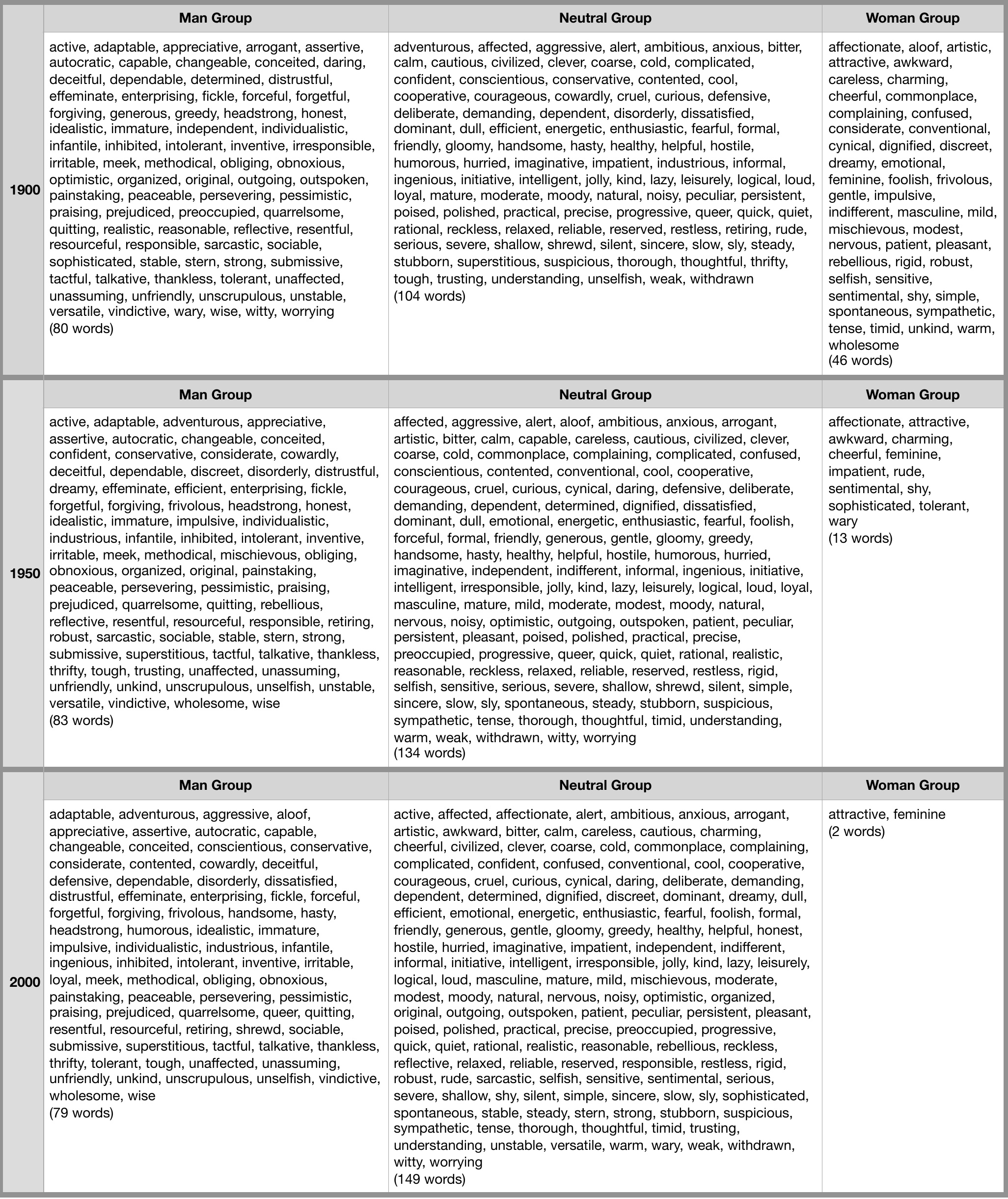}
\end{center}
\caption{\scriptsize{Merged groups for the adjective bias data. Groups obtained by merging the estimated clusters from Table~\ref{fig:adjtable}.}}
\label{fig:adjtable2}
\end{table}

\newpage
\section*{APPENDIX A: Details of the Gibbs sampler}

Here, we provide some details about sampling from the full-conditional $p( \bw_1,\ldots, \bw_T\mid \bs_1 \ldots, \bs_T,\psi)=:  p_\psi( \bw_1,\ldots, \bw_T\mid \bs_1 \ldots, \bs_T)$
of the Gibbs Sampler in Section  \ref{sec:mixturemodel}.
Observe that the model specified in \eqref{eq:pesetti} - \eqref{eq:trans} defines a Bayesian nonlinear temporal dynamic model \citep[see, for instance][]{PradoWest2010}.

Standard algorithm such as the Forward Filtering and Backward Sampling \citep{baum1967inequality,baum1970maximization} are not applicable due to the fact that 1-step predictive distributions $p_{\psi}(\bw_t \mid \bw_{t-1})$ cannot be derived analytically.  

%
%
Therefore, we employ  
particle MCMC \citep{andrieu2010particle}, using the prior to generate samples $\be_t$, where $\be_t=(\epsilon_{1t},\ldots,\epsilon_{t J-1})$.  In what follows for ease of notation we use an index to denote the conditioning to $\psi$ and we exploit the fact that 
$$ \mathcal{L}_{\psi} (\bw_t\mid\bw_{t-1})= \mathcal{L}_{\psi} (\be_t\mid\be_{t-1})
$$

Resampling strategies can be employed to improve the efficiency of the algorithm.

Let us denote the number of particles by  $R$. The SMC approximation becomes:
$$ \widehat{p}_{\psi}(\bw_1,\ldots, \bw_T\mid \bs_1 \ldots, \bs_T) = \sum_{r=1}^R \omega_T^r\delta_ {\boldsymbol{\epsilon}_{1:T}} . $$

\bigskip
\noindent\textbf{Conditional SMC Algorithm}\\
The standard multinomial resampling procedure below is the operation by which offspring particles at time $t$ choose their ancestor particles at time $t-1$, according to the weights of ancestor particles.
In what follows $r$ denotes the particle, $i$ the stick breaking r.v., $t$ the time, $\be_{1:T} = (\be_1^{B_1}, \be_2^{B_2}, \ldots,\be_{T-1}^{B_{T-1}}, \be_T^{B_T})$ is a path that is associated with the ancestral `lineage' $\boldsymbol{B}_{1:T} = (B_1,...,B_T)$, where $B_{t}, t = 1,...,T$ is the index which defines the ancestor particle of $\be_{1:T}$ at generation $t$.
\begin{itemize}
\item At time $t=1$, for $r\neq B_1$, sample $\be^r_1 \sim N(0,1)$. Compute weights $\omega_1$ and normalised weights $\widetilde{\omega}_1$:
\begin{align}
 \omega_1(\be^r_1) & = \frac{\left(\prod_{i}^J N(\epsilon_{1i};0,1)\right) \text{Multinomial}(\bs_1\mid \bw_1(\be^r_1)) }{   \left(\prod_{i}^J N(\epsilon_{1i};0,1)\right)  }\\
 & = \text{Multinomial}(\bs_1\mid \bw_1(\be^r_1))\\
 \widetilde{\omega}_1^r &= \frac{\omega_1(\be^r_1)}{\sum_{r=1}^R \omega_1(\be^r_1)}
\end{align} 
\item At time $t=2, \ldots, T$, let $\bo_t=(\omega^{1}_t , \ldots, \omega_{t}^R)$
\begin{itemize}
\item[-] Choose the parent at time $t-1$. For $r \neq B_t$, sample  $A^r_{t-1}\sim \text{Discrete} (\widetilde{\bo}_{t-1})$. In this step we choose the parent of each particle at time $t$ that needs to be generated.  
\item For $r \neq B_t$, sample $\be^r_t \sim \prod_{i}^J N(\epsilon_{it}; \psi \epsilon_{it-1}^{A^r_{t-1}} , 1- \psi^2)$. Set $\be^r_{1:t} = (\be_{1:t-1}^{A^r_{t-1}},\be_t^r)$.
\item Compute and normalise weights:
\begin{align}
 \omega_t(\be^r_{1:t} ) & = \text{Multinomial}(\bs_t\mid \bw_t(\be^r_t))\\
 \widetilde{\omega}_t^r &= \frac{\omega_t(\be^r_{1:t})}{\sum_{r=1}^R \omega_1(\be^r_{1:t})}
\end{align}
\end{itemize}
\item At time $T$ we can approximate the conditional posterior distribution of the weights by  
$$ \widehat{p}_{\psi}(\bw_1,\ldots, \bw_T\mid \bs_1 \ldots, \bs_T) =
\sum_{r=1}^R \widetilde{\omega}_T^r \delta_{\be_{1:T}^r} $$
from which we can obtain approximate samples by drawing an index from the discrete distribution with parameter $\widetilde{\omega}_T$, i.e. pick a trajectory $\be_1^r, \ldots , \be_T^r$ with probability $\widetilde{\omega}_T$. Note that we sample from the prior as proposal. Briefly, at time $T$, we have $R$ particles, each particle $j$ represents a trajectory $\epsilon_{1j}, \dots,\epsilon_{Tj}$ ($j = 1,\dots,R$). For each particle, there is a weight $\omega_j$ ($j = 1,\dots,R$).
We sample one particle out of $R$ particles with a probability proportional to the  weight $\omega_j$.
After sampling $\epsilon_1, \dots,\epsilon_T$, we are able to calculate the corresponding $\xi_1,\dots,\xi_T$ and $w_1,\dots,w_T$. See also \cite{andrieu2010particle}, Section 2.2.2.  The weights at time $T$ do not depend on those computed at previous times.

\item We can now approximate the conditional distribution of $\bs_{1:T}$: 
\begin{align}
\widehat{p}_{\psi} (\bs_{1:T} \mid rest) &= \widehat{p}_{\psi} (\bs_1) \prod_{t=2}^T \widehat{p}_{\psi} (\bs_t \mid \bs_{1:t-1})
\\
\widehat{p}_{\psi} (\bs_t \mid \bs_{1:t-1}  )& = \frac{1}{R} \sum_{r=1}^R \omega_t(\be_{1:t}^r) = \frac{1}{R} \sum_{r=1}^R \text{Multinomial}(\bs_t\mid \bw_t(\be^r_t))\
\end{align}

\end{itemize}

To update $\psi$ and $\be_{1:T} $ we employ Particle Marginal M-H Sampler \citep[see][Sect. 2.4.2]{andrieu2010particle}:

\begin{description}
\item[Step 1: initialization, $j = 0$.]
Initialize $\psi$ as $\psi^0$, $\be_{1:T}(0)$ and $\boldsymbol{B}_{1:T}(0)$ arbitrarily.

\item[Step 2: for iteration $j \geq 1$,] 
\mbox{} \newline
\begin{itemize}
\item sample $\psi^{\star} \sim TN(\psi^{j-1}, \tau^2 , -1,1)$ (truncated Gaussian proposal). 
\item with probability 
$$ \min\left\{1, \alpha \right\}$$
where 
$$ \alpha = \frac{\text{prior}(\psi^{\star}) }{\text{prior}(\psi^{j-1} ) } \frac{p(\be_{1:T}|\psi^{\star}) }{p(\be_{1:T}|\psi^{j-1}) } \frac{ TN(\psi^{j-1}; \psi^{\star}, \tau^2 , -1,1)   }{TN(\psi^{\star};\psi^{j-1}, \tau^2 , -1,1)}
$$
accept  $\psi^{\star}$. In other words, $\psi^{j} = \psi^{\star}$. Otherwise, $\psi^{j} = \psi^{j-1}$.
\item run a conditional SMC algorithm targeting $p_{\psi^{j}}(\be_{1:T}|\bs_{1:T})$ conditional on $\be_{1:T}(j-1)$ and $\boldsymbol{B}_{1:T}(j-1)$, and
\item sample $\be_{1:T}(j)\sim\widehat{p}_{\psi^{j}}(\cdot|\bs_{1:T})$  (and hence $\boldsymbol{B}_{1:T}(j)$ is also implicitly sampled).
\end{itemize}

\end{description}

\bigskip
\section*{APPENDIX B: Data and further posterior inference for the gender stereotypes example - word emebeddings}

All word lists  for this application are from \cite{Garg_etal_2018}, available on their 
GitHub page at \texttt{https://github.com/nikhgarg/EmbeddingDynamicStereotypes}. In this Appendix  we explain how the two  variables of interest, gender embedding bias referring to occupational and adjective lists, respectively, have been derived. 
%

First of all, we consider four collated work lists from \cite{Garg_etal_2018} to represent each gender (men, women) and neutral words (occupations and adjectives), which we denote here by  $W_{man}$, $W_{woman}$, $W_{occu}$ and $W_{adj}$. The lists follow here: 
\begin{itemize}
\item \textit{Man words} ($W_{man}$):  he, son, his, him, father, man, boy, himself, male, brother, sons, fathers, men, boys, males, brothers, uncle, uncles, nephew, nephews.
\item \textit{Woman words} ($W_{woman}$): she, daughter, hers, her, mother, woman, girl, herself, female, sister, daughters, mothers, women, girls, femen, sisters, aunt, aunts, niece, nieces.
\item \textit{Occupations} ($W_{occu}$): janitor, statistician, midwife, bailiff, auctioneer, photographer, geologist, shoemaker, athlete, cashier, dancer, housekeeper, accountant, physicist, gardener, dentist, weaver, blacksmith, psychologist, supervisor, mathematician, surveyor, tailor, designer, economist, mechanic, laborer, postmaster, broker, chemist, librarian, attendant, clerical, musician, porter, scientist, carpenter, sailor, instructor, sheriff, pilot, inspector, mason, baker, administrator, architect, collector, operator, surgeon, driver, painter, conductor, nurse, cook, engineer, retired, sales, lawyer, clergy, physician, farmer, clerk, manager, guard, artist, smith, official, police, doctor, professor, student, judge, teacher, author, secretary, soldier.
\item \textit{Adjectives} ($W_{adj}$): headstrong, thankless, tactful, distrustful, quarrelsome, effeminate, fickle, talkative, dependable, resentful, sarcastic, unassuming, changeable, resourceful, persevering, forgiving, assertive, individualistic, vindictive, sophisticated, deceitful, impulsive, sociable, methodical, idealistic, thrifty, outgoing, intolerant, autocratic, conceited, inventive, dreamy, appreciative, forgetful, forceful, submissive, pessimistic, versatile, adaptable, reflective, inhibited, outspoken, quitting, unselfish, immature, painstaking, leisurely, infantile, sly, praising, cynical, irresponsible, arrogant, obliging, unkind, wary, greedy, obnoxious, irritable, discreet, frivolous, cowardly, rebellious, adventurous, enterprising, unscrupulous, poised, moody, unfriendly, optimistic, disorderly, peaceable, considerate, humorous, worrying, preoccupied, trusting, mischievous, robust, superstitious, noisy, tolerant, realistic, masculine, witty, informal, prejudiced, reckless, jolly, courageous, meek, stubborn, aloof, sentimental, complaining, unaffected, cooperative, unstable, feminine, timid, retiring, relaxed, imaginative, shrewd, conscientious, industrious, hasty, commonplace, lazy, gloomy, thoughtful, dignified, wholesome, affectionate, aggressive, awkward, energetic, tough, shy, queer, careless, restless, cautious, polished, tense, suspicious, dissatisfied, ingenious, fearful, daring, persistent, demanding, impatient, contented, selfish, rude, spontaneous, conventional, cheerful, enthusiastic, modest, ambitious, alert, defensive, mature, coarse, charming, clever, shallow, deliberate, stern, emotional, rigid, mild, cruel, artistic, hurried, sympathetic, dull, civilized, loyal, withdrawn, confident, indifferent, conservative, foolish, moderate, handsome, helpful, gentle, dominant, hostile, generous, reliable, sincere, precise, calm, healthy, attractive, progressive, confused, rational, stable, bitter, sensitive, initiative, loud, thorough, logical, intelligent, steady, formal, complicated, cool, curious, reserved, silent, honest, quick, friendly, efficient, pleasant, severe, peculiar, quiet, weak, anxious, nervous, warm, slow, dependent, wise, organized, affected, reasonable, capable, active, independent, patient, practical, serious, understanding, cold, responsible, simple, original, strong, determined, natural, kind.
\end{itemize}

For each word in those lists, we download (from https://nlp.stanford.edu/projects/histwords)  the corresponding embeddings  
from previously trained Genre-Balanced American English embeddings from Corpus of Historical American English (COHA) \citep{coha} for three of the ten decades available, specifically referring to year 1900, year 1950 and year 2000. Following \cite{Garg_etal_2018}, from the embeddings and word lists, we are able to measure the embedding bias (i.e. strength of association) between words that represent gender groups (i.e. women and men) and neutral words such as Occupations and Adjectives. Here subscript $t$ represents the year (decade), and in the analysis in Section \ref{sec:genderbias} we consider $t=1900, 1950, 2000$. 

\begin{itemize}
\item For each time $t$, we compute $\mathbf{m}_t$, the average embedding vector for $W_{man}$, averaged over the 20 words in the list $W_{man}$; similarly, $\mathbf{w}_t$ denotes the average embedding for $W_{woman}$, averaged over all the word embeddings in the list $W_{woman}$.

\item We define the embedding bias $Y_{tj}'$ for (or against)  women of the $j$th occupation word in $W_{occu}$ at time $t$ as the difference of the Euclidean distances between the $j$th word and the men representative and the difference of the Euclidean distances between the $j$th word and the women average vector, i.e. 
$$\textrm{occupation bias}_{tj} = Y'_{tj}=  ||\mathbf{o}_{tj} - \mathbf{m}_t||_2 - ||\mathbf{o}_{tj} - \mathbf{w}_t||_2,$$
where $\mathbf{o}_{tj}$ is the word embedding vector of the $j$th occupation word at time $t$ and $||\cdot||_2$ is the Euclidean norm of a finite-dimensional vector. We  standardize the occupation embedding bias for women by considering $Y_{tj}$ as $Y_{tj}'$ minus the overall mean and then divide it by the overall standard deviation.

\item Similarly, we define the adjective embedding bias $Z'_{tl}$ for (or against) women of the $l$th adjective word in $W_{adj}$ at time $t$,  as the difference 
$$\textrm{adjective bias}_{tl} = Z'_{tl}=||\mathbf{a}_{tl} - \mathbf{m}_t||_2 - ||\mathbf{a}_{tl} - \mathbf{w}_t||_2,$$
where $\mathbf{a}_{tl}$ is the word embedding vector of the $l$th adjective word at time $t$. Here each $Z'_{tl}$ has been standardized subtracting from $Z'_{tl}$ the overall mean and then divide the difference by the overall standard deviation to get $Z_{tl}$ as for the occupational bias for women.
\end{itemize}
Note that, if the bias value is negative, then the embedding more closely associates the occupation (or adjective) word with men, because the distance between the occupation (or adjective) word is closer to men than women.  Other norm definitions could be used here, as, for instance, cosin similarity.  Hence, gender bias \textit{against women} corresponds to negative values of the embedding bias.

\bigskip
\section*{APPENDIX C: Simulation Study} 
\label{app:Simulations}
In order to check the flexibility of the model described in Section~\ref{sec:mixturemodel}, as well as  the performance of the algorithm, we consider many simulation scenarios. For all of them, we simulate $n=100$ subjects for $T=4$ times, unless otherwise specified.  The examples are designed to investigate the  ability of the model and the algorithm to detect the true clustering of the data under different scenarios as follows:
\begin{itemize}
\item in scenario 1, for all $t$, we simulate data from only one single cluster; 

\item in scenario 2 we have two single clusters, with all patients  in the same cluster over time and the cluster distribution not changing over time, while scenario 3  is the same as scenario 2 but  the cluster distributions change over time;

\item in scenario 4 data are generated  at the initial time from two clusters, but at the successive times, the subjects change cluster with probability 50\%, and the cluster distributions change as well; scenario 5 is as scenario 4 but the probability of changing cluster is increased to 80\%;

\item scenarios 6 and 7 are aimed to detect clustering with a number of clusters that changes over time, i.e. from 1 to 2 and from 2 to 1. In these cases, we simulate only $n=100$ data for each of $T=2$ times.  
 
\end{itemize}

We fit model \eqref{eq:trunc_model}-\eqref{eq:trans} to all the simulated datasets, i.e. we assume
\begin{align*}
& Y_{tj}|\mu_{tj}, \tau_{tj} \sim \text{N}(\mu_{tj}, (\tau_{tj})^{-1}),  \quad j = 1,...,n, \\
& \mu_{tj}, \tau_{tj} |G_t \iid G_t, \quad \textrm{ for each } t = 1,...,T
\end{align*} 
Instead of using the sequence of infinite-dimensional r.p.m.s $(G_t)_t$, we assume its finite-dimensional counterpart in \eqref{eq:trunc_model} with $ J = 50$, and
\begin{equation*}
 G_0 \sim \text{N}(0,100)\times \text{Gamma}(2,2) \quad
 \psi \sim \text{U}(-1,1) \quad
 M \sim \text{Gamma}(4,4).
\end{equation*}
In this case, the prior mean of the number of clusters $\E(K)$ is equal to 4 and the number of particles in the algorithm (see Appendix~A) is $R = 500$.

For all simulations, we run the MCMC algorithm described in Section \ref{sec:mixturemodel}  for 50,000 iterations, discarding the first 25,000 iterations as burn-in. Unless otherwise stated, we check through standard
diagnostics criteria such as those available in the R package CODA \citep{CODA} that convergence of the chain is satisfactory for most of the parameters.

\subsection*{Scenario 1}
\begin{figure}[H] %
\begin{center}
\includegraphics[width=0.45\textwidth]{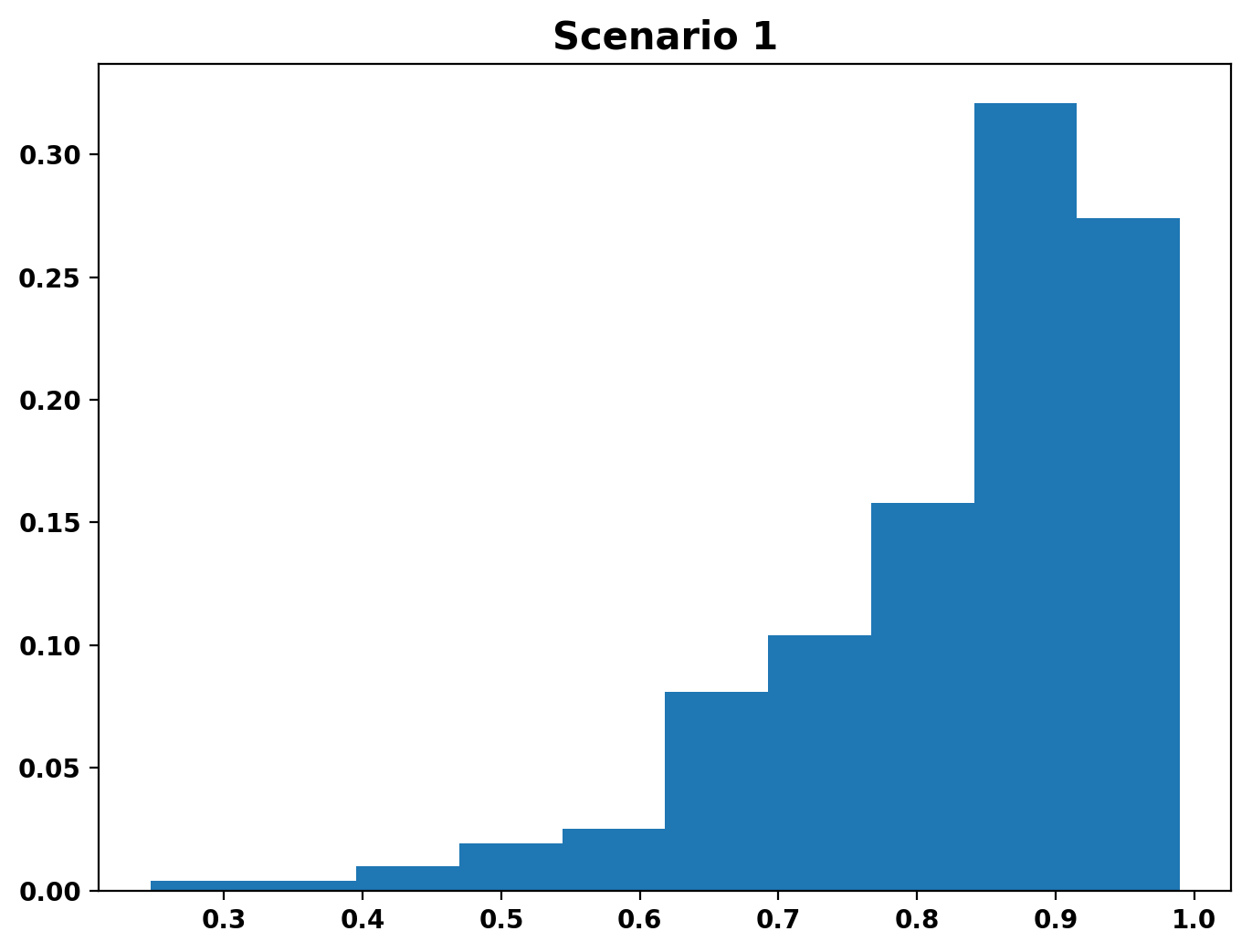}
\includegraphics[width=0.45\textwidth]{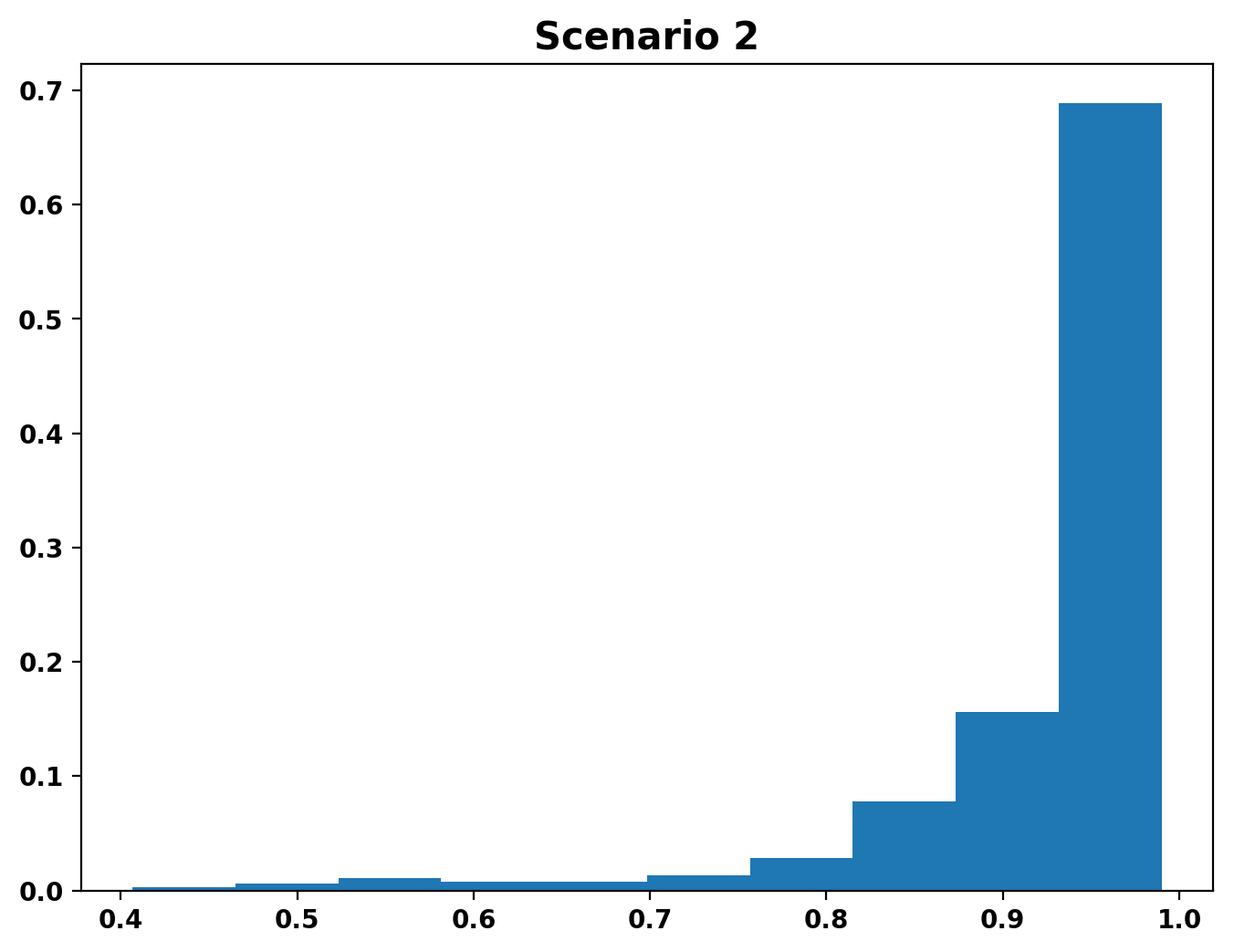}
\includegraphics[width=0.45\textwidth]{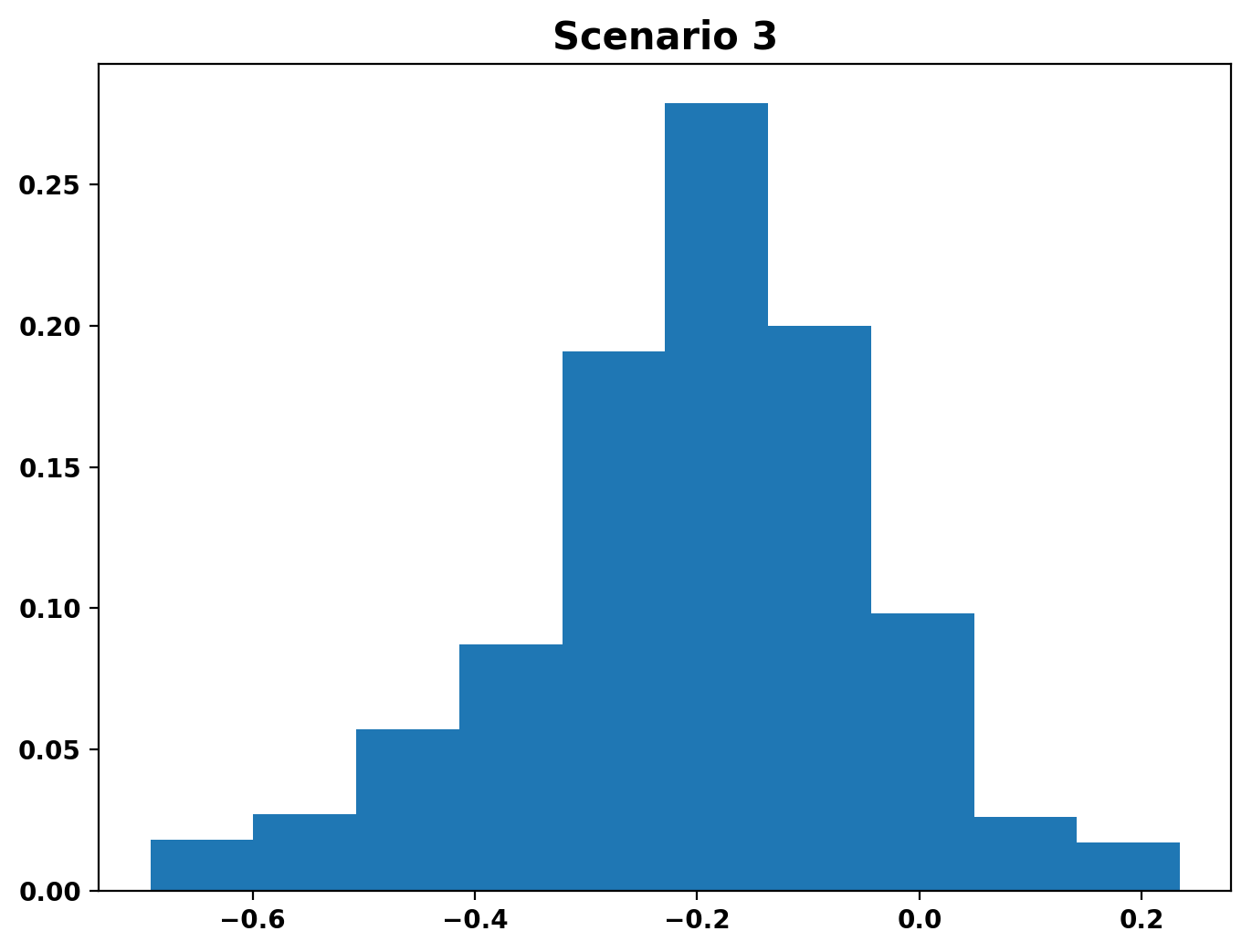}
\includegraphics[width=0.45\textwidth]{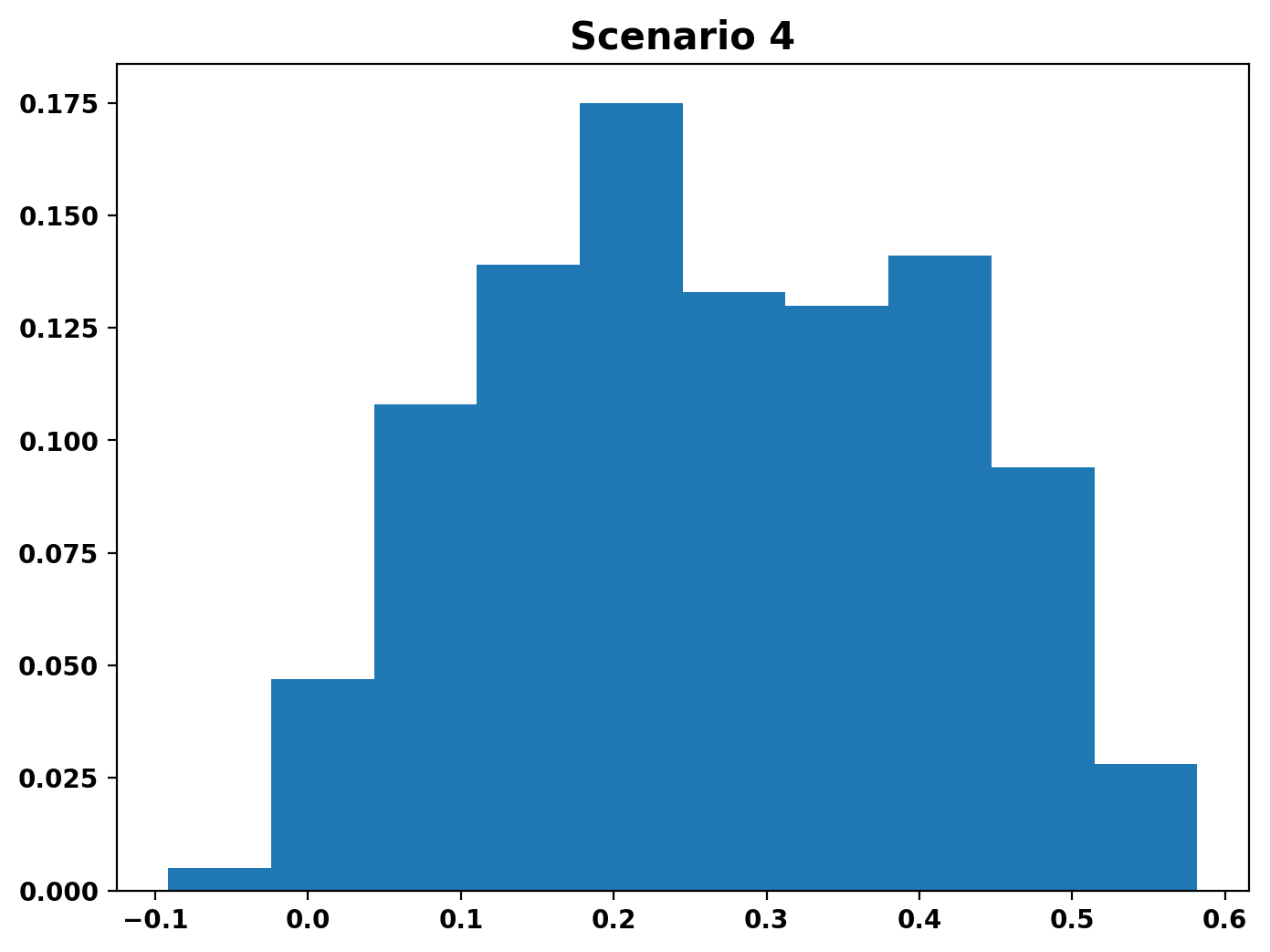}
\includegraphics[width=0.45\textwidth]{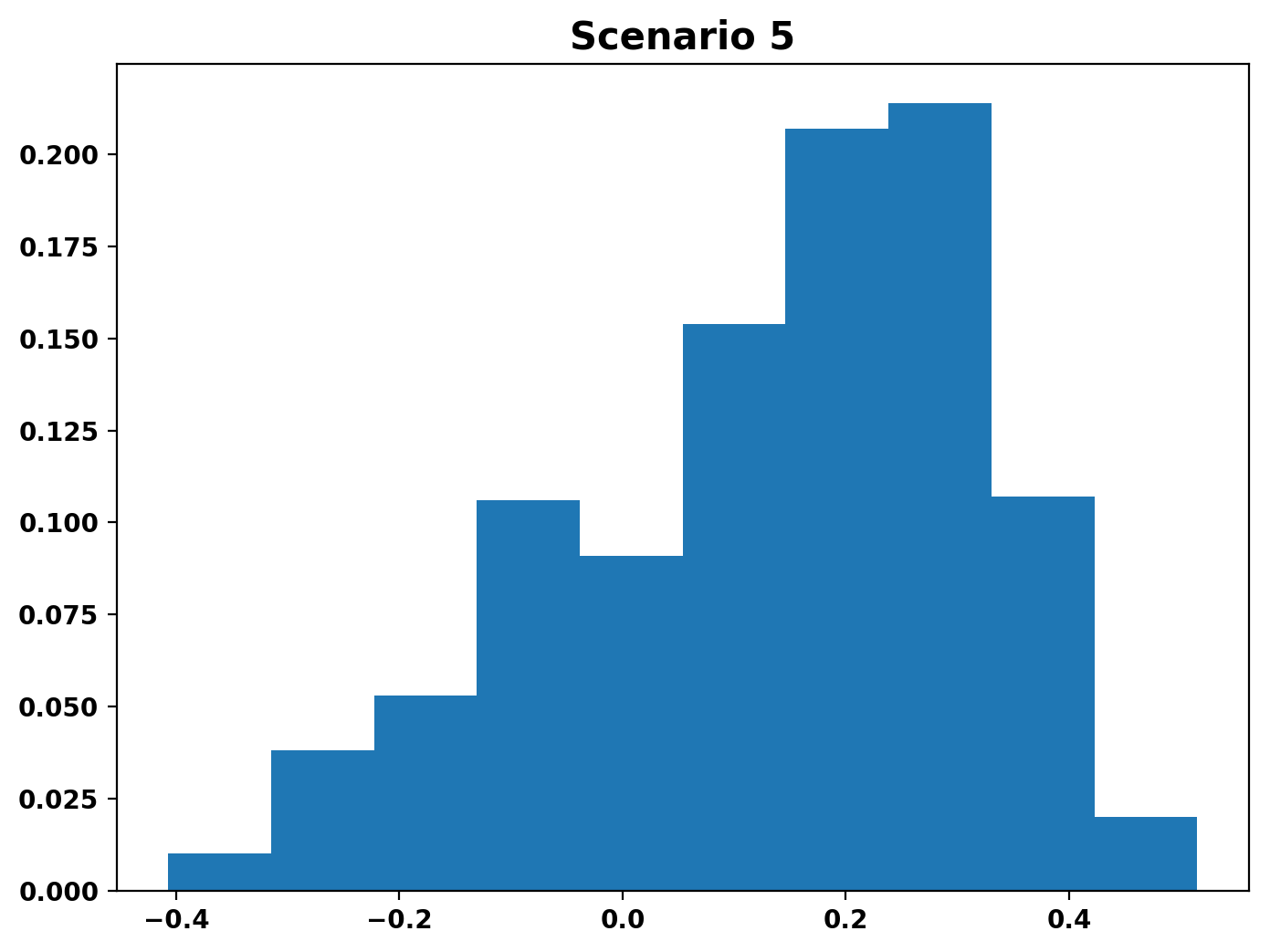}
\includegraphics[width=0.45\textwidth]{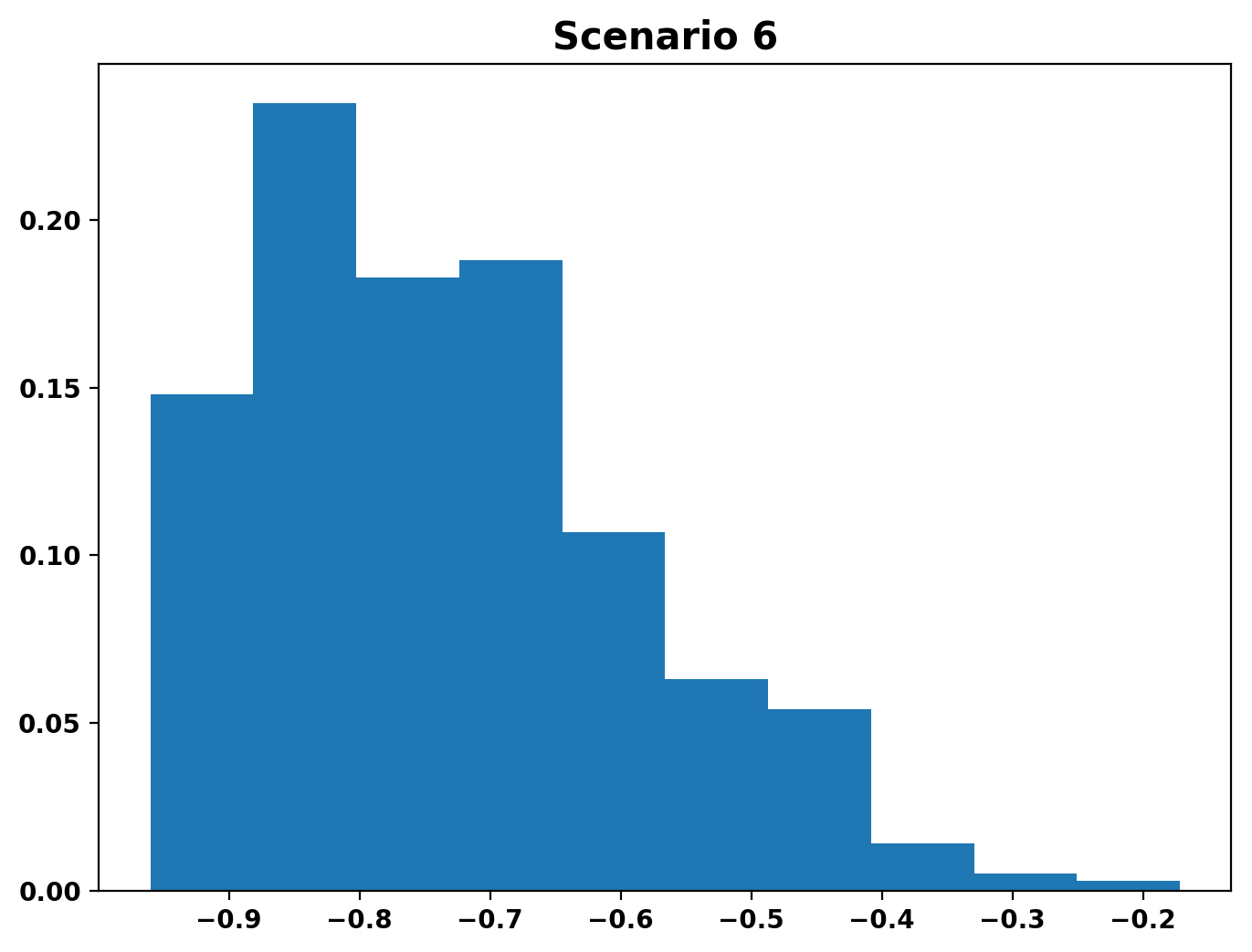}
\includegraphics[width=0.45\textwidth]{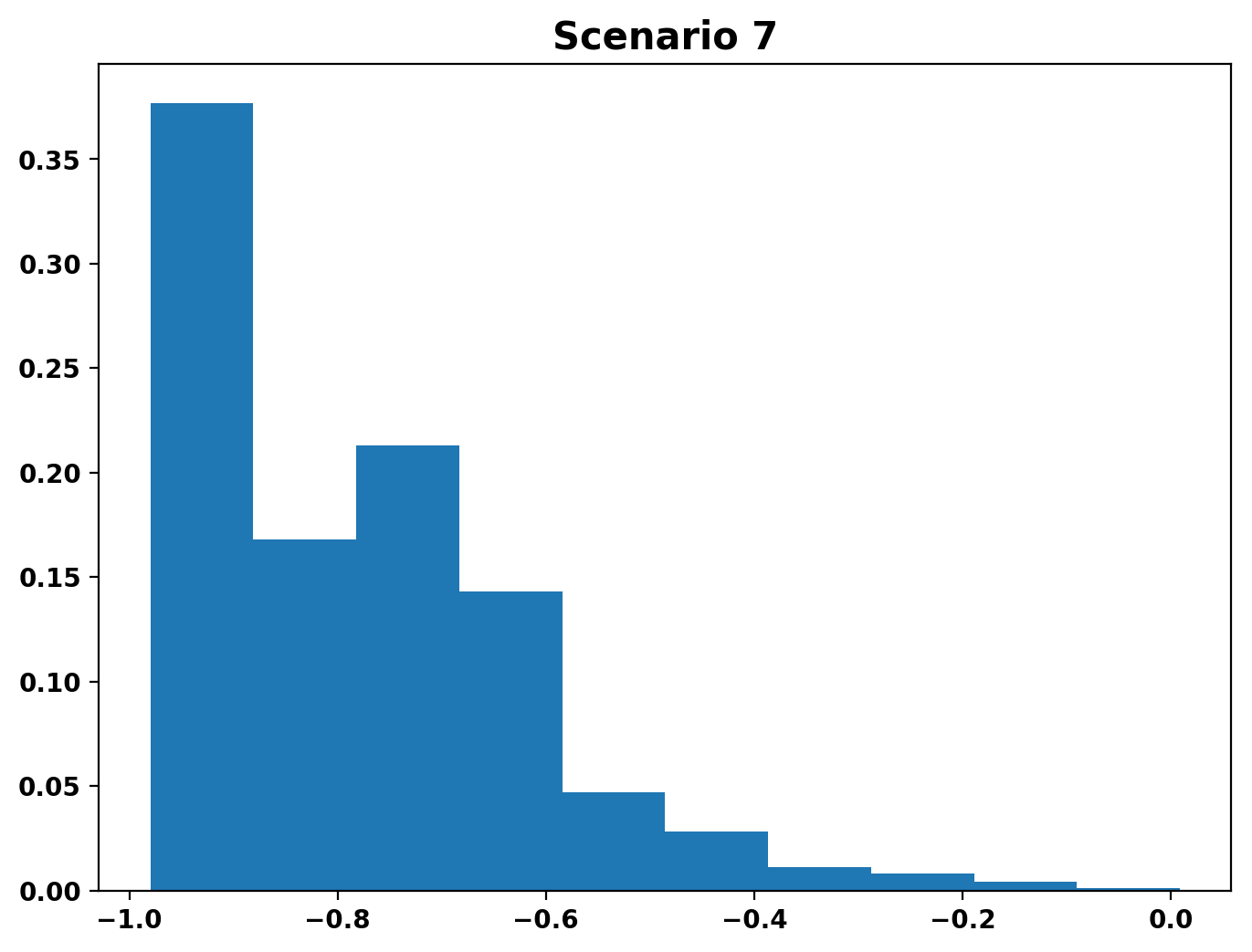}
\end{center}
\vspace{-1cm}
 \caption{Marginal posterior distribution of parameter $\psi$ of all scenarios.}
 \label{fig:simulated_psi}
\end{figure}

\begin{figure}[H]
\begin{center}
\includegraphics[width=.45\textwidth]{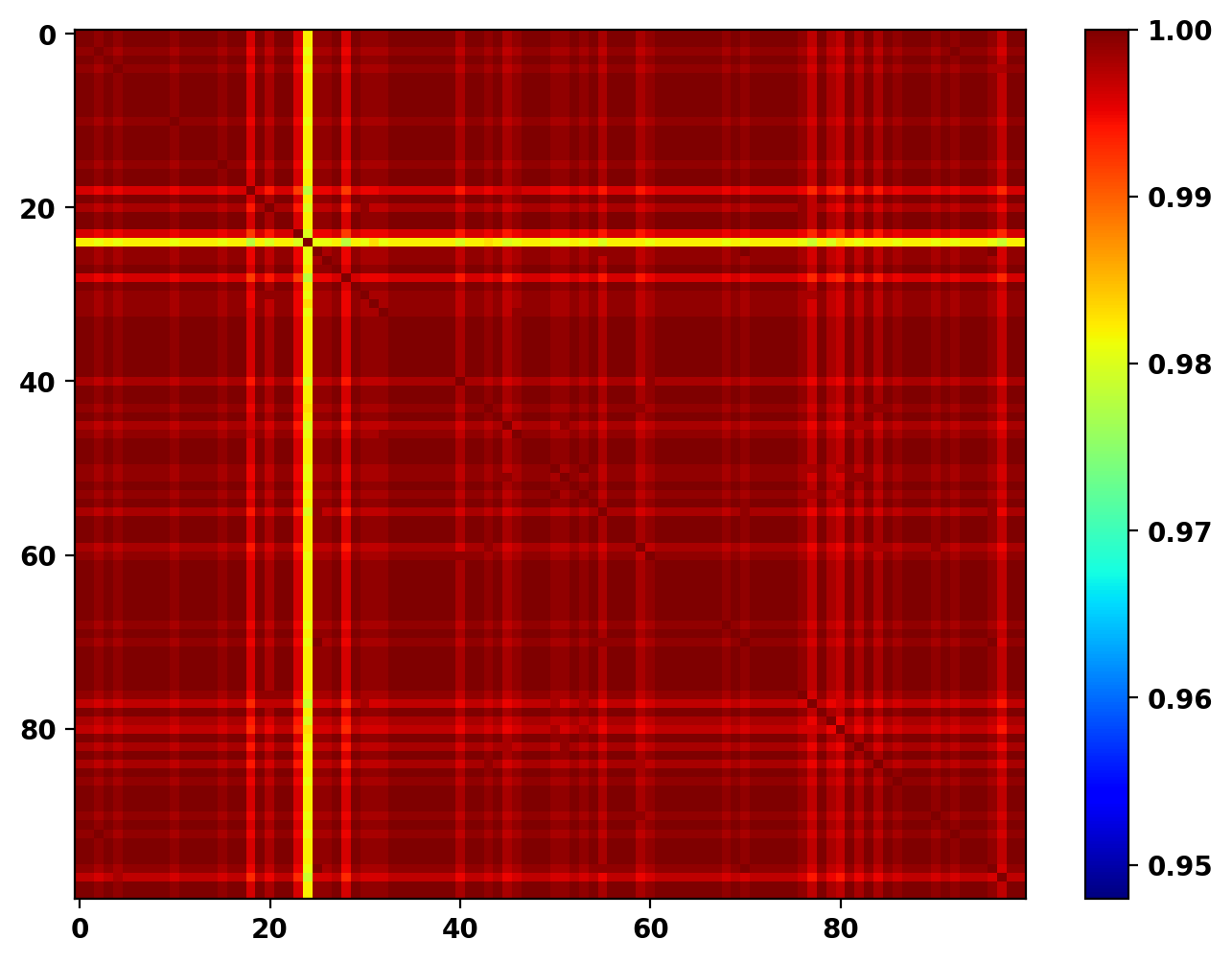}
\includegraphics[width=.45\textwidth]{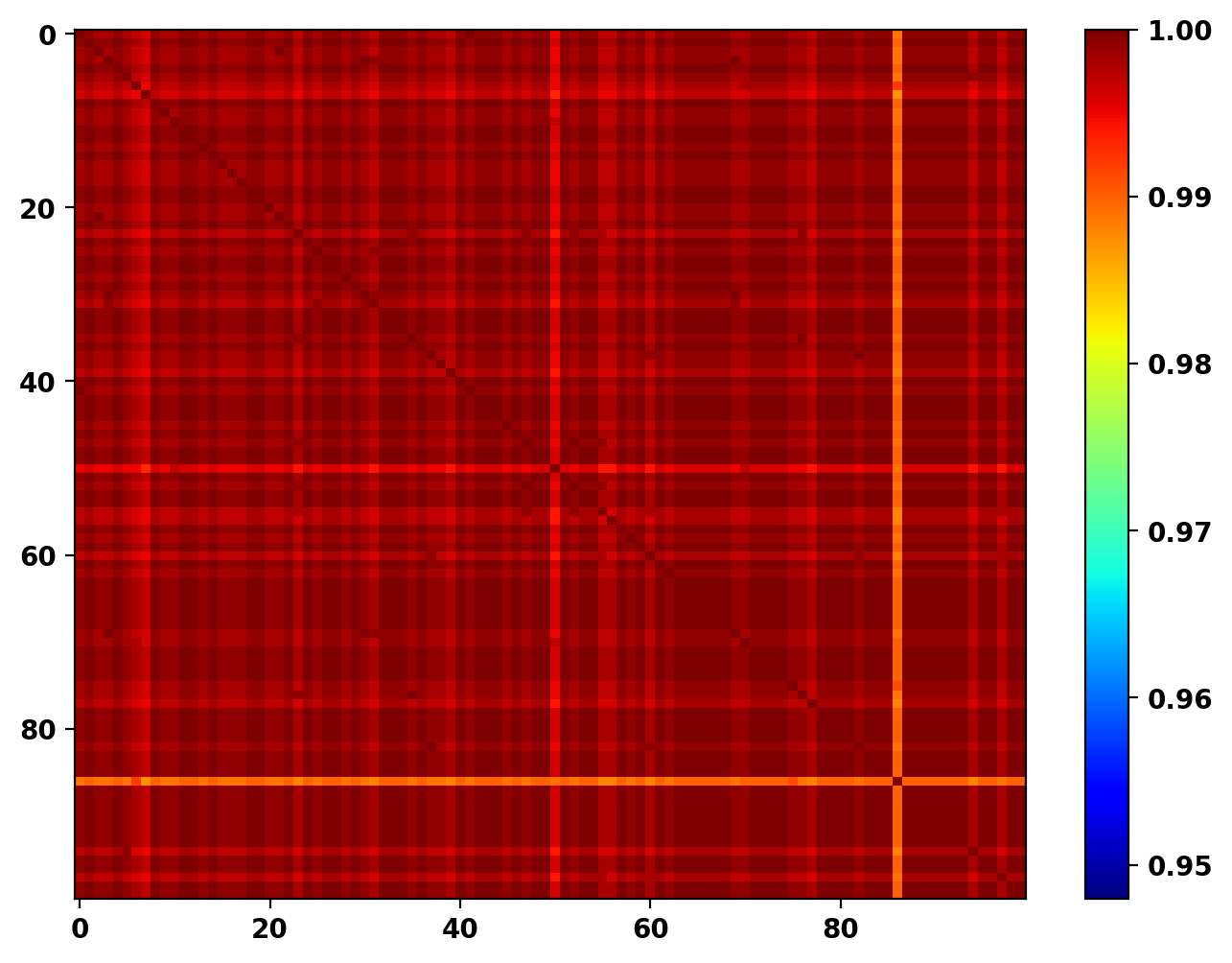}
\includegraphics[width=.45\textwidth]{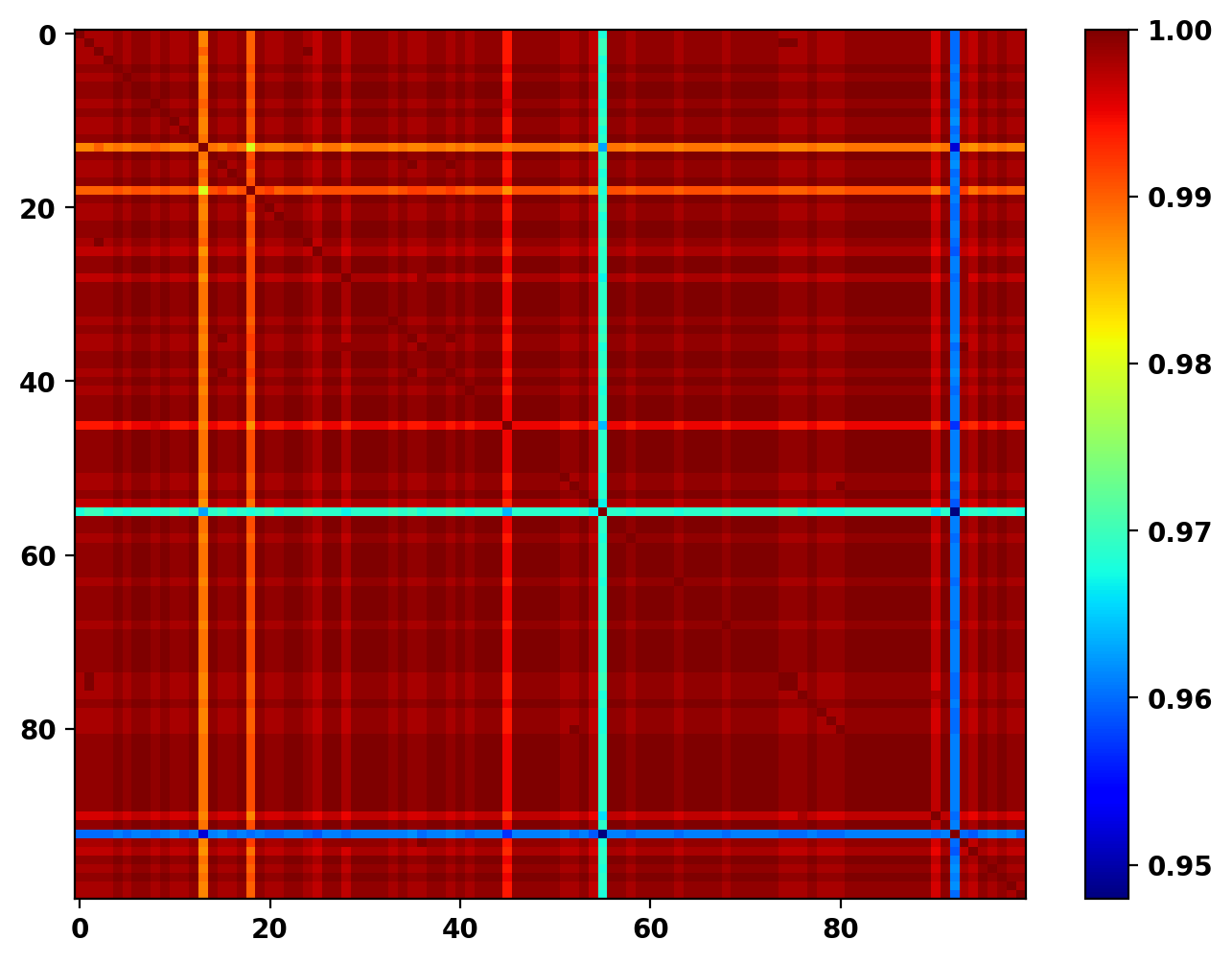}
\includegraphics[width=.45\textwidth]{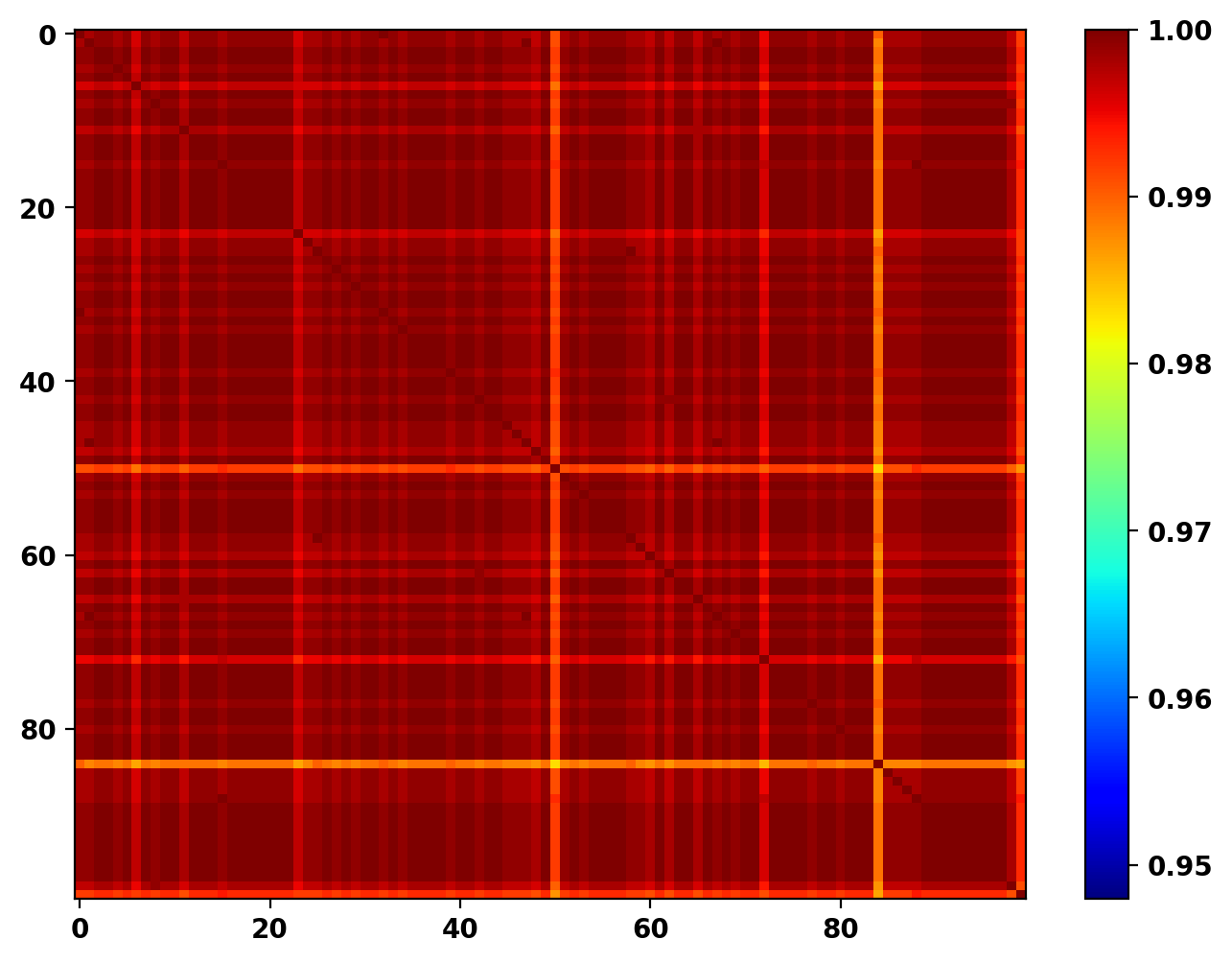}
\end{center}
\vspace{-1cm}
 \caption{Simulation scenario 1, posterior co-clustering probability plots, i.e. posterior probability that two items in the sample are assigned to the same cluster, for all pairs of items in the sample, for $t=1,2,3,4$.}
 \label{fig:coclust1}
\end{figure}

In this case, data are generated independently from a standard Gaussian distribution, so that the (only) true cluster is the same for each time $t$ and the true correlation parameter $\psi$ is equal to 1. 
Figure~\ref{fig:simulated_psi} (topleft panel) shows the marginal posterior distribution of $\psi$, where the estimated posterior mean of $\psi$ is 0.832. The posterior co-clustering probabilities, i.e. the probability that the two item sin the sample are assigned to the same cluster, for each $t=1,2,3,4$, are displayed in Figure~\ref{fig:coclust1}. This confirms that there is only one estimated cluster for each time $t$. It is clear that our model is able to recover correctly the data generation process and detect the unique true cluster in all time periods. 

\subsection*{Scenario 2}
In this case, at $t=1$ we simulate $n=100$ observations  from two groups of equal size, using two different Gaussian distributions ($\text{N}(-80,1)$ and $\text{N}(-40,4)$). For $t =  2,3,4$, we assume that people will remain in the same cluster as at time $t-1$. In practice,  
for $t= 1, 2, 3, 4$, we have independently simulated data of first 50 subjects from $\text{N}(-80,1)$, while those of the remaining 50 subjects are simulated from $\text{N}(-40,4)$. 

\begin{figure}[H]
\begin{center}
\includegraphics[width=.45\textwidth]{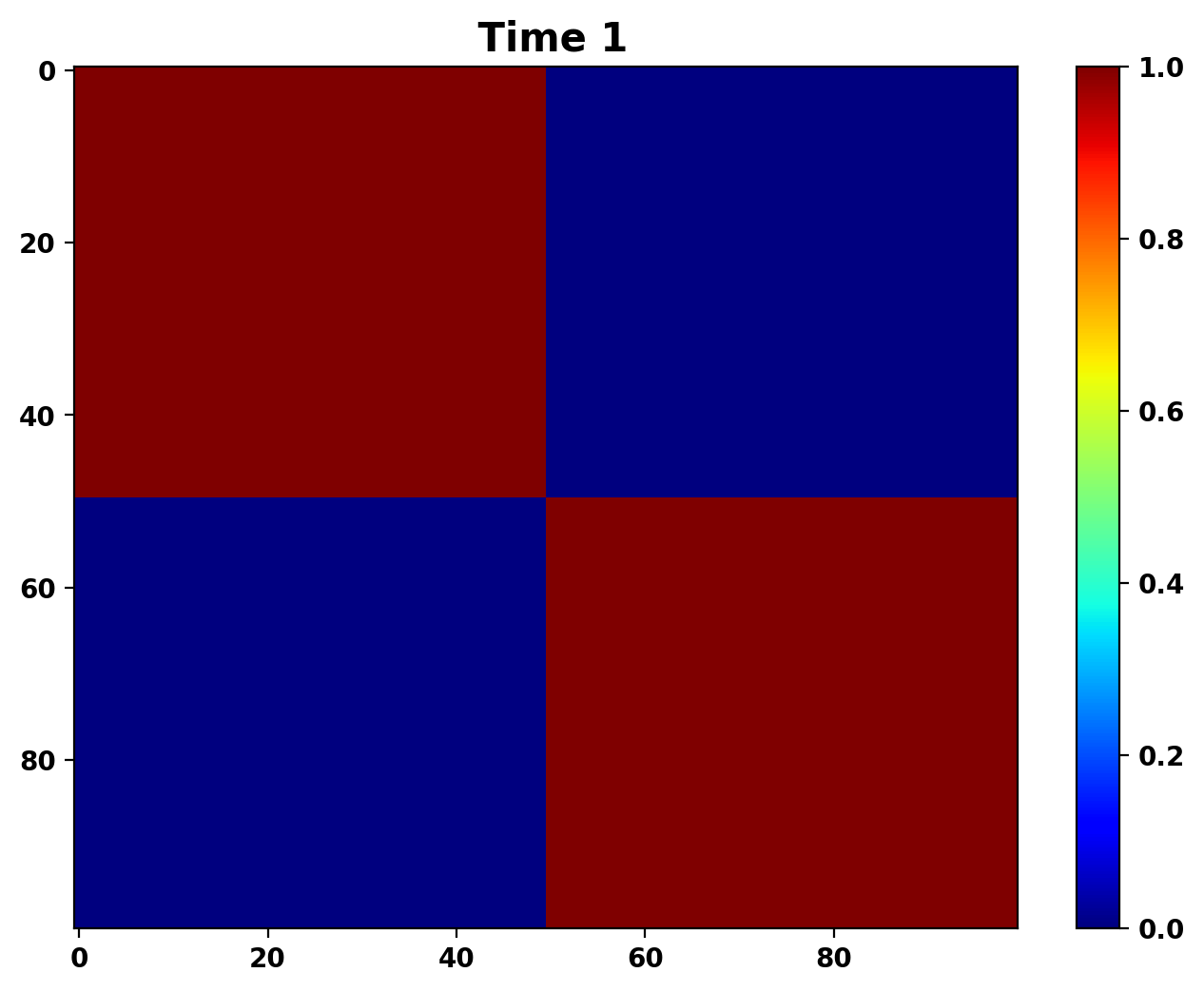}
\includegraphics[width=.45\textwidth]{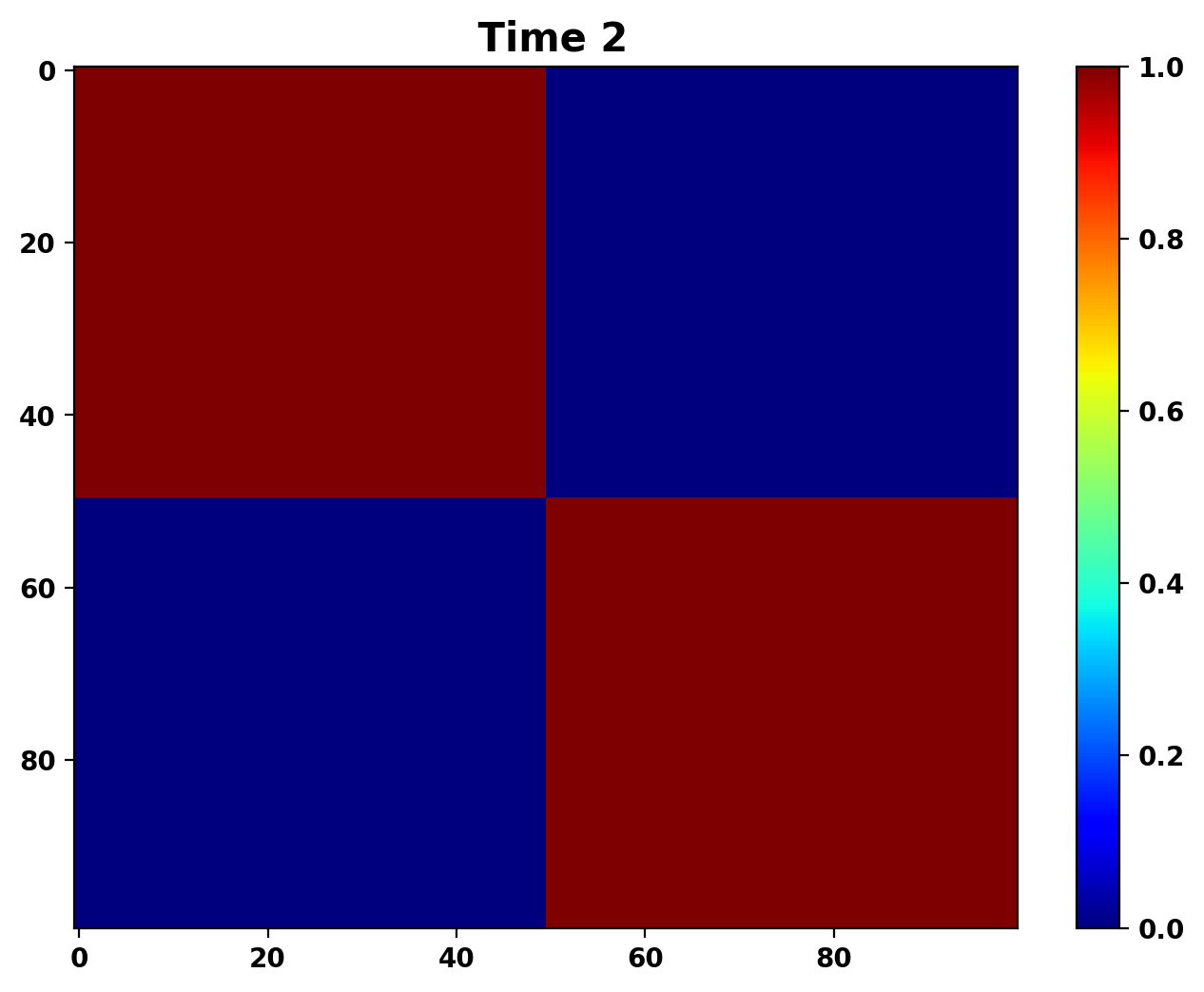}
\includegraphics[width=.45\textwidth]{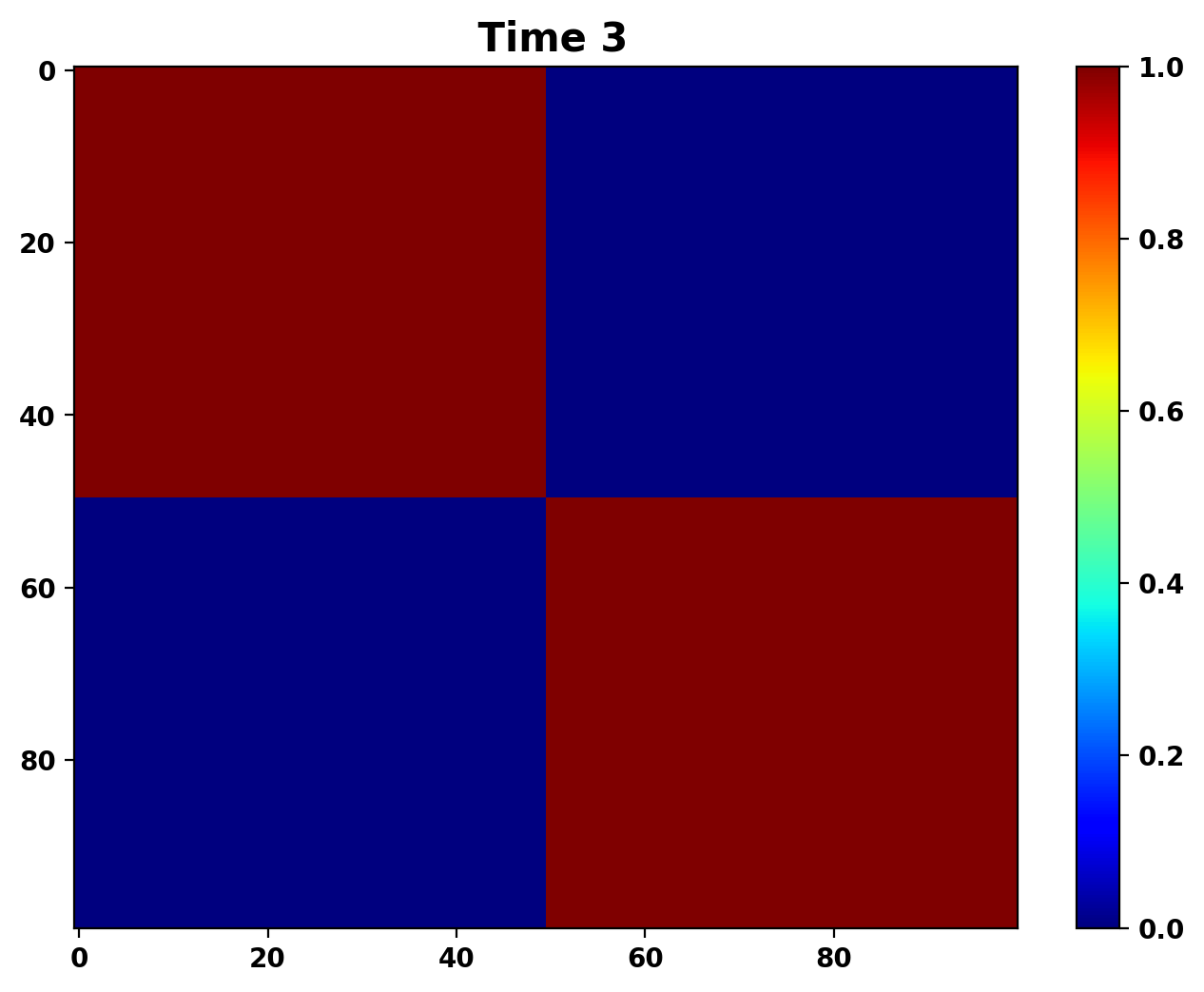}
\includegraphics[width=.45\textwidth]{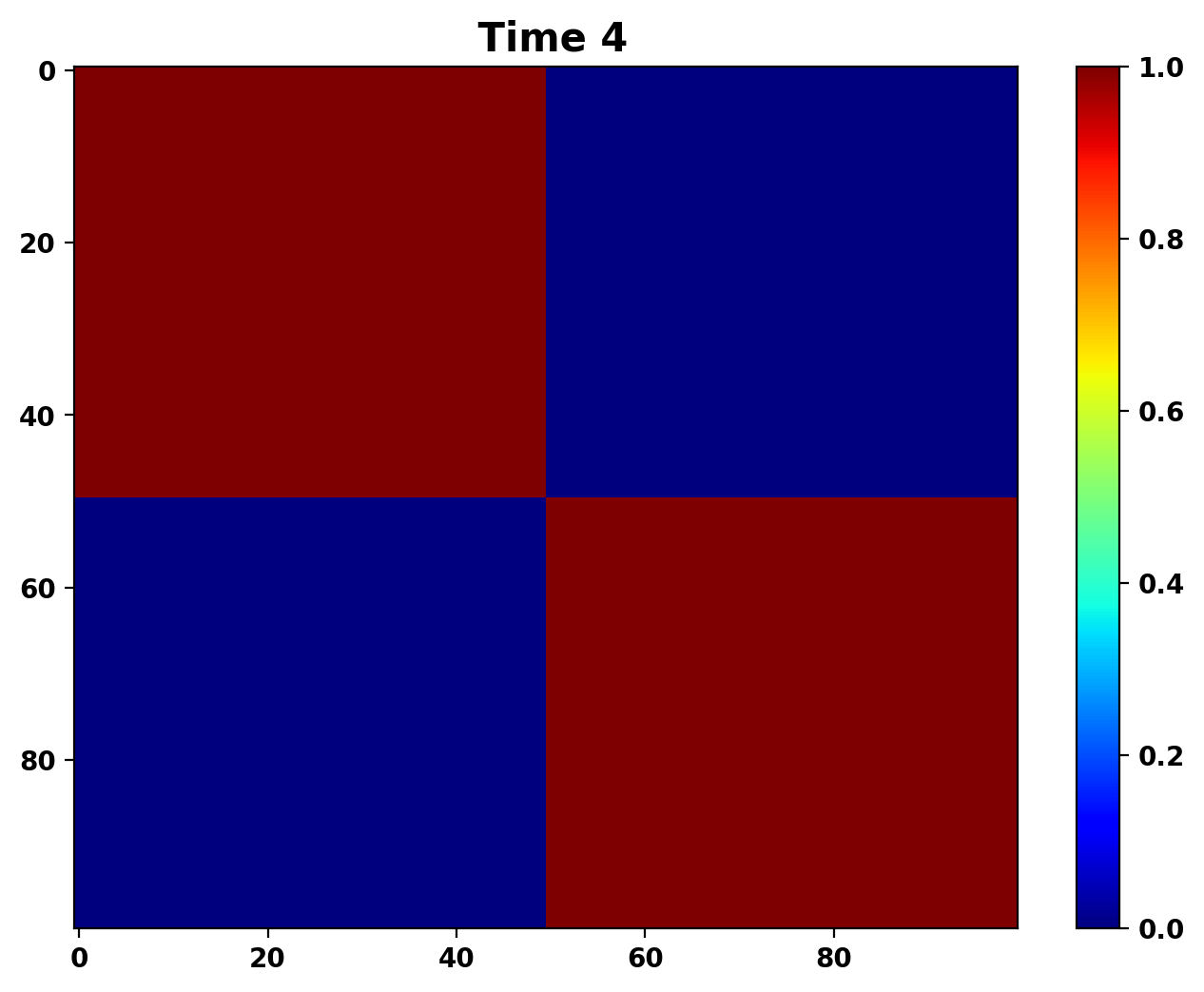}
\end{center}
\vspace{-1cm}
 \caption{Posterior co-clustering probability plots for simulation scenario 2 over time.}
 \label{fig:coclust2}
\end{figure}

The estimated posterior mean of $\psi$ under our model is 0.926, while the true value of $\psi$ is 1, which is on the boundary of the support of the posterior of $\psi$ itself (Figure~\ref{fig:simulated_psi}, topright panel). The trace-plot of $\psi$ (not reported here) shows a bad mixing of the chain, but the model is still able to detect the high correlation between the time periods. 
It is clear from Figure~\ref{fig:coclust2}, which displays the posterior co-clustering plots over time, that our model is able to detect the true value of the number of clusters.


\subsection*{Scenario 3}
At time $t = 1$,  we simulate 50 observations iid from a Gaussian distribution ($\text{N}(-80,1)$), and the remaining  50  from a different Gaussian distribution ($\text{N}(80,1)$). For $t =2,3,4$, we design the scenario in such a way that individuals remain in the same cluster they belong  at time $t - 1$, but the simulation distributions in the two clusters change over time. Specifically, the simulation distributions when $t=2$ are $\text{N}(-60,1)$ and $\text{N}(20,1)$, when $t=3$ are $\text{N}(-40,1)$ and $\text{N}(40,1)$, while for $t=4$ we simulate from $\text{N}(-20,1)$ and $\text{N}(60,1)$. 

\begin{figure}[H]
\begin{center}
\includegraphics[width=.45\textwidth]{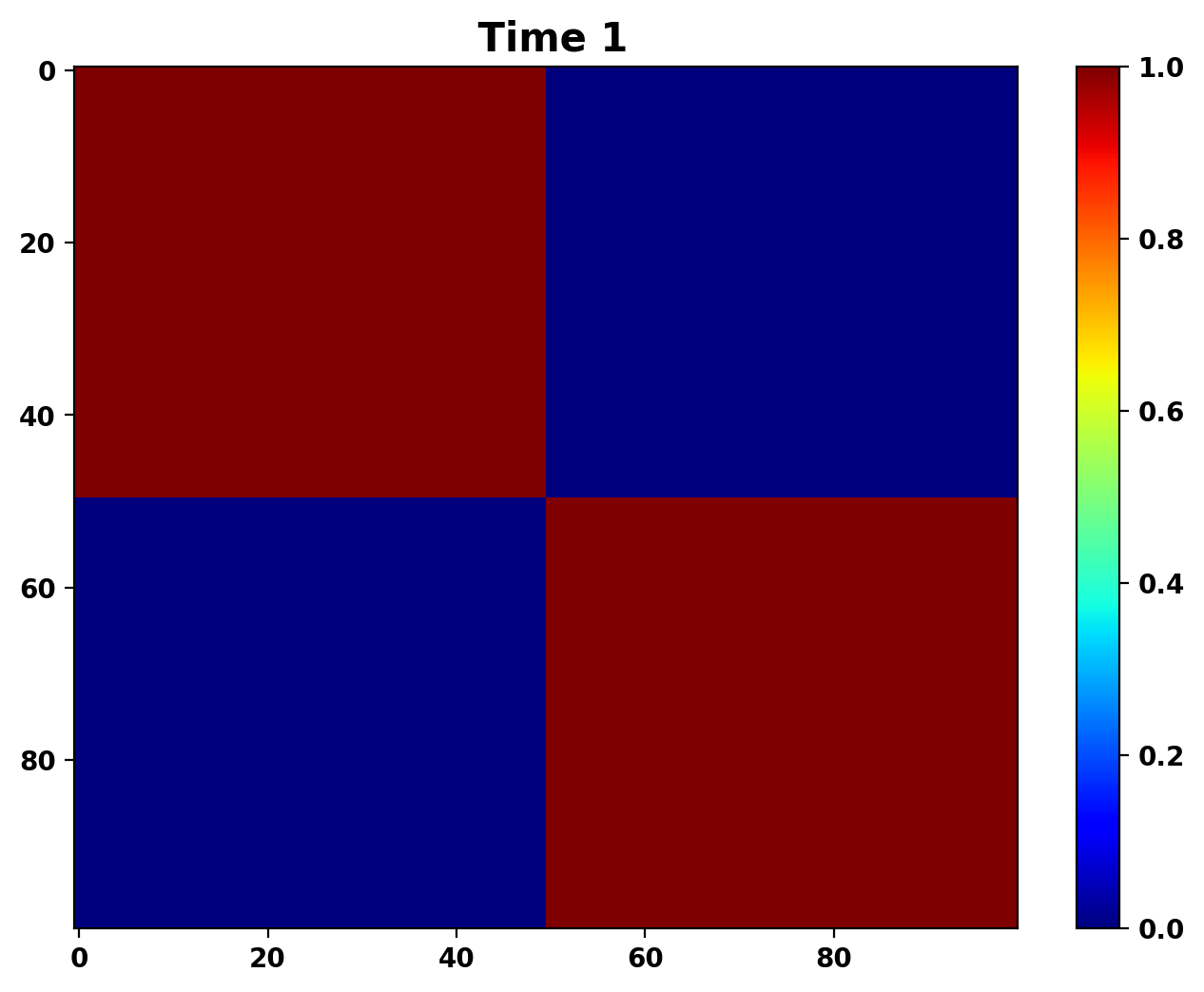}
\includegraphics[width=.45\textwidth]{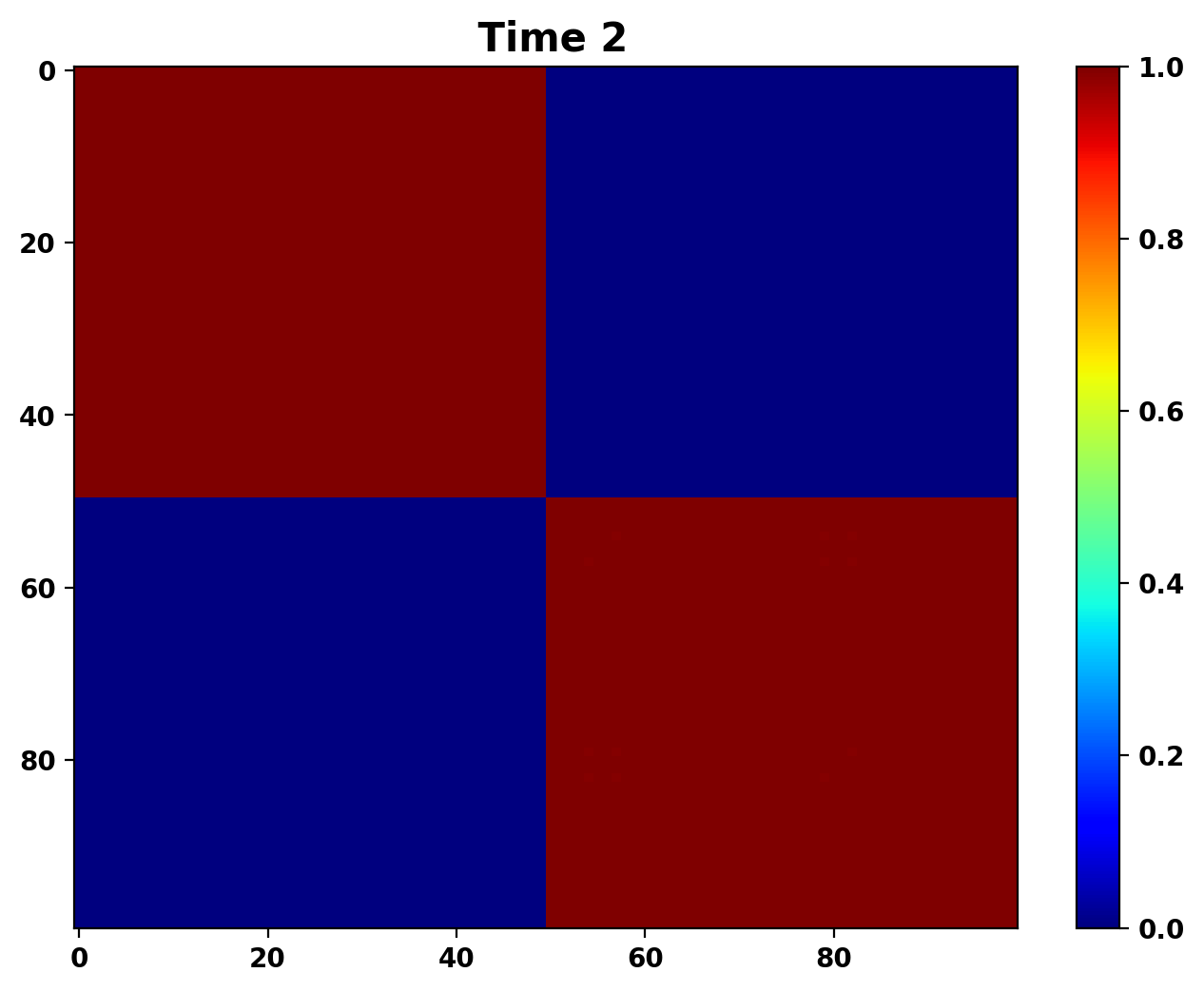}
\includegraphics[width=.45\textwidth]{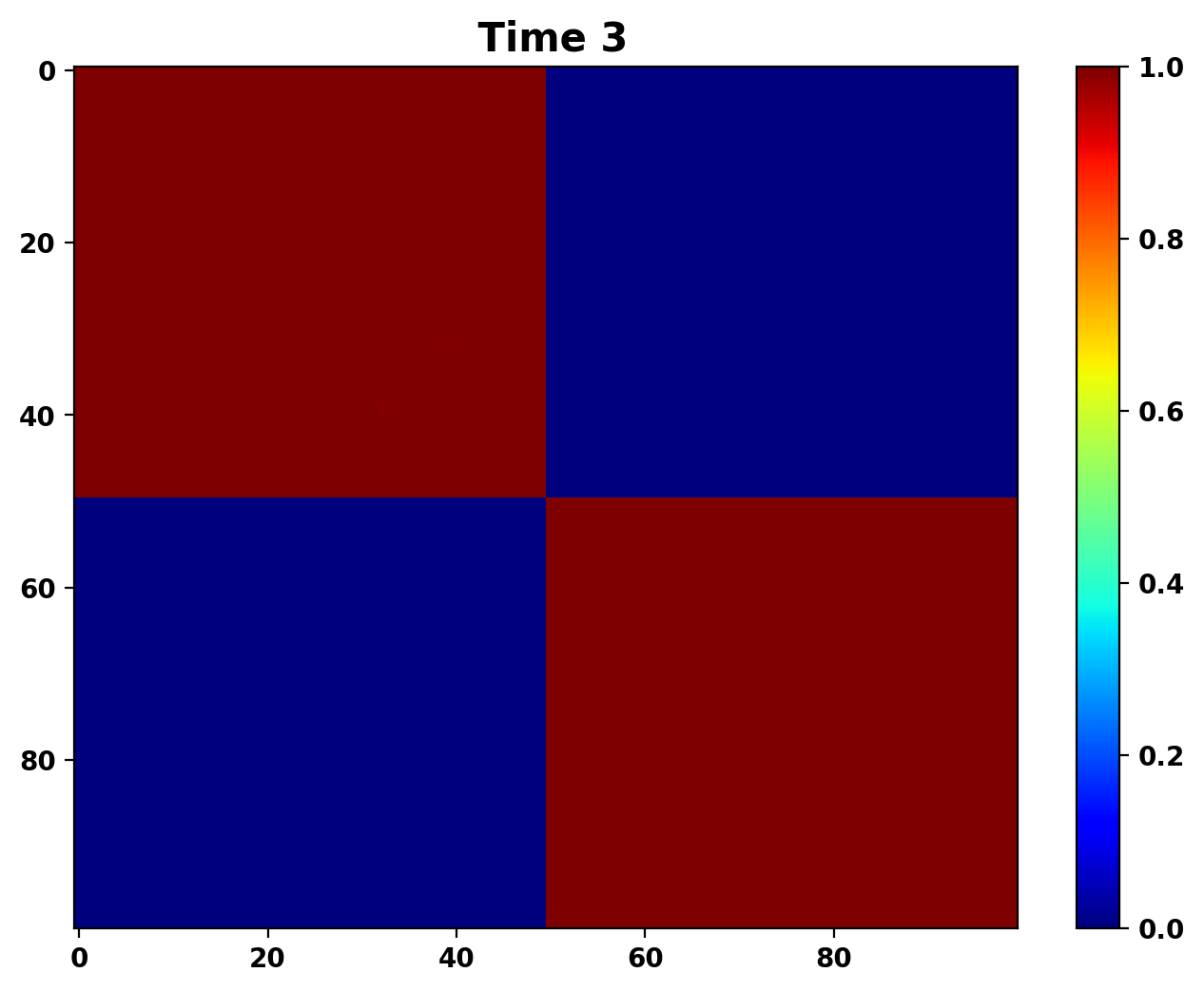}
\includegraphics[width=.45\textwidth]{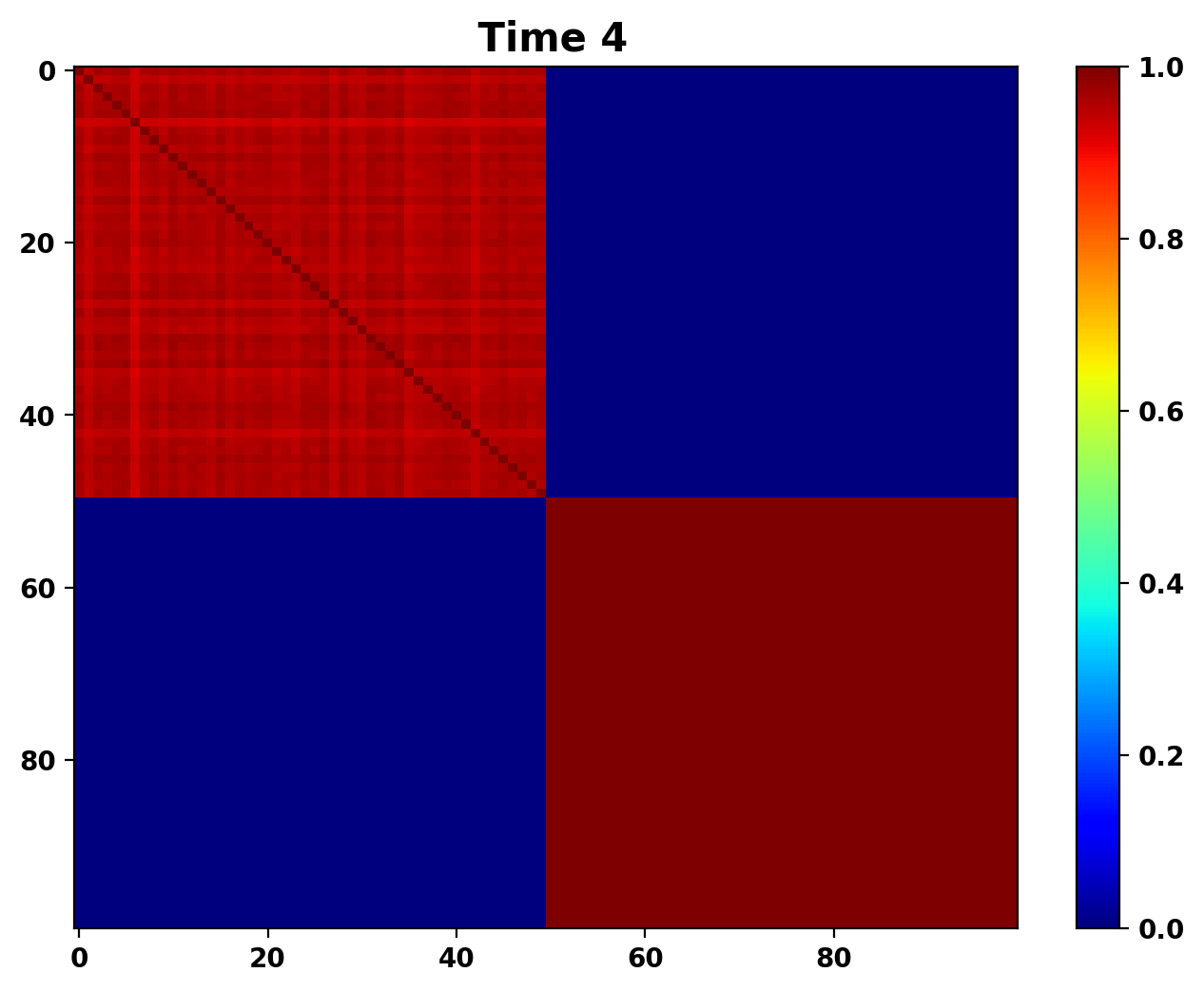}
\end{center}
\vspace{-1cm}
 \caption{Posterior co-clustering probability plots for simulation scenario 3  over time.}
 \label{fig:coclust3}
\end{figure}

We find that the posterior mean of $\psi$ is -0.200, while the marginal posterior distribution is shown in Figure~\ref{fig:simulated_psi}  (second row first column). Those negative values for $\psi$ may be explained by the change of the location of the clusters over time. 
The co-clustering probabilities over time (Figure~\ref{fig:coclust3}) are correctly estimated.

\subsection*{Scenario 4} 
We simulate 50 observations from each of two different Gaussian distributions ($\text{N}(-80,1)$ and $\text{N}(80,1)$) at time $t=1$. Differently from the previous scenario, for $t =  2,3,4$, individuals remain in the same cluster they belong to at time $t-1$ with probability 50\%, and, with probability 50\%, individuals switch to the other cluster. Note that, in this case,  the simulation distribution for each cluster  changes over time. Specifically, the simulation distributions when $t=2$ are $\text{N}(-60,1)$ and $\text{N}(20,1)$; when $t=3$ are $\text{N}(-40,1)$ and $\text{N}(40,1)$; when $t=4$ are $\text{N}(-20,1)$ and $\text{N}(60,1)$.  

\begin{figure}[H]
\begin{center}
\includegraphics[width=.45\textwidth]{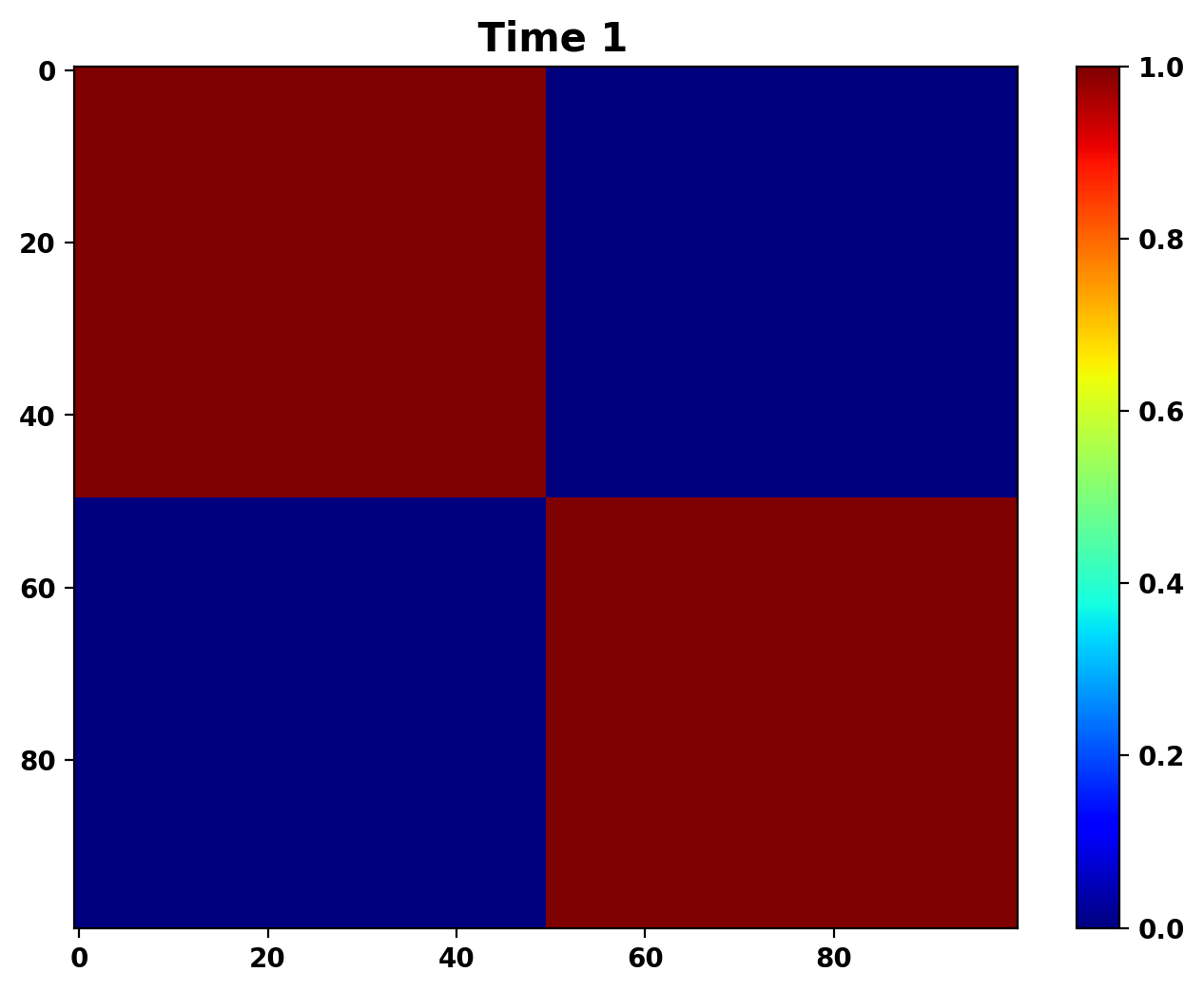}
\includegraphics[width=.45\textwidth]{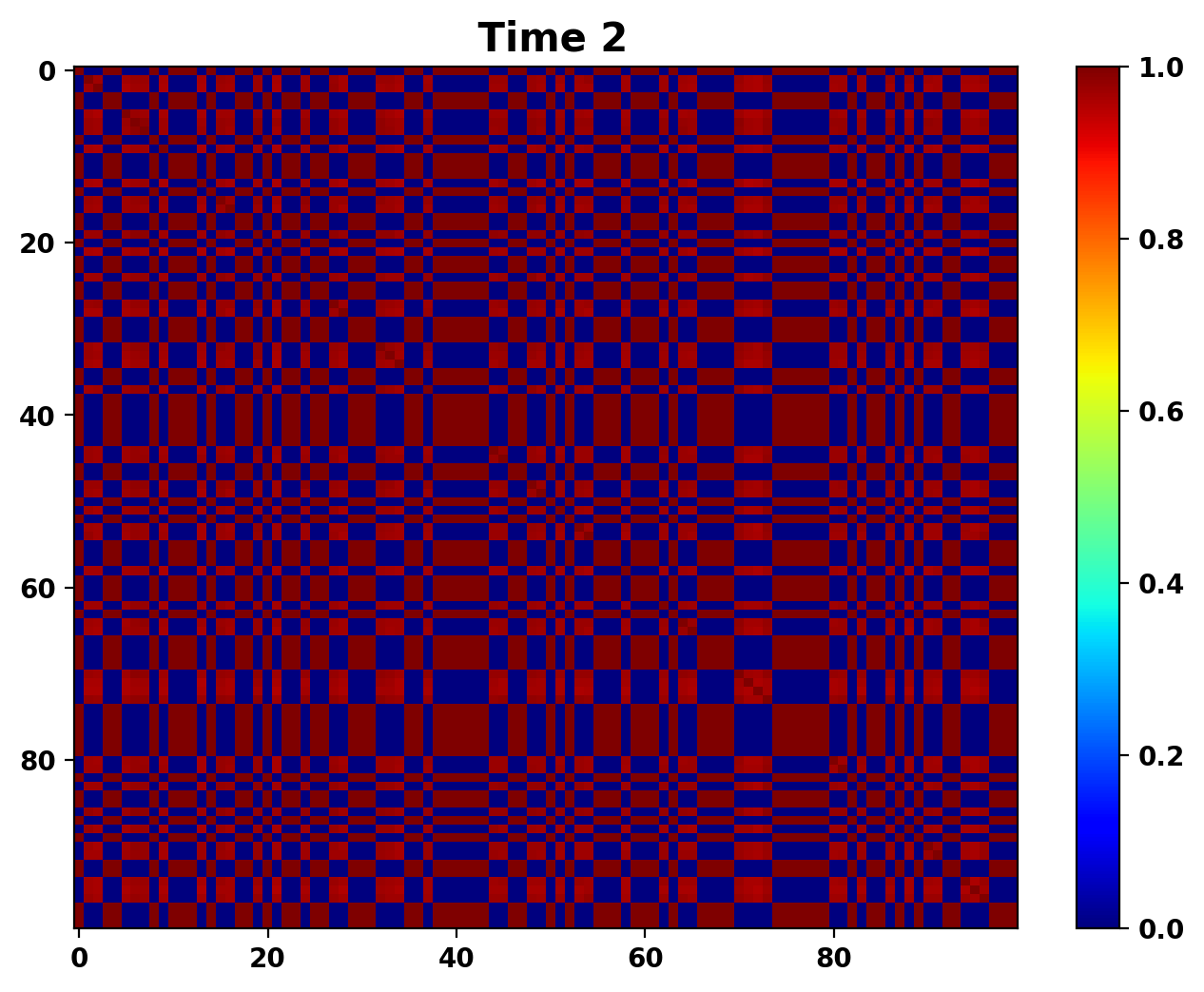}
\includegraphics[width=.45\textwidth]{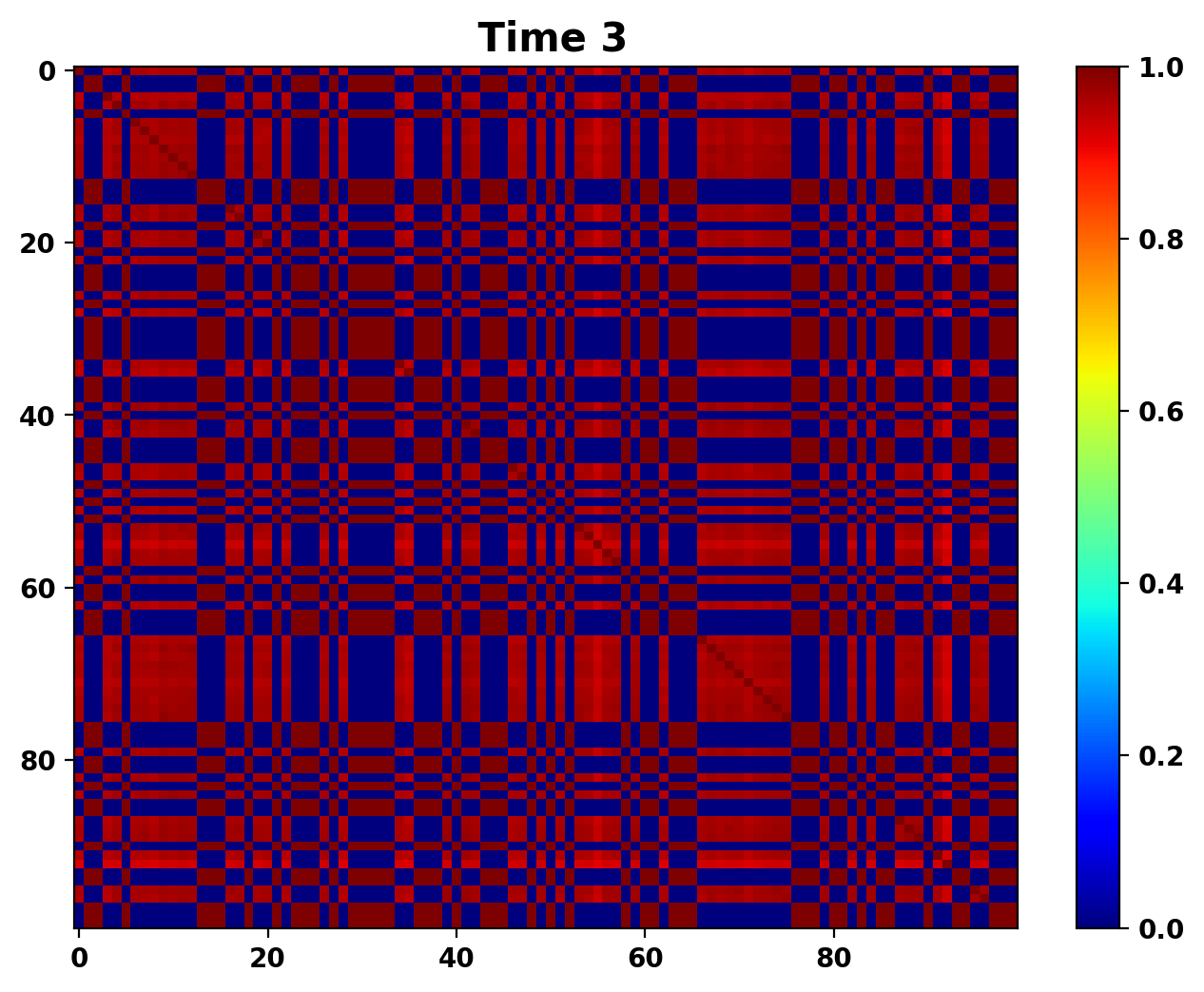}
\includegraphics[width=.45\textwidth]{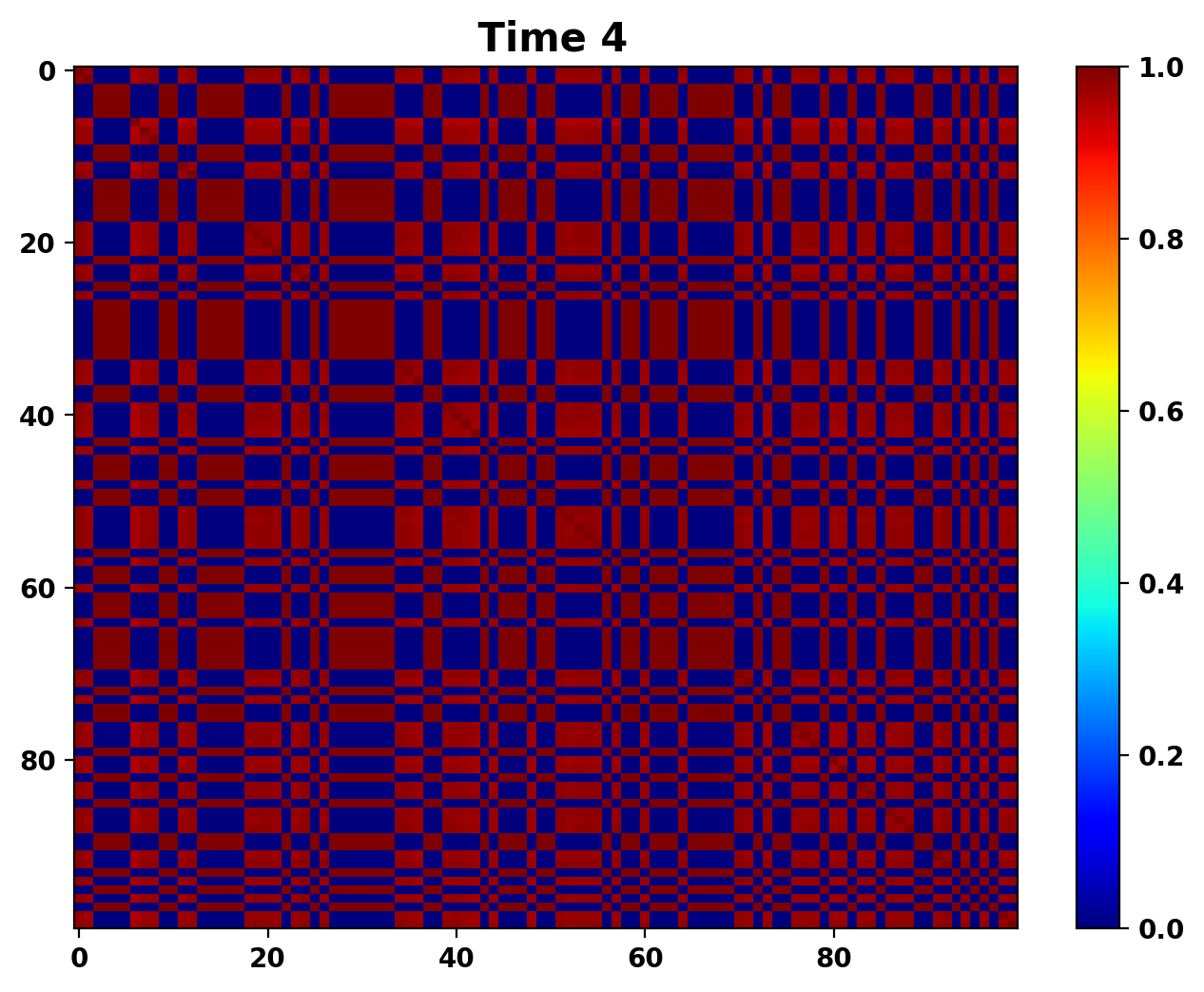}
\end{center}
\vspace{-1cm}
 \caption{Posterior co-clustering probability plots for simulation scenario 4 over time.}
 \label{fig:coclust4}
\end{figure}

We find that the posterior mean of $\psi$ is 0.267, while the marginal posterior distribution is mostly concentrated on positive values (see 
Figure~\ref{fig:simulated_psi}, second row second column),  i.e. we detect  positive correlation in the AR-process. 
Again, we are able to correctly detect the two clusters at time $t=1$; see the co-clustering probability plots in Figure~\ref{fig:coclust4}.  However, for $t=2,3,4$, the plots show more variation in agreement with data changing cluster with probability 0.5. 

\subsection*{Scenario 5}
\begin{figure}[H]
\begin{center}
\includegraphics[width=.45\textwidth]{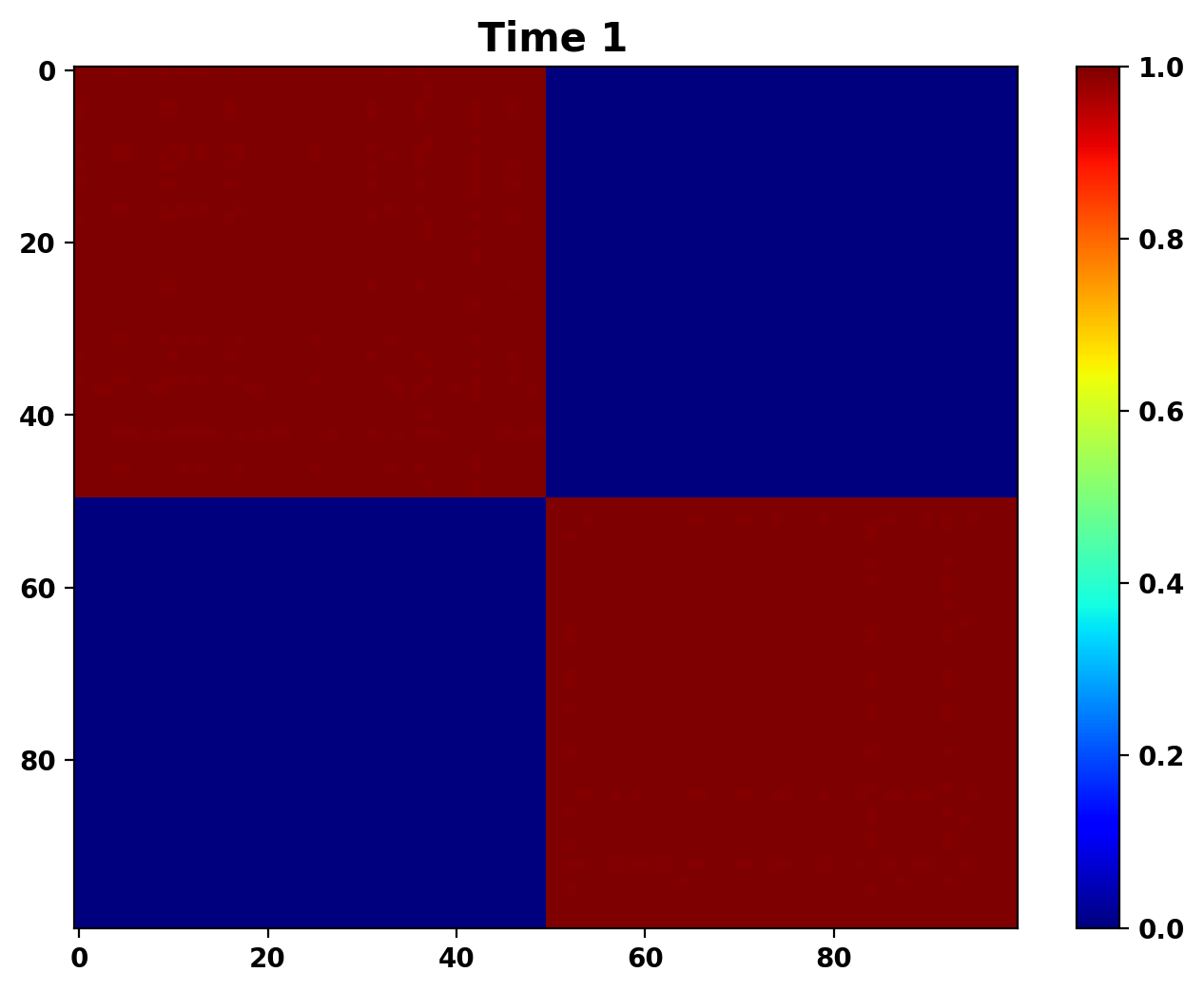}
\includegraphics[width=.45\textwidth]{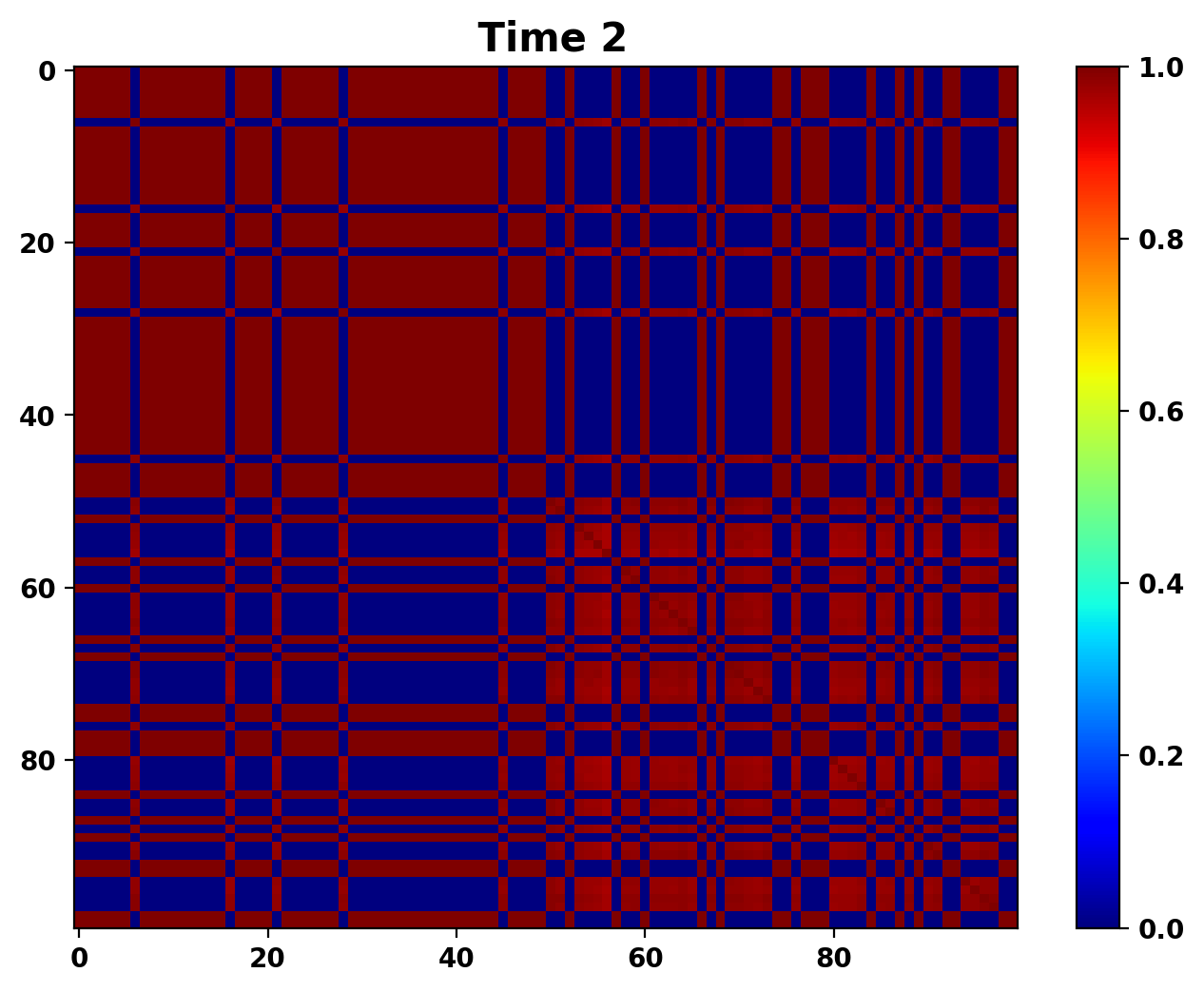}
\includegraphics[width=.45\textwidth]{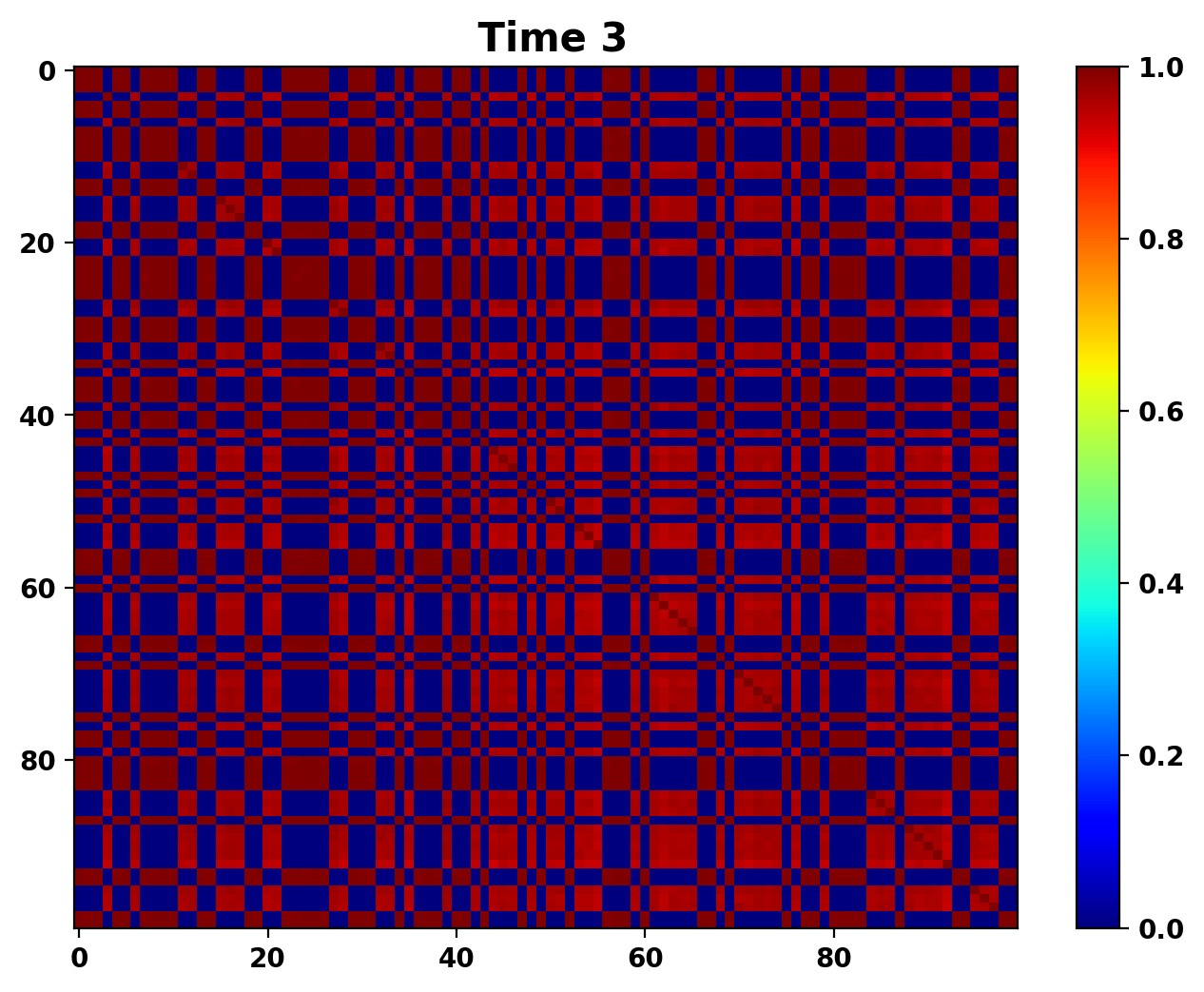}
\includegraphics[width=.45\textwidth]{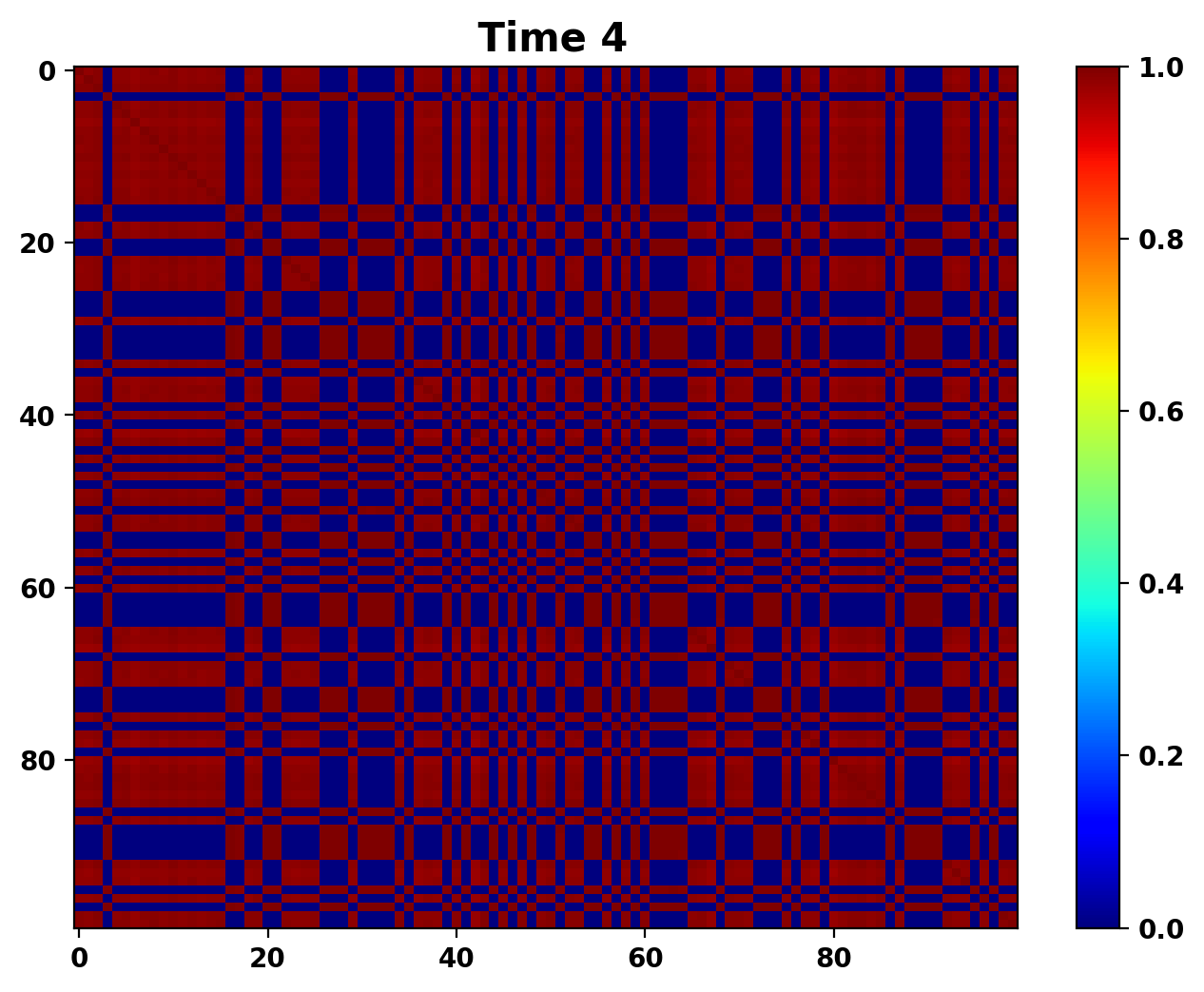}
\end{center}
\vspace{-1cm}
 \caption{Posterior co-clustering probability plots for simulation scenario 5 over time.}
 \label{fig:coclust5}
\end{figure}

This scenario differs from the previous one only because we increase the probability of remaining in the same cluster at time $t-1$ from 50\% to 
80\%. 
The posterior mean of $\psi$ is smaller than before (equals 0.134), and the  marginal posterior distribution  is shifted to the left (Figure~\ref{fig:simulated_psi}, third row first column). 
The co-clustering probability plots over time (Figure~\ref{fig:coclust5}) differ from those of scenario 4 mostly for the plot at time $t=2$, where 
the posterior probability of correctly identifying the true two clusters increases, since we have simulate data that remain in the same cluster with higher (.8) probability that in scenario 4 (.5).

\subsection*{Scenario 6}
Here and in the next scenario, we simulate data only for $t=1,2$, where the number of clusters changes with $t$. The true number of clusters is 1 at time $t=1$ and 2 at time $t=2$, while in the next example we have the opposite situation. Specifically,  at time $t = 1$, we simulate 100 observations from a single Gaussian distribution ($\text{N}(-80,1)$). For $t =2$, the unique cluster is separated into two clusters, that is  50 observations are generated from a Gaussian distribution $\text{N}(-40,1)$) and the other 50 are generated from a  Gaussian distribution $\text{N}(40,1)$. 

\begin{figure}[H]
\begin{center}
\includegraphics[width=.45\textwidth]{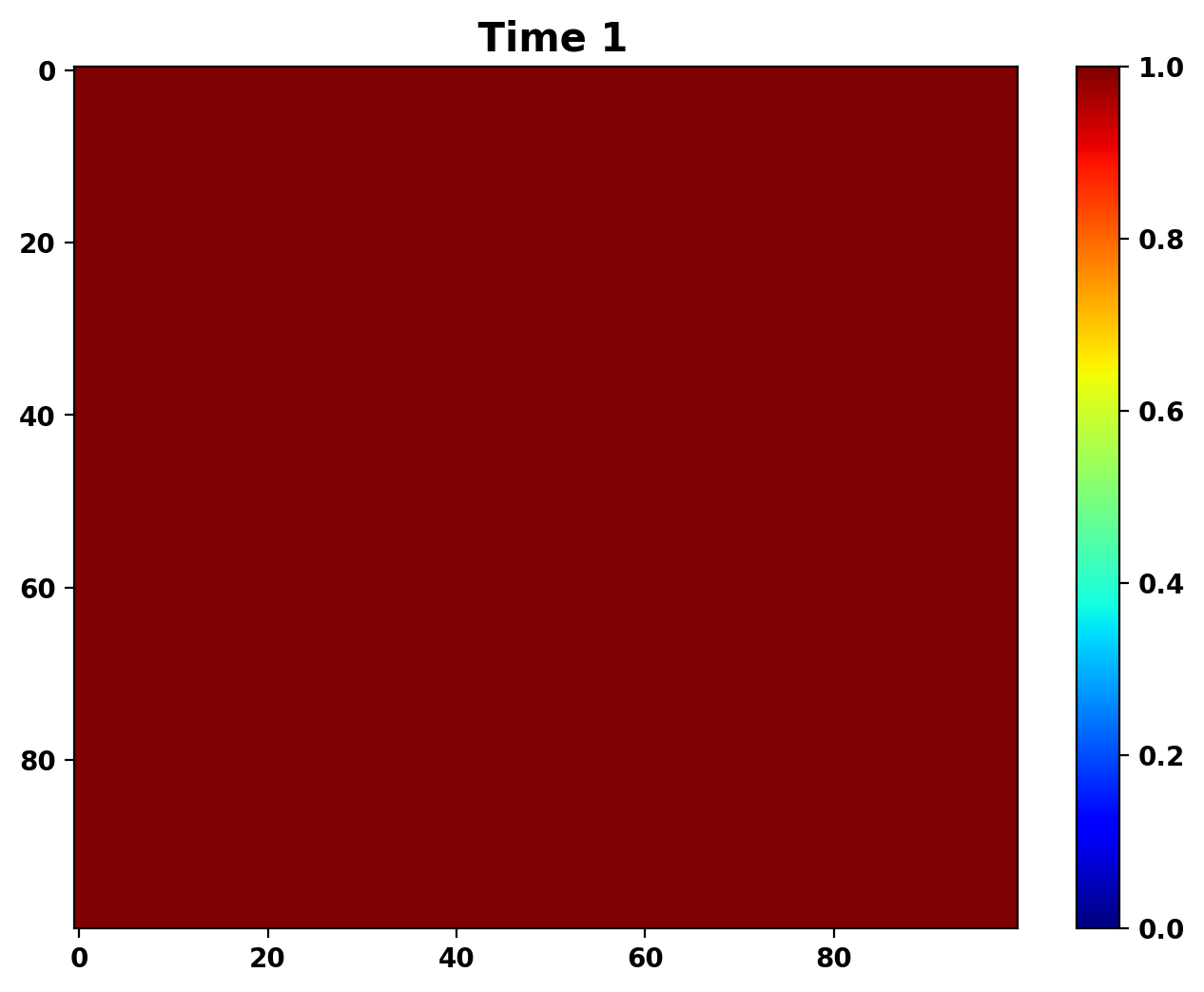}
\includegraphics[width=.45\textwidth]{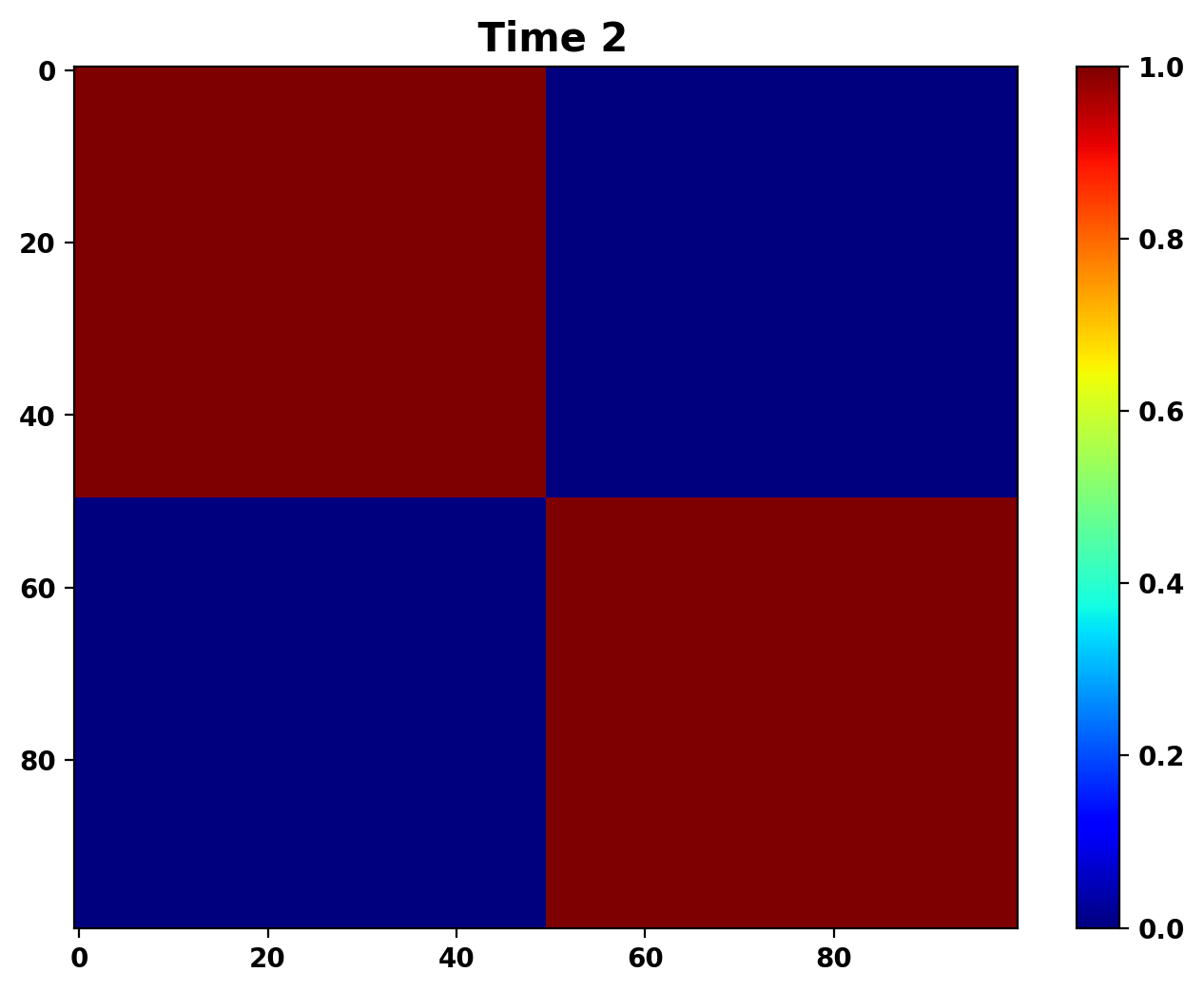}
\end{center}
\vspace{-1cm}
 \caption{Posterior co-clustering probability plots for simulation scenario 6 at times $t=1,2$.}
 \label{fig:coclust6}
\end{figure}

In this case, we find that the posterior mean of $\psi$ is -0.734, and the marginal posterior distribution is mostly concentrated on negative values (see Figure~\ref{fig:simulated_psi}, third row second column). 
From the co-clustering probability plots over time  in Figure~\ref{fig:coclust6}, it is clear that our model is able to detect the change of clusters over time.  

\subsection*{Scenario 7}
At time $t = 1$, we generate  50 observations from a Gaussian distribution $\text{N}(-40,1)$ and the other 50  from a different Gaussian $\text{N}(40,1)$). Instead, at time $t =2$, the two clusters are merged into a single one, and we simulate all 100 observations from a Gaussian distribution $\text{N}(-80,1)$). 

\begin{figure}[H]
\begin{center}
\includegraphics[width=.45\textwidth]{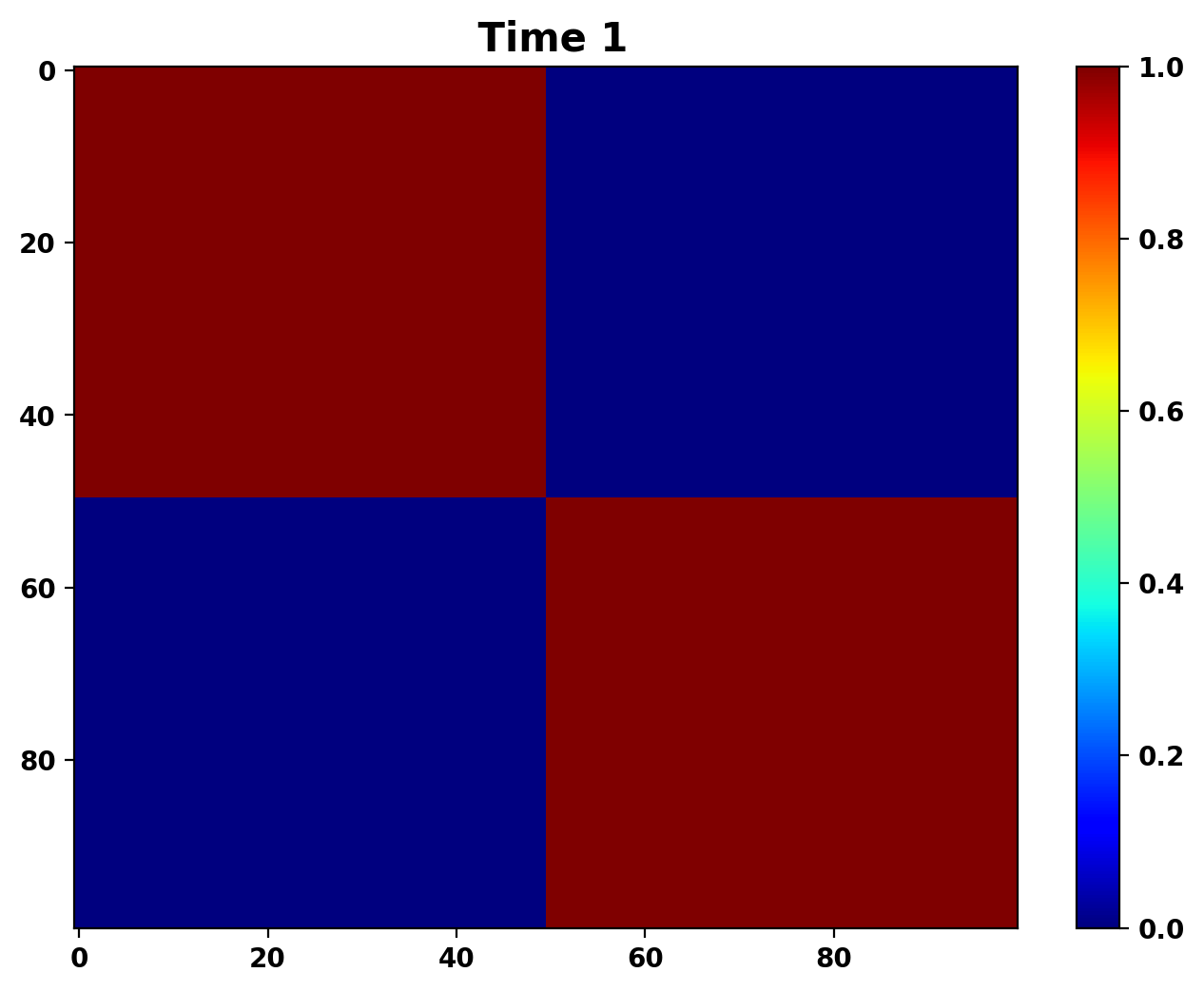}
\includegraphics[width=.45\textwidth]{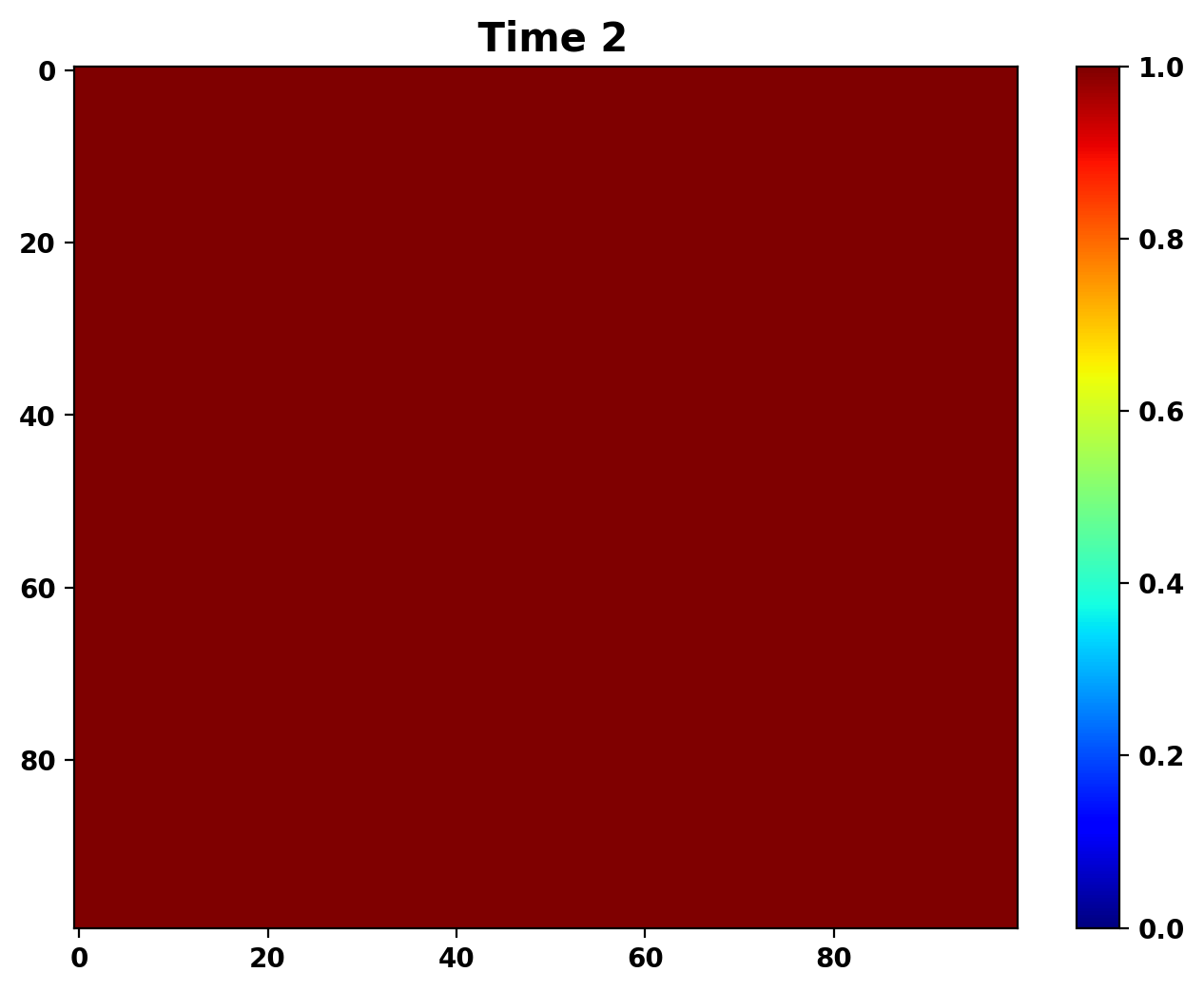}
\end{center}
\vspace{-1cm}
 \caption{Posterior co-clustering probability plots for simulation scenario 7 at times $t=1,2$.}
 \label{fig:coclust7}
\end{figure}

The posterior mean of $\psi$ is -0.783, while the marginal posterior distribution is more shifted towards -1 than in scenario 6 (Figure~\ref{fig:simulated_psi}, fourth row). 
The change of clustering structure over time is correctly identified; see Figure~\ref{fig:coclust7}.  

\section*{APPENDIX D: Extra plots} 
\label{app:extraplots}

\begin{figure}[H]
\begin{center}
\includegraphics[width=.30\textwidth]{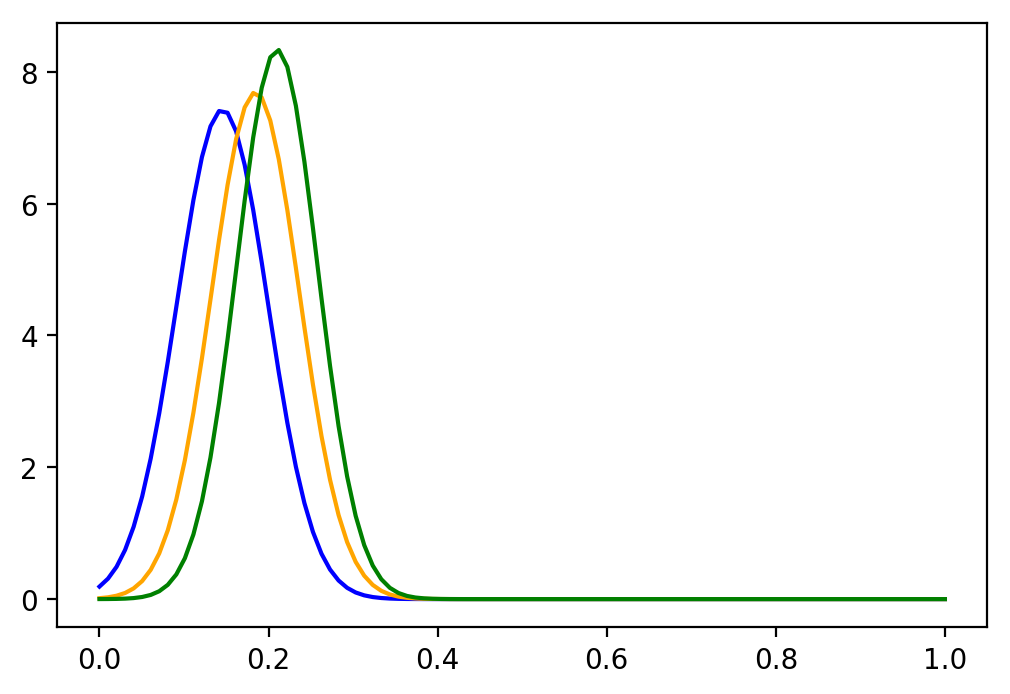}
\includegraphics[width=.30\textwidth]{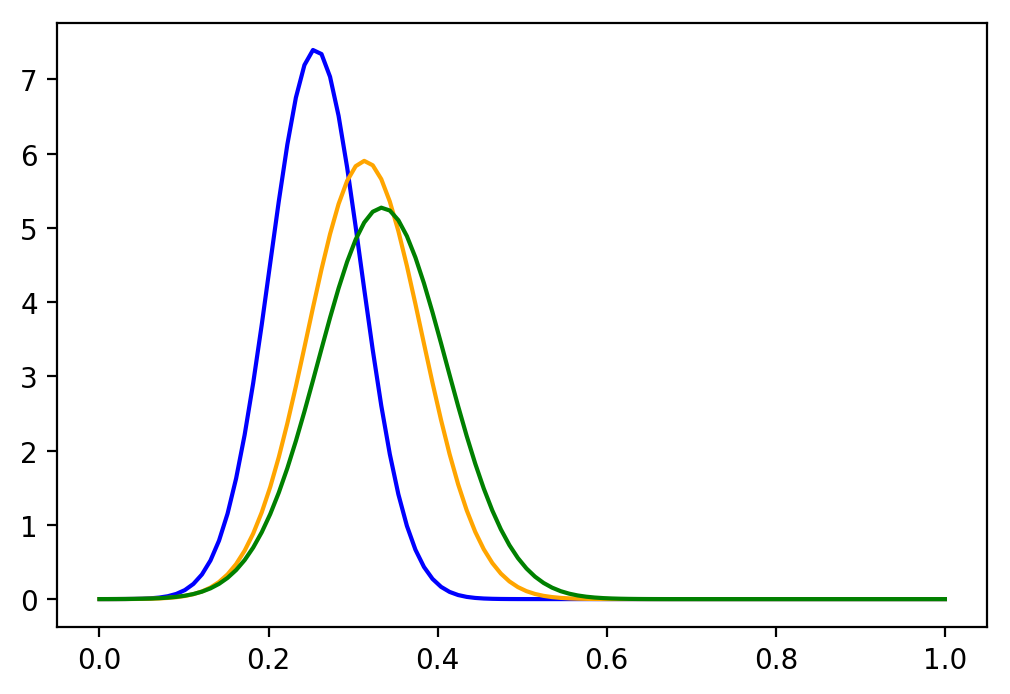}
\includegraphics[width=.30\textwidth]{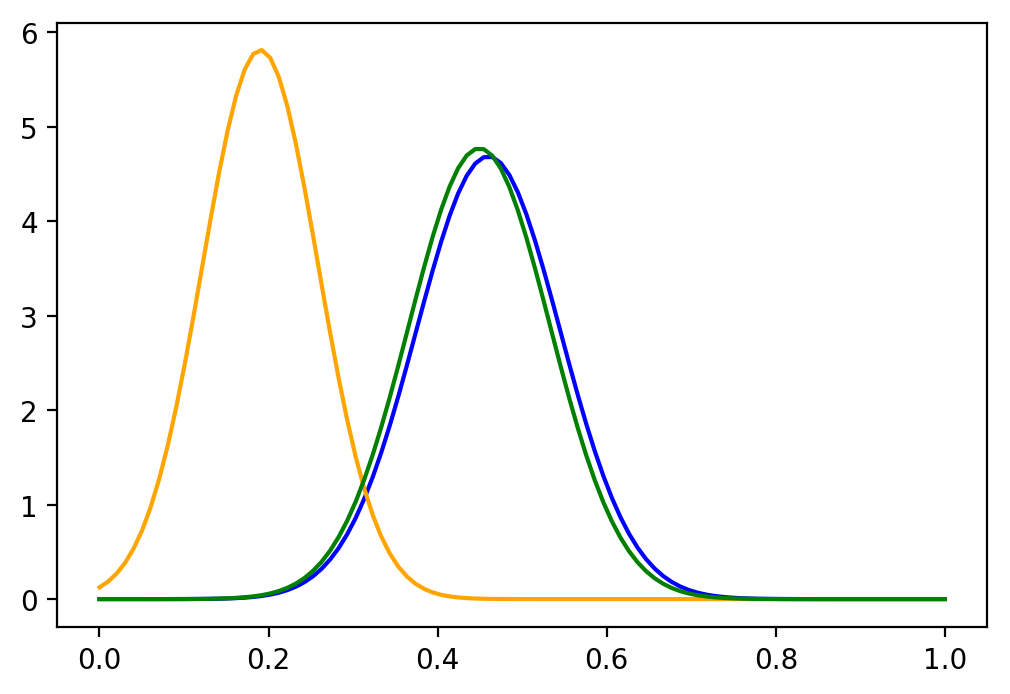}
\end{center}
\vspace{-0.5cm}
 \caption{{Hellinger distances between the density functions $f_{1}$ and $f_t$, for $t=2,3,4$, (blue line: $d_H(f_2,f_1)$, 
 orange line: $d_H(f_3,f_1)$, green line: $d_H(f_4,f_1)$) and for $\psi=0.9$ (left panel), $\psi=0.5$ (center panel) and $\psi=-0.9$ (right panel).}}
 \label{fig:hellinger}
\end{figure}

Figure \ref{fig:hellinger} shows the prior distribution of the Hellinger distances between  $f_t, t=2,3,4$ and $f_1$, defined as in \eqref{eq:rand_dens} as location mixtures of Gaussian kernels with variance equal to 1, for $\psi=-0.9, 05, 0.9$. We set $M=10$ and $G_0$ is the Uniform (-30,30) distribution, and we assume $J = 20$.  It is clear that if $\psi$ is positive (left and center panels), the Hellinger distance increases with $t$ (i.e. the support is pushed to higher values), and that this increase is smaller for large (positive) $\psi$. On the contrary, when $\psi$ is negative, the Hellinger distance shows a time-dependent behaviour which is what we expect from an AR1 process.

\begin{figure}[h]
\begin{center}
\includegraphics[width=.4\textwidth]{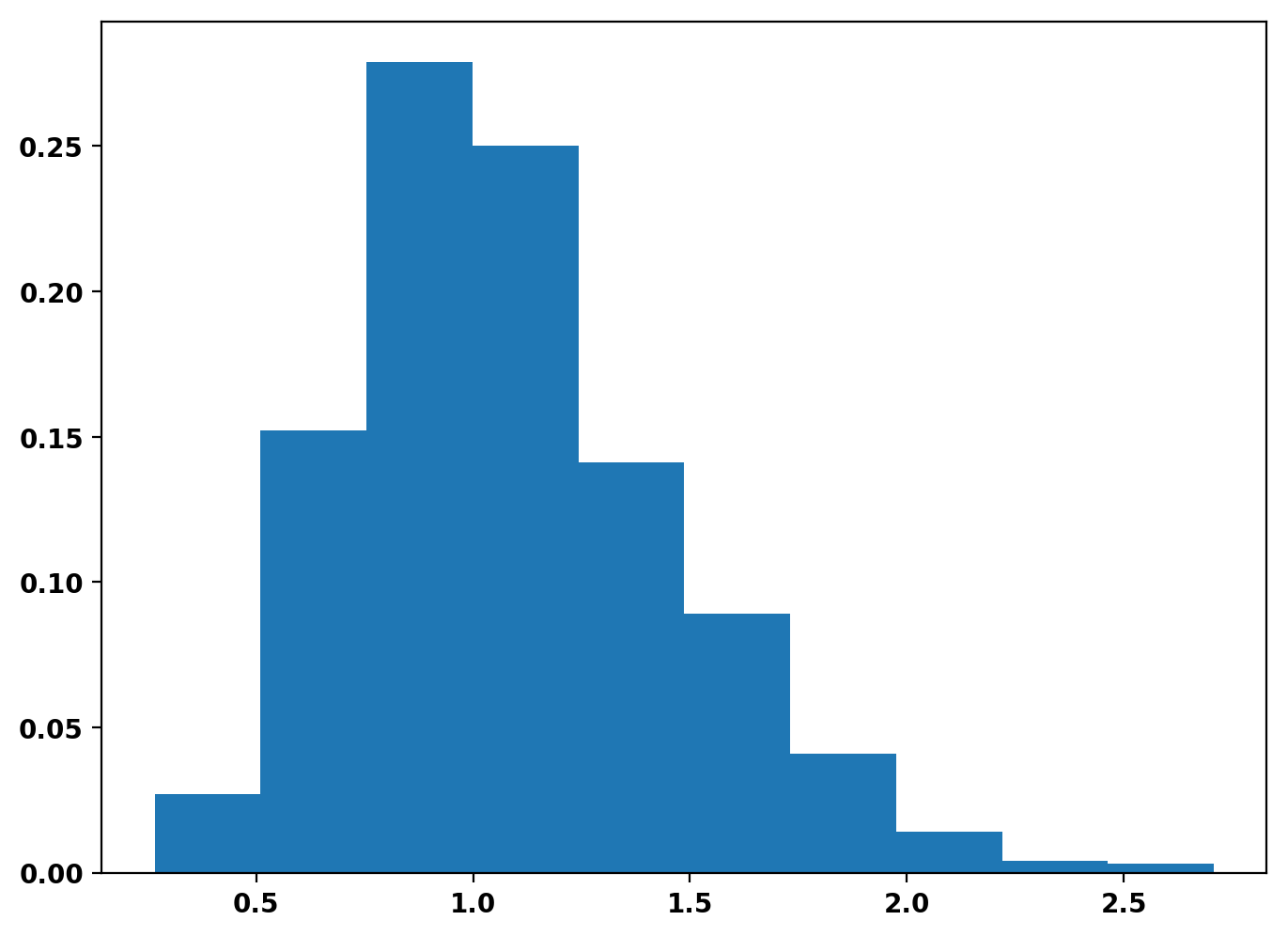}
\includegraphics[width=.4\textwidth]{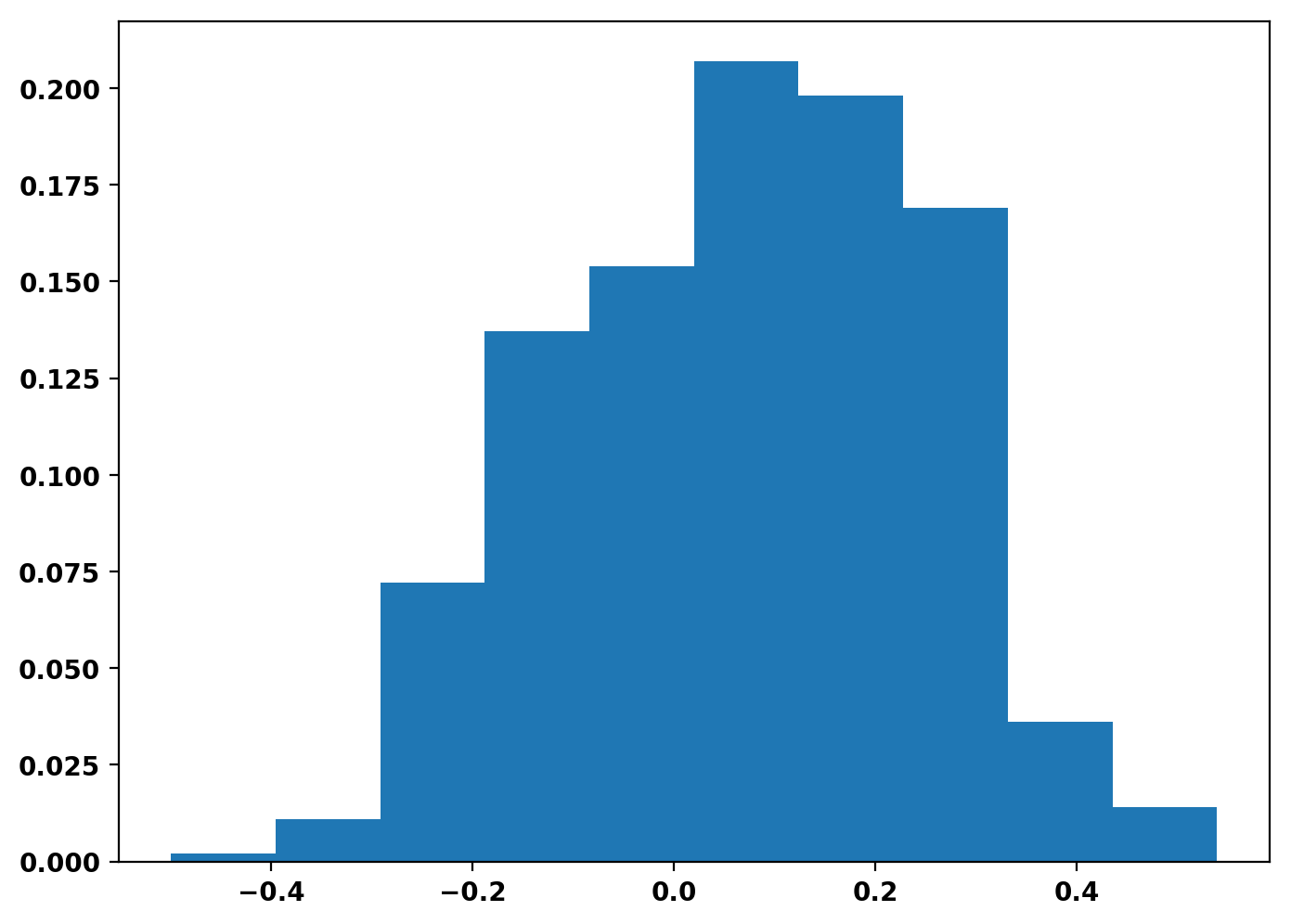}
\end{center}
\vspace{-0.5cm}
\caption{{Marginal posterior distributions of $M$ (left panel) and $\psi$ (right panel) for adjectives bias data.} }
\label{fig:adjectivebias_Mpsi}
\end{figure}

\begin{figure}[h]
\begin{center}
\includegraphics[width=.30\textwidth]{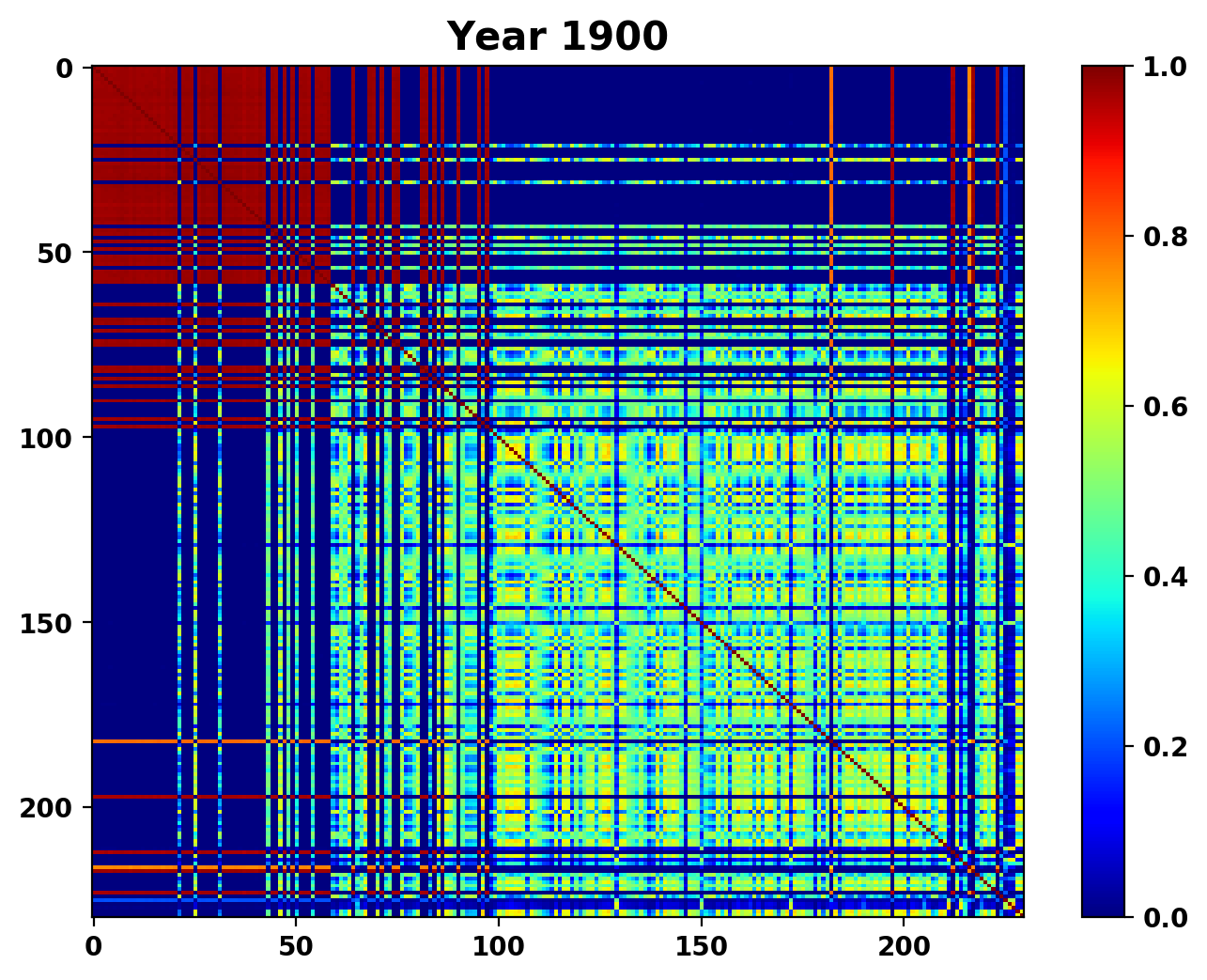}
\includegraphics[width=.30\textwidth]{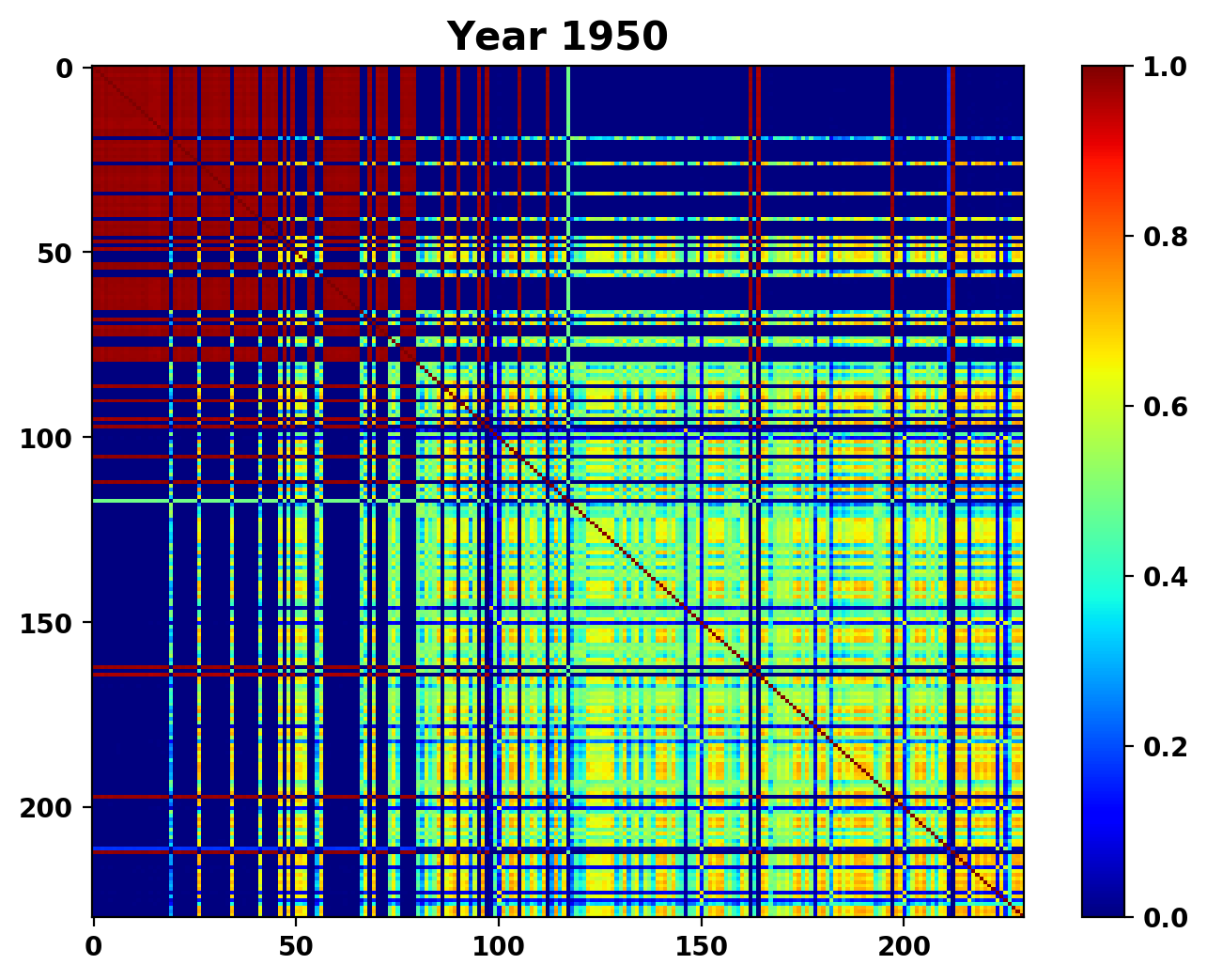}
\includegraphics[width=.30\textwidth]{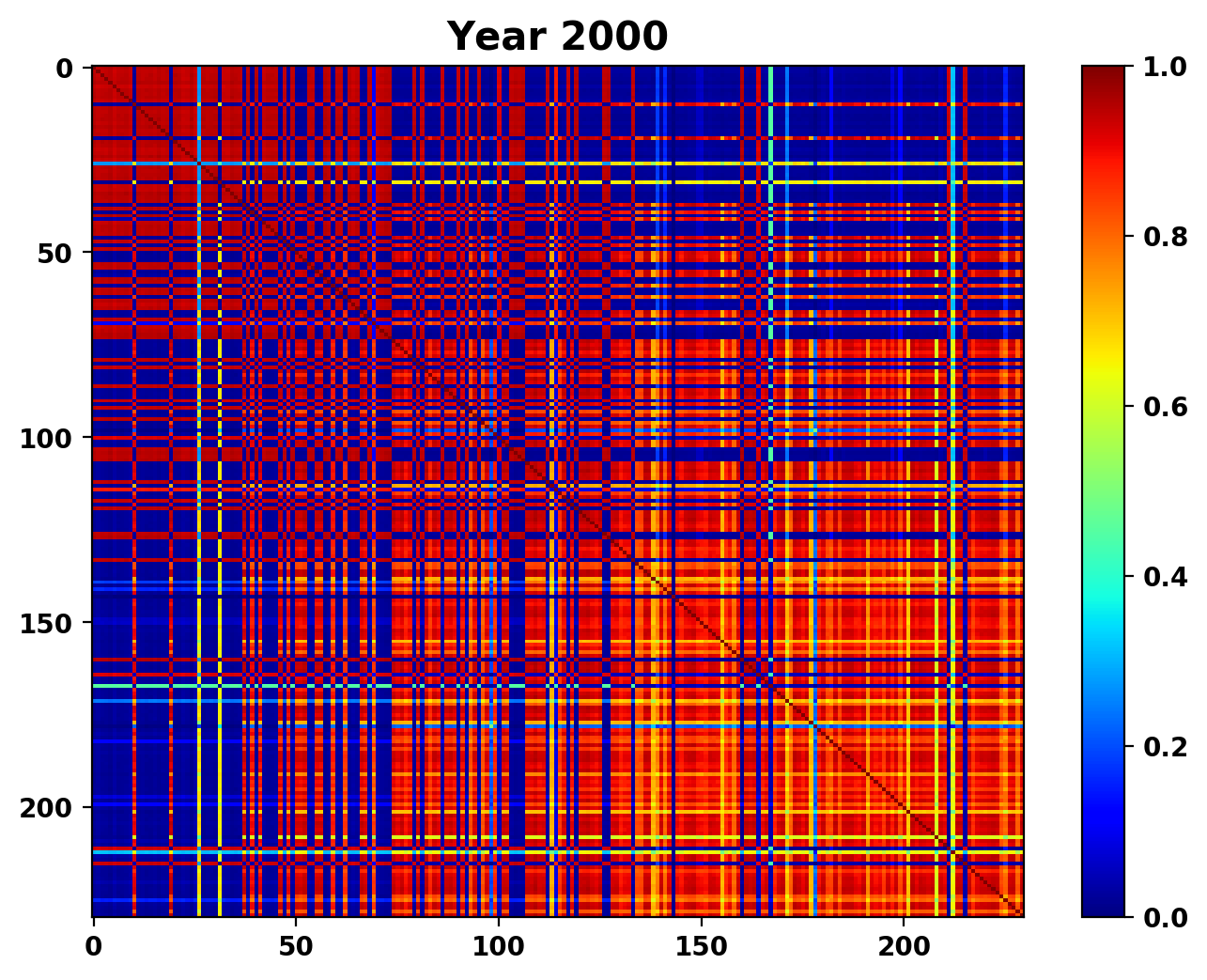}
\end{center}
\vspace{-0.5cm}
\caption{{Posterior co-clustering for adjectives bias data for $t=1990$ (left panel), $t=1950$ (center panel) and $t=2000$ (right panel).} }
\label{fig:adjectivebias_coclust}
\end{figure}

\bigskip
The right panel of Figure~\ref{fig:adjectivebias_Mpsi} shows the marginal posterior distribution of $\psi$, with posterior mean  equal to $0.072$, in the case of 
the adjective embedding bias data in Section \ref{sec:adjective}.
The posterior co-clustering probabilities in this case are shown in Figure~\ref{fig:adjectivebias_coclust}. The predominant colors are red, green and blue, and the left and middle panel (1900 and 1950) are very similar. Instead the right panel (2000) has very little green and most blue (low co-clustering probability) and red (high  co-clustering probability).

\end{document}